# Regulation of the Pho Regulon and Reconstruction of the Photosynthetic Apparatus under Phosphate Limitation in *Synechocystis* sp. PCC 6803

**Jacob Ryan Irby**

A thesis submitted for the degree of

Master of Science

University of Otago

Dunedin, New Zealand

**Abstract**

For millennia, cyanobacteria have evolved strategies for acclimatization to dynamic environments – this requires a molecular response system. Phosphate is essential for cellular integrity and metabolism, yet inorganic phosphate is often the limiting-nutrient for most aquatic environments. This study characterizes two periplasmic phosphate-binding proteins, namely SphX and SphZ, which are essential for the functional response of the SphS-SphR signal transduction system in *Synechocystis* sp. PCC 6803 and incorporates both into a model. SphX offers a competitive role so as to ensure inorganic phosphate is not depleted from the environment at a faster rate than is necessary for metabolism and, therefore, hinders luxury uptake; whereas, SphZ, encoded by *sll0540*, is the auxiliary sensor necessary for the SphS response under phosphate limitation in tandem with a protein complex association with the phosphate-specific transport 2 system and the negative regulator, SphU. Cross-regulation between the pho regulon and the photosynthetic apparatus is presented as well as a parameterized analysis for low-temperature fluorescence emission spectroscopy.

# Acknowledgements

Our species is equipped with the cognition to recognize that none of us chose this life, but we choose what we do with the limited time and actively shape reality to address problems so as to leave this world a little better than in which we had found it. I am grateful for the privilege to have continued my studies in New Zealand: life lessons were embedded throughout each day, for learning never truly stops. I will forever be indebted to Navient and Sallie Mae for this endeavor.

To my grandparents, I thank each of you for your patience and guidance throughout this life. To my mother and father, I thank you for the gift that is life; for the boundless love; and for your unwavering encouragement. To the extended family, I thank you for the courage and strength you show in everything that you do. To Tommy, Jordan, Angela, Kelsey, and Cassidy, I thank you for the laughs and pure joy with which you brighten every day. To Thomas and Krystal, I thank you for the pride in which fills my heart and wish the utmost happiness in this life as you follow your dreams. Persistence is key: even when it feels as if you cannot succeed, please hold your head up high and strive toward those dreams and purpose which drive you. In the words of Thomas Jefferson, "[be] bold in the pursuit of knowledge, never fearing to follow truth and reason to whatever results they [lead]." To my AVBrothers, I thank you the least because y'all suck. LYLAS.

To the JER 308 and Bannister lab members over the past 4 years, the Department of Botany and the Department of Biochemistry (Afreen Saeed, Alex Charlton, Andreas Joe, Andrew Douglas, Andy Nilsen, Beatrice Koh, Bharat Kumar, Carlo Luis Lee, Cédric Viry, Chandra Rodriguez, Claudia Rossig, Cornelius Fischer, David Burritt, David Orlovich, Devi Ratna Asih, Ei Phyo Khaing, Elisa Saager, Faiza Arshad, Felix Tosend, Grace Lim, Hannah Heynderickx, Jack Forsman, Jackie Daniels, James Gorrie, Janice Lord, Jasmine Divinagracia, Jaz Morris, Josh McCluskey, Jullieta Bohórquez, Katharine Dickinson, Katherine van der Vliet, Katja Schweikert, Kevin Sheridan, Laura van Galen, Layne Kay, Liam Le Lievre, Liangliang Hui, Linn Hoffmann, Lisa Biiri, Matthew Larcombe, Merissa Strawsine, Miguel Desmarais, Pamela Cornes, Paul Guy, Peixi Choo, Qinsi Niu, Rebecca Macdonald, Ro Allen, Sam Gifkins, Sandeep Biswas, Shannon White, Susan Mackenzie, Toni Renalson, Victor Zhong, Ziva Louisson), I thank each of you for the laughs, the tears, the help, the guidance, and the excitement that you brought into the collective space.

To Tina and Julian: I thank you for your thought-provoking insight; I thank you for your breadth of knowledge; I thank you for holding the space for me to make mistakes and then holding me accountable; I thank you for broadening my perspective; I thank you for the many lessons you have imprinted; and, most importantly, I thank you for your time. It truly was an honor and privilege to work alongside both of you in this life. I will never forget the experience. Thank you.

softly as i leave you

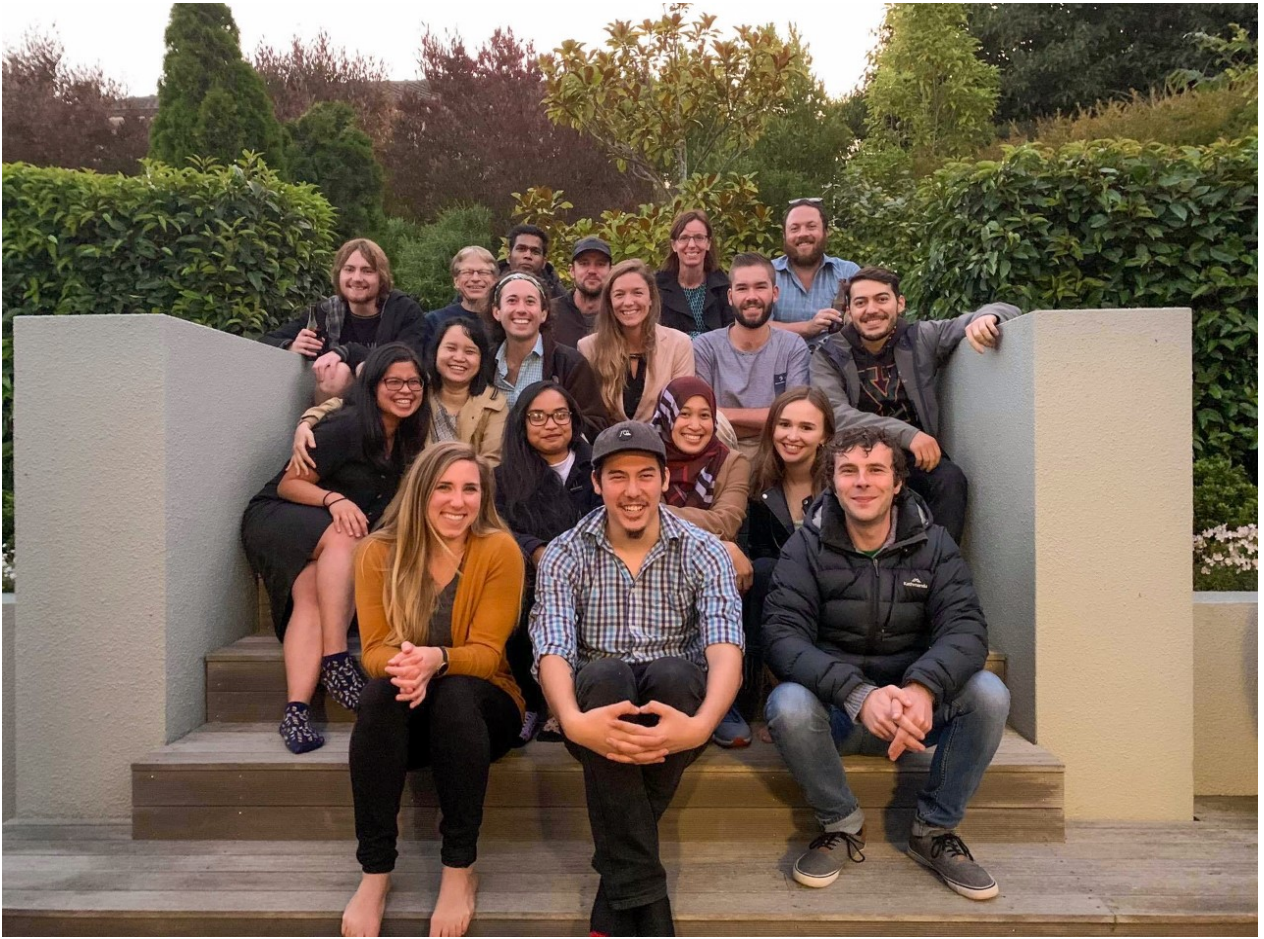

# Table of Contents

# List of Figures

# List of Tables

# Abbreviations

| | |
|---|---|
| ABC | ATP-Binding Cassette |
| ADP | Adenosine Diphosphate |
| APC | Allophycocyanin |
| ATP | Adenosine Triphosphate |
| $A_{Xnm}$ | Absorbance at X nm |
| BN | Blue Native |
| Chl | Chlorophyll |
| Cyt | Cytochrome |
| $e^-$ | Electron |
| E | Einstein, one mole of Photons |
| ETC | Electron Transport Chain |
| Fd | Ferredoxin |
| FNR | Ferredoxin:$NADP^+$ Oxidoreductase |
| g | Grams or Gravity |
| h | Hours |
| $H^+$ | Proton |
| K | Kelvin |
| Kb | Kilobase Pair |
| min | Minute |
| NADP | Nicotinamide Adenine Dinucleotide Phosphate |
| $OD_{Xnm}$ | Optical Density At X Nm |
| OEC | Oxygen Evolving Complex |
| $P_{680}$ | Chlorophyll Cluster embedded in PS II RC |
| $P_{700}$ | Chlorophyll Cluster embedded in PS I RC |
| PAGE | Polyacrylamide Gel Electrophoresis |
| PAS | Per-Arnt-Sim |
| PBS | Phycobilisome |
| PC | Phycocyanin or Plastocyanin |
| PCC | Pasteur Culture Collection |
| PCR | Polymerase Chain Reaction |
| PDB | Protein Data Bank |
| Pheo | Pheophytin |
| $P_i$ | Inorganic Phosphate |
| polyPs | Inorganic Polyphosphates or Polyphosphate Storage Bodies |
| PQ | Plastoquinone |
| $PQH_2$ | Plastoquinol |
| PS I | Photosystem I |
| PS II | Photosystem II |
| Pst | Phosphate-Specific Transporter |
| PVDF | Polyvinylidene Fluoride |
| RC | Reaction Center |
| Rpm | Revolutions Per Minute |
| SDS | Sodium Dodecyl Sulfate |
| *Synechocystis* 6803 | *Synechocystis* Sp. PCC 6803 |
| TE | Terminal Emitters |
| Tris | Tris(Hydroxymthyl)Aminomethane |
| v/v | Volume per Volume |
| w/v | Weight per Volume |

# Chapter One: Introduction

## 1.1 Context of Earth Sustainability

The earth is considered a closed system by which stochastic interactions contribute to material being cycled between the lithosphere, atmosphere, hydrosphere, and biosphere. Every closed system reaches a capacity by which resources are depleted at a rate in which equilibria cannot be maintained at a steady state. However, there is a constant influx of energy from a nearby star, which emits electromagnetic radiation produced via the proton-proton chain reaction. Visible light comprises only a minute portion of the electromagnetic spectrum, which is classified by wavelength, and consists of quanta, or discrete quantities of energy, called photons. Since matter can neither be created nor destroyed, Earth is sustained through a process by which energy is transferred from photons to a chemical form – this process is called photosynthesis – and, due to this process, the natural world has maintained adequate resources for energetic transference between the trophic levels. Yet, photosynthesis is reliant upon nutrients supplied from the extracellular environment and an organism's ability to acclimate to dynamic environments. Photosynthetic bacteria offer the simplest context for such a negative feedback loop.

**1.2 Cyanobacteria and Environmental Impacts**

Cyanobacteria are gram-negative photosynthetic prokaryotes and are among the oldest life forms on Earth. Cyanobacteria are credited with the rise of oxygenic photosynthesis, up to 2.4 Ga during the early Paleoproterozoic Era, transforming the Earth's atmosphere (Catling and Claire, 2005; Kopp *et al.*, 2005). This period of increasing atmospheric and oceanic oxygen concentrations is referred to as the Great Oxidation Event, which ultimately led to the present composition of Earth's atmosphere (Bekker *et al.*, 2004). The global supply of oxygen relies heavily on cyanobacteria, particularly those in open oceans (Field *et al.*, 1998; Percival and Williams, 2014). In addition, these microscopic organisms form the base of food webs, providing energy to higher trophic levels within an ecosystem (Iturriaga and Mitchell, 1986; Burkill *et al.*, 1993). This ancient and diverse bacterial phylum inhabits a range of environments from aquatic to terrestrial rock and soil to extreme habitats, such as hot springs, frozen lakes, and hypersaline waters (Whitton, 1992; Whitton and Potts, 2012). Cyanobacteria have evolved strategies to enhance their fitness, for example the ability to rise and sink in the water column by varying the amount of air within the cell or scavenging for nutrients such as inorganic phosphate (Grossman *et al.*, 1994; Brookes *et al.*, 1999; Schwarz and Forchhammer, 2005; Su *et al.*, 2007; Santos-Beneit, 2015). However, the rapid growth of cyanobacteria may form unwanted "algal blooms" within nutrient-rich aquatic environments, causing adverse effects.

Cyanobacterial harmful blooms are dense accumulations of cyanobacteria within aquatic systems and are commonly visible as surface water discoloration (Heisler *et al.*, 2008). These blooms usually form due to an excess of nutrients, particularly phosphates, and decrease the potential of sunlight to penetrate the water column to reach underlying aquatic vegetation, directly impeding their photosynthetic capability. As a result, the trophic structure may be altered as organisms compete for available resources and the function composed of associations within the ecosystem may be changed (Havens, 2008). Moreover, each cyanobacterium is short-lived, so the concentration of dead organic matter increases. Dissolved oxygen is consumed during the decomposition of organic matter, causing hypoxic conditions and leading to dead zones where aerobic life is unsustainable (Robarts *et al*., 2005). Studies have indicated that the intensity and frequency of cyanobacterial blooms are likely to increase due to anthropogenic change (Davis and Koop, 2006; Paerl and Huisman, 2008; Davis *et al.*, 2009; O'neil *et al.*, 2012; Michalak *et al*., 2013; Paerl and Otten, 2013; Taranu *et al*., 2015; Huisman *et al.*, 2018; Ullah *et al*., 2018). Importantly, some of these bloom-forming cyanobacterial species produce toxins which impact not only the inhabitants within aquatic ecosystems but also terrestrial organisms.

Cyanotoxins are secondary metabolites produced by several species of cyanobacteria. The role of cyanotoxins within cyanobacterial blooms may be to impede predation or could function as signaling compounds within the community (DeMott *et al.*, 1991; Codd, 1995; Penn *et al.*, 2014; Rastogi *et al.*, 2014; Huisman *et al.*, 2018). As stated, some of these by-products have quite damaging effects on aquatic life; however, their effects on mammalian health have drawn the most attention (Carmichael *et al.*, 2001; Azevedo *et al.*, 2002; Funari and Testai, 2008). Two well-known classes of these cyanotoxins are anatoxin-a and microcystins which target the nervous system and liver, respectively (Carmichael, 1992; Beltran and Neilan, 2000). Anatoxin-a is commonly referred to as the "very fast death factor" due to respiratory arrest within minutes or hours; whereas, the symptoms of microcystins are latent, as they target the liver, with death occurring within a few days (Carmichael, 1992). Recreational activities within lakes during these cyanobacterial harmful blooms have resulted in death for a range of vertebrates after ingestion and, while the water after filtration should not be lethal to humans, reports have been made of illness due to the consumption of drinking water (Falconer and Buckley, 1989; Falconer, 1991; Carmichael, 1992; Codd, 1995; Carmichael, 2001; Hudnell, 2010). Due to these adverse effects, it is clear that these bloom-forming cyanobacterial species need to be regulated and monitored (Backer, 2002; Funari and Testai, 2008; Delpla *et al.*, 2009; Hudnell, 2010; Huisman *et al.*, 2018). It was proposed that by setting bioavailable nutrients, particularly phosphate, below a given threshold would be the most effective way to circumvent these blooming-forming conditions (Chambers *et al.*, 2012).

**1.3 Inorganic Phosphate and Regulatory Role**

Phosphorus is an essential element required for life on Earth, yet it is rarely found on its own due to phosphorus readily reacting with oxygen. Its fully oxidised form, phosphate, is a water-soluble inorganic salt and is considered a limiting nutrient for most ecosystems (Schindler, 1977; Schweitzer and Simon, 1995; Thingstad *et al.*, 1998; Correll, 1999; Hudson *et al.*, 2000; Wu *et al.*, 2000). In its free form, the phosphate ion is characterized with a central phosphorus atom surrounded by four oxygen atoms in a tetrahedral arrangement and denoted with the empirical formula $(PO_4)^{3-}$. In nature, the phosphate ion also binds with hydrogen as two species, namely $H_2PO_4^-$ and $HPO_4^{2-}$, in solution with a pH ranging from four to nine. Phosphate is needed to produce biomolecules, such as complex carbohydrates, membrane lipids, nucleotides, and nucleic acids. Therefore, its bioavailability can limit the productivity of producers, such as cyanobacteria, even though other nutrients are readily available (Su *et al.*, 2007). Phosphate can be stable on its own, monophosphate, but also has the ability to bind to another phosphate ion, pyrophosphate, or create phosphate ion chains, polyphosphate. These various forms are referred to as inorganic phosphate ($P_i$).

Though there is an abundance of phosphates found throughout environments, inorganic phosphates are the only bioavailable forms (Wanner, 1996; Wu *et al.*, 2000; Sundareshwar *et al.*, 2003; Su *et al.*, 2007). Since cells possess an intricate protective coating usually comprised of a lipopolysaccharide layer, S-layer, outer membrane and peptidoglycan layer, cyanobacteria have developed a process by which $P_i$ uptake is optimized through this compartmentalization (Hoiczyk and Hansel, 2000; Mohamed *et al.*, 2005). Monophosphates readily enter the cell through transporter proteins; whereas, chains of phosphate ions must be separated by an enzyme called alkaline phosphatase by which $P_i$ moieties are cleaved from polyphosphate chains and then enter the cell through phosphate-specific transport systems (Hirani *et al.*, 2001; Morohoshi *et al.*, 2002; Suzuki *et al.*, 2004; Pitt *et al.*, 2010). This process happens in the gel-

**5**

like matrix space between the inner membrane and peptidoglycan layer and is referred to as the periplasmic space. In a similar process, organophosphates, organic compounds containing phosphorus, pass the outer membrane via a porin protein and enter the periplasmic space where compounds are then broken down into chemical constituents by alkaline phosphatase or other periplasmic phosphatases (Tiwari *et al*., 2015). Once inside the cells, inorganic phosphate can be utilized for cellular metabolism or stored as linear polymers referred to as polyphosphate storage bodies or inorganic polyphosphates (polyPs), linked with high-energy phosphoanhydride bonds, and primarily neutralized by $Mg^{2+}$, $Ca^{2+}$, and $K^+$ cations (van Groenestijn *et al.*, 1988; Voronkov and Sinetova, 2019).

In biological systems, inorganic phosphate is crucial for cellular integrity due to its role of phosphorylation, the attachment of a phosphoryl group to molecules, to aid protein function. For example, the attachment of a phosphoryl group to a specific amino acid residue of a transcription factor causes the upregulation of specific genes. Since concentrations of $P_i$ in aquatic systems vary between the nanomolar to picomolar ranges and limiting for these environments, organisms have evolved scavenging strategies (Wu *et al.*, 2000; Sundareshwar *et al.*, 2003; Thingstad *et al.*, 2005, Su *et al.*, 2007).

## 1.4 Two-component Signal Transduction System

Bacteria monitor external conditions in order to adapt to changing environments via two-component signal transduction (Parkinson and Kofoid, 1992). Examples of these environmental parameters include pH, temperature, and nutrient concentrations (Murata and Suzuki, 2005; Schwarz and Forchhammer, 2005). Two-component signal transduction systems usually consist of a histidine kinase and a response regulator (Parkinson and Kofoid, 1992). In the simplest context, a histidine residue of the sensory protein, the histidine kinase, is phosphorylated due to an environmental stimulus (Figure 1.1). The phosphoryl group is then transferred to an aspartic acid residue on the cognate response regulator, a transcription factor for the upregulation of certain genes, ultimately giving rise to a molecular and physiological response (Figure 1.1).

The histidine kinase usually detects environmental fluctuations though its N-terminus input domain (Figure 1.1). This input domain is usually comprised of hydrophobic membrane-spanning domains which are separated by a variable periplasmic domain. The functionality of this protein comes from the C-terminus transmitter domain, which usually lies within the cytoplasm and interacts with the receiver domain of the cognate response regulator. C-terminal transmitter domains are quite conserved and may be identified by primary sequence motifs which are designated the H, N, G1, F, and G2 boxes (Figure 1.1; Parkinson and Kofoid, 1992). The H motif and surrounding residues are involved in autophosphorylation activity, forming the dimerization subdomain (Figure 1.1; Parkinson and Kofoid, 1992). The specificity of phosphotransfer from the histidine kinase to the cognate response regulator is controlled by a cluster of residues within this dimerization subdomain (Figure 1.1; Parkinson and Kofoid, 1992; Skerker *et al.*, 2008) The phosphorylated response regulator is then able to bind to specific sequences on the DNA to activate or repress the transcription of genes. As the first cyanobacterial genome to be sequenced, *Synechocystis* sp. PCC 6803 has provided insight in the field of molecular biology, particularly regarding two-component signal transduction (Kaneko *et al*., 1996; Mizuno *et al*., 1996). In particular, prokaryotes use a specialized response strategy to increase the bioavailability of inorganic phosphate by upregulating the expression of transport systems and phosphatases – these genes comprise the phosphate regulon, which is under the control of a two-component signal transduction system.

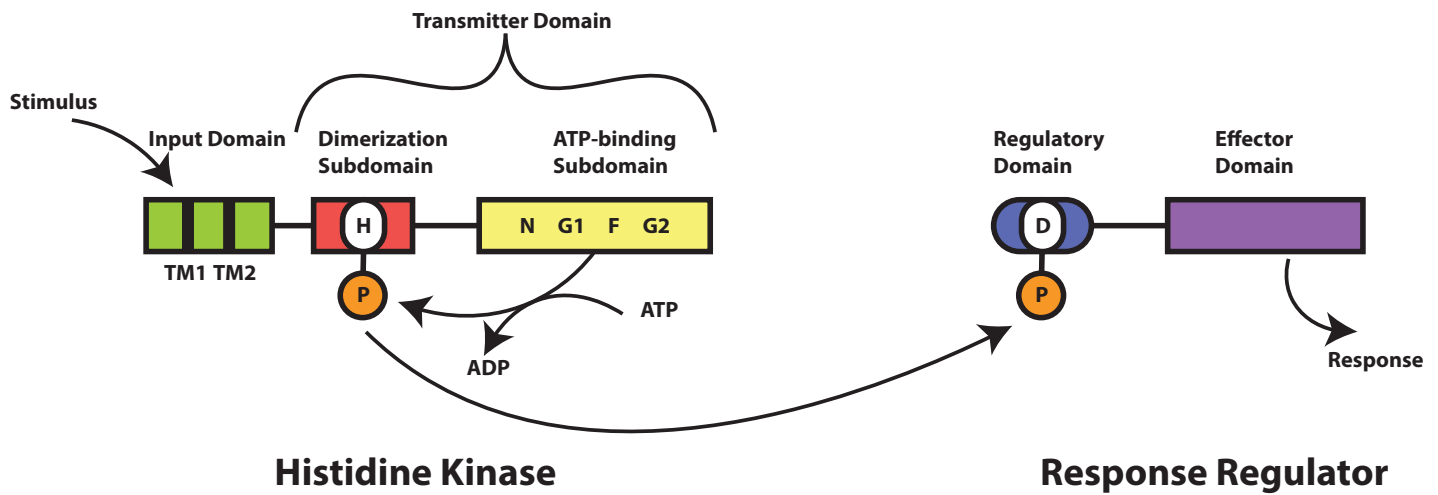

**Figure 1.1 Schematic Representation of Two-Component Signal Transduction System.** The histidine kinase (left) is comprised of an input and transmitter domain and anchored to the inner membrane via hydrophobic membrane-spanning domains (TM1; TM2). The stimulus is received by the input domain (green) and a histidine residue (H) located in the dimerization subdomain (red) is autophosphorylated (orange) due to the ATP-binding subdomain (yellow). The phosphoryl group (orange) is then transferred to an aspartic acid residue (D) in the regulatory domain (blue) of the response regulator (right). A specific molecular response is then elicited from the effector domain (purple). Adapted from Parkinson and Kofoid (1992).

## 1.5 *Synechocystis* sp. PCC 6803

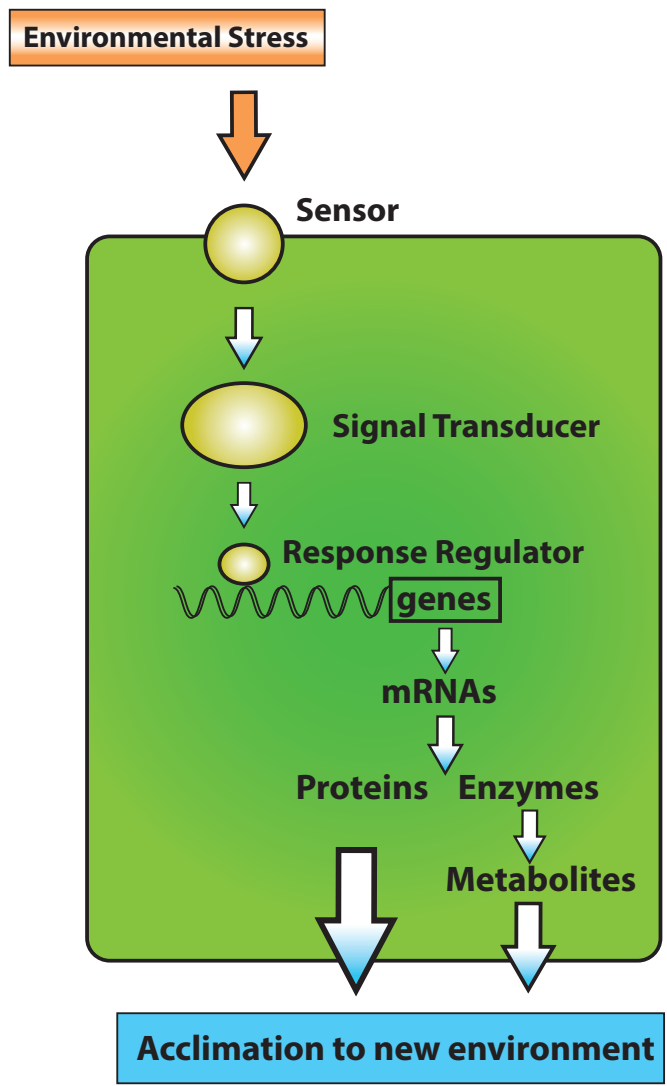

*Synechocystis* sp. PCC 6803 (hereafter *Synechocystis* 6803) has been used as a model organism for various metabolic studies for decades and has contributed to the understanding of photosynthesis and bacterial responses (Figure 1.2). This species is easily cultured and has the ability to uptake foreign DNA which is then incorporated into its own genome via homologous recombination as well as established targeted mutagenesis approaches (Grigorieva and Shestakov, 1982; Williams, 1988; Vermaas, 1998; Ikeuchi and Tabata, 2001; Yoshihara *et al.*, 2001). During short periods of starvation, *Synechocystis* 6803 accesses phosphate from reserves of polyPs, while the onset of $P_i$-limiting conditions result in the synthesis of an alkaline phosphatase as well as an increased capacity for phosphate uptake via phosphate-specific transport systems (Allen 1984; Grillo and Gibson, 1979; Ray *et al.*, 1991; Moore *et al.*, 2005). Additionally, *Synechocystis* 6803 reduces the amount of absorbed photosynthetically active radiation, range of visible light wavelength acquired for photosynthesis, which in turn protects the cells from photodamage (Collier and Grossman, 1992; Grossman *et al.*, 2003). *Synechocystis* 6803 has 42 histidine kinases and 38 response regulators: the identification of both proteins responsible for two-component signal transduction in phosphate-limiting conditions and the development of assays for the analysis of this pathway have proven useful for understanding its molecular mechanism (Ray *et al.*, 1991; Mizuno *et al.*, 1996; Hirani *et al.*, 2001).

**Figure 1.2 Generalized Response of Cyanobacteria to Environmental Stress.** Environmental changes are detected via a sensor protein. Information is then relayed to a signal transducer. Genes are regulated via binding of a response regulator due to transduction. The response alters mRNA transcript levels and, by extension, functional proteins in order to acclimate to a dynamic environment. Adapted from Los *et al.* (2008).

### 1.5.1 The SphS-SphR Two-Component Signal Transduction System

During phosphate-limiting conditions, *Synechocystis* 6803 responds via a two-component signal transduction system comprised of the histidine kinase, SphS, and a transcription factor, SphR (Figure 1.3; Hirani *et al.*, 2001; Suzuki *et al.*, 2004; Juntarajumnong *et al.*, 2007). SphS is localized to the inner membrane via a hydrophobic membrane-associated domain at the N-terminus (Figure 1.3; Burut-Archanai *et al.*, 2009; Kimura *et al.*, 2009). SphS is autophosphorylated when $P_i$ in the environment is low: how the sensing of extracellular phosphate occurs remains elusive though the hydrophobic membrane-associated domain at the N-terminus of SphS is essential (Figure 1.3; Juntarajumnong *et al.*, 2007; Burut-Archanai *et*

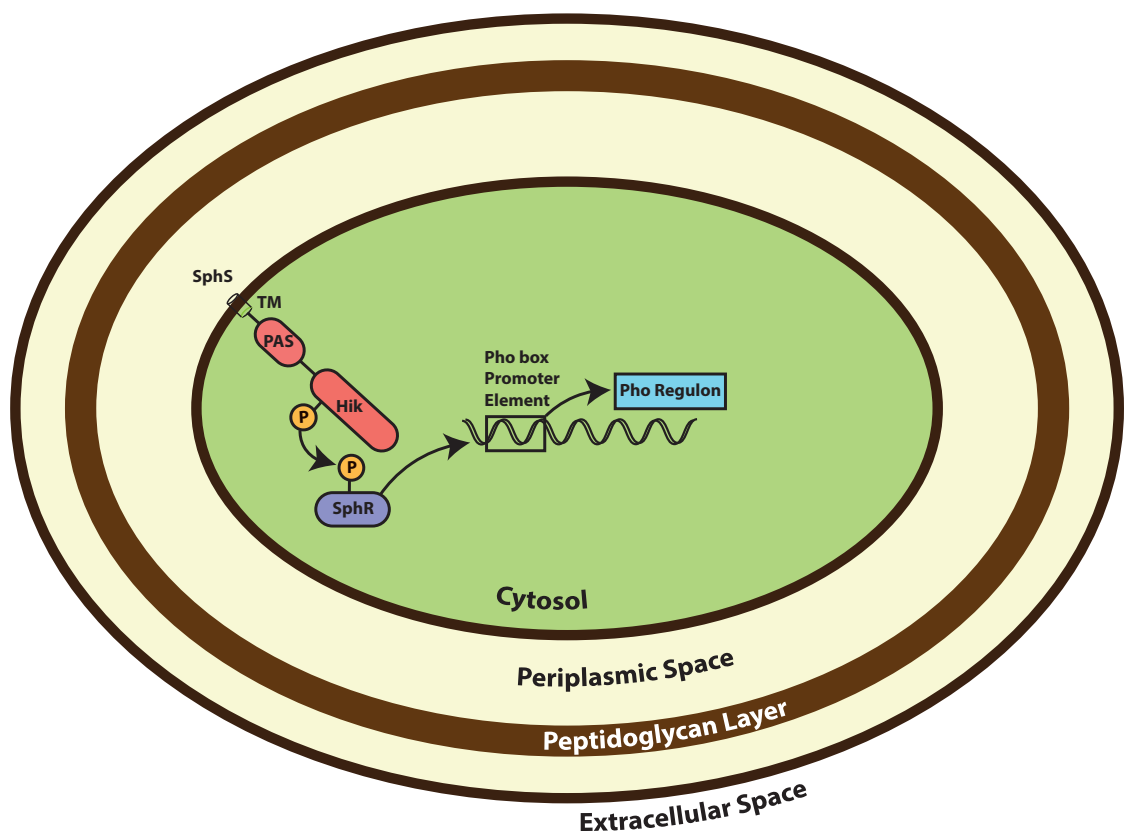

**Figure 1.3 SphS-SphR Two-Component Signal Transduction System within *Synechocystis* 6803.** The SphS histidine kinase autophosphorylates under low $P_i$ concentrations and relays a phosphoryl group to the response regulator SphR, which then binds to specific gDNA sequence motifs referred to as the pho box promoter elements and, ultimately, upregulate genes which constitute the pho regulon.

*al.*, 2009; Pitt *et al.*, 2010). SphR activates the transcription of genes by binding to at least three tandem repeats, or pho boxes, of 8 nucleotide base pairs, namely PyTTAAPyPy(T/A), in the promoter region of these genes (Figure 1.3; Wanner, 1996; Suzuki *et al.*, 2004; Su *et al.*, 2007; Juntarajumnong *et al.*, 2007). These upregulated genes have been termed the phosphate, or pho, regulon and encode proteins involved in increasing bioavailability and subsequent uptake of $P_i$ from the environment.

### 1.5.2 The Pho Regulon and $P_i$ Uptake Systems

The pho regulon is a set of 12 non-contiguous genes involved in inorganic phosphate uptake (Suzuki *et al.*, 2004). It is comprised of three main operons: *sphX-pstS1-C1-A1-B1-B1'* and *pstS2-C2-A2-B2*, which encode the high affinity phosphate-specific transport (Pst) systems, Pst1 and Pst2, and an operon which contains the *phoA* and *nucH* genes, which encode alkaline phosphatase and an extracellular nuclease, respectively (Table 1.1; Figure 1.4; Ray *et al.*, 1991; Falkner *et al.*, 1998; Suzuki *et al.*, 2004; Scanlan *et al.*, 2009).

*Table 1.1 List of Pho Regulon Genes with Corresponding Proteins.*

| ORF | Gene | Product |
| --- | --- | --- |
| *sll0679* | *sphX* | ABC-type phosphate transporter phosphate-binding protein |
| *sll0680* | *pstS1* | ABC-type phosphate transporter phosphate-binding protein |
| *sll0681* | *pstC1* | ABC-type phosphate transporter permease protein |
| *sll0682* | *pstA1* | ABC-type phosphate transporter permease protein |
| *sll0683* | *pstB1* | ABC-type phosphate transporter ATP-binding protein |
| *sll0684* | *pstB1'* | ABC-type phosphate transporter ATP-binding protein |
| *slr1247* | *pstS2* | ABC-type phosphate transporter phosphate-binding protein |
| *slr1248* | *pstC2* | ABC-type phosphate transporter permease protein |
| *slr1249* | *pstA2* | ABC-type phosphate transporter permease protein |
| *slr1250* | *pstB2* | ABC-type phosphate transporter ATP-binding protein |
| *sll0654* | *phoA* | Alkaline phosphatase |
| *sll0656* | *nucH* | Extracellular nuclease |

The activation of the *phoA-nucH* operon in *Synechocystis* 6803 under $P_i$ limitation is of particular importance for understanding the regulatory process. NucH cleaves $P_i$ moieties from nucleic acids; whereas, alkaline phosphatase cleaves $P_i$ from polyphosphate chains: both of which increase $P_i$ bioavailability to the cell (Figure 1.5; Ray *et al.*, 1991; Suzuki *et al.*, 2004). An assay has been designed to measure the activity of alkaline phosphatase and, thereby, is used as an indicator for upregulation of the pho regulon itself (Figure 1.5; Ray *et al.*, 1991;

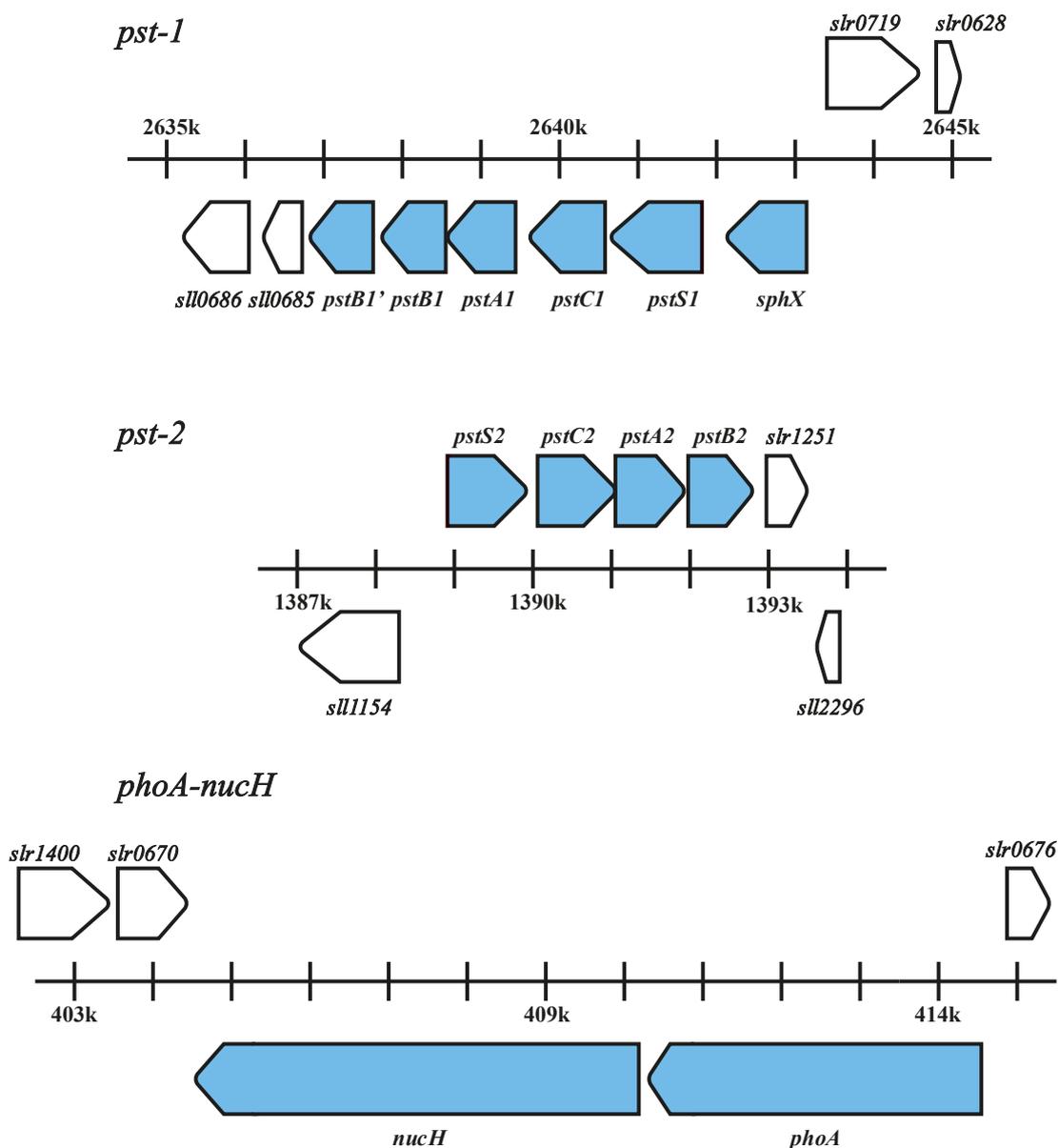

**Figure 1.4 Genomic Context of the Pho Regulon.** Three operons comprise the pho regulon, namely *pst-1, pst-2,* and *phoA-nucH*. Distinct PstSCAB transporter systems are encoded by the *pst-1* and *pst-2* operons. The *phoA-nucH* operon encodes scavenger proteins which cleave $P_i$ moieties in order to increase bioavailability.

Aiba *et al.*, 1993; Wanner, 1996; Hirani *et al.*, 2001). The transport of $P_i$ from the periplasmic space relies on ATP-binding cassette (ABC) transporter systems. The *pst-1* and *pst-2* operons encode the PstSCAB-transport systems, which actively transport phosphate moieties from the periplasmic space into the cytoplasm (Table 1.1; Figure 1.4; Figure 1.5).

Both of these Pst systems are comprised of an association and network of four proteins: two membrane-spanning subunits, PstA and PstC, an ATPase, PstB, and a periplasmic phosphate-binding protein, PstS. The membrane channel is formed by PstA and PstC and directly transport $P_i$ across the inner membrane by utilizing the energy supplied from PstB dimerized subunits, which are cytosolic and utilize energy supplied via ATP (Table 1.1; Figure 1.5; Hirani *et al.*, 2001). The presence of two high affinity ABC-type transport systems specific for $P_i$ uptake is probably an environmental adaptation and selective advantage due to the optimization of uptake at low $P_i$ concentrations in environments (Hirani *et al.*, 2001). Yet, these two Pst systems are differentially expressed based on external $P_i$ concentrations as well as differentiable kinetics with a high-velocity, low-affinity Pst1 transporter system compared to that of a low-velocity, high-affinity Pst2 transporter system (Pitt *et al.*, 2010; Burut-Archanai *et al.*, 2011). The *pst-1* operon in *Synechocystis* 6803 was highly expressed during sufficient $P_i$ conditions and facilitated in a high rate of $P_i$ uptake as well as high growth rate and storage of polyPs; whereas, a substantial increase of *pst-2* and *phoA-nucH* gene expression levels were observed under $P_i$-limiting conditions (Figure 1.4; Figure 1.5; Pitt *et al.*, 2010). In order to reach the cytosol, the inner membrane barrier must be crossed, yet this is contingent upon $P_i$ being donated to the transporter systems. The periplasmic space and its constituents require further analysis.

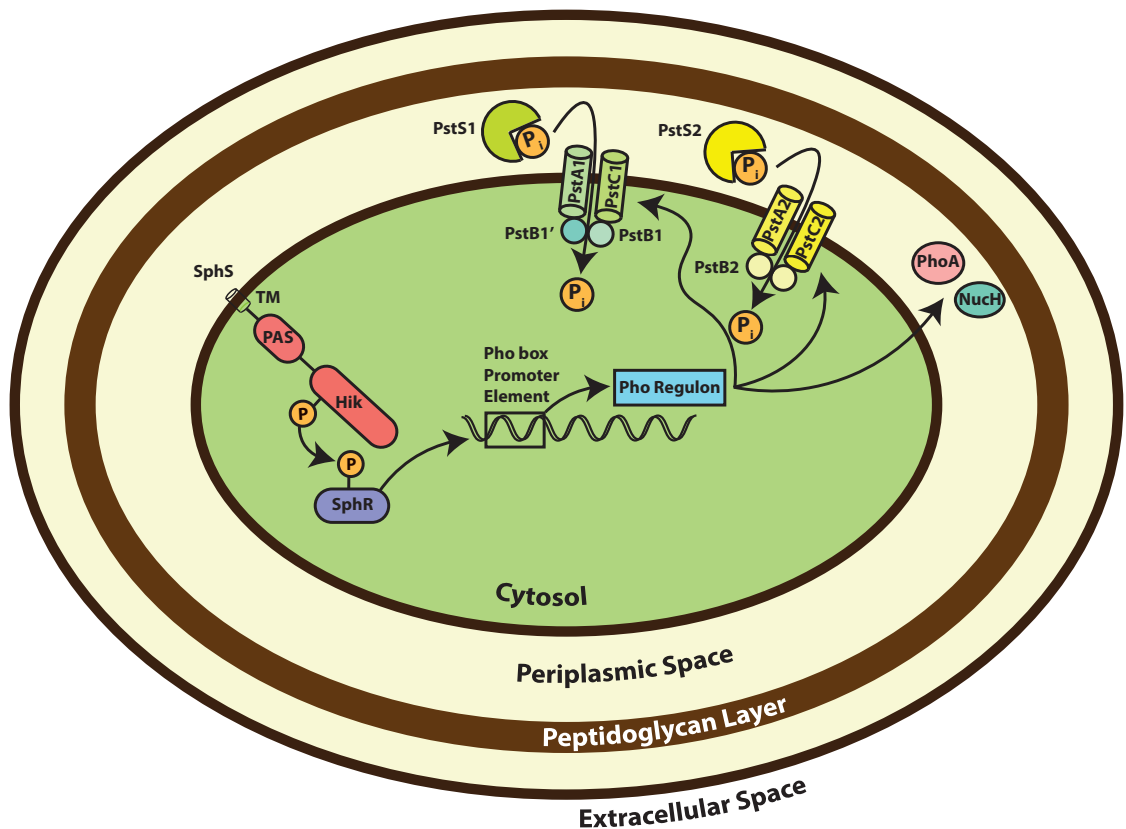

**Figure 1.5 P$_i$ Limitation Response within *Synechocystis* 6803.** The SphS-SphR signal transduction system (red and purple, respectively) is utilized to acclimate to low P$_i$ concentrations with the translation of Pst1 and Pst2 transporter systems (green and yellow, respectively), alkaline phosphatase (PhoA; pink), and an extracellular nuclease (NucH; blue-green).

### 1.5.3 Periplasmic Phosphate-Binding Proteins

*Synechocystis* 6803 encodes four different periplasmic phosphate-binding proteins (PBPs). Three of these periplasmic PBPs are derived from the *pst1* and *pst2* operons, namely *sphX*, *pstS1*, and *pstS2* (Figure 1.4; Aiba and Mizuno, 1994; Scanlan *et al.*, 2009). As previously stated, PstS1 and PstS2 are protein subunits for the ABC-type phosphate transporters (Table 1.1; Scanlan *et al.*, 2009). Although the functionality of SphX in *Synechocystis* 6803 has yet to be resolved, inactivated *sphX* in *Synechococcus 7942* resulted in the inability for this cyanobacterium to adapt to varying P$_i$ concentrations compared to the wild-type strain (Falkner *et al.*, 1998). The *sphX* gene is considered a constituent of the *pst-1* operon yet there are two pho box promoter elements within the operon: one prior to *sphX* and the other prior to *pstS1* (Pitt *et al.*, 2010). It was shown that transcripts are differentially expressed and argued that the ratio between SphX:PstS1 increased under $P_i$ limitation (Pitt *et al.*, 2010). On the other hand, the fourth putative periplasmic PBP lies outside of the pho regulon completely and is known simply as its annotated gene name *sll0540*. As implied, the transcription of *sll0540* does not increase under $P_i$-limiting conditions, but this protein is homologous to other cyanobacterial phosphate-binding proteins and should be analyzed accordingly (Scanlan *et al.*, 2009; Pitt *et al.*, 2010). Phylogenetic analysis has revealed that these periplasmic phosphate-binding proteins extracted from cyanobacterial genomes form two clades: a PstS-like and an SphX-like clade (Pitt *et al.*, 2010). The PstS-like clade is more conserved compared to that of the SphX-like clade, which indicates variation throughout its evolution and differentiation in metabolic role (Pitt *et al.*, 2010). Different cyanobacterial species are well represented in the SphX-like clade, which also contained *sll0540* from *Synechocystis* 6803, and theoretically must be fulfilling a specific metabolic role within cyanobacteria, such as interacting with a negative regulator for the SphS-SphR two-component transduction system (Scanlan *et al.*, 2009; Fuszard *et al.*, 2010; Muñoz-Martín *et al.*, 2011).

### 1.5.4 The SphS-SphR Negative Regulator, SphU

SphU is the negative regulator under $P_i$-sufficient conditions for the SphS-SphR two-component transduction system in *Synechocystis* 6803 through its interaction with the Per-Arnt-Sim (PAS) domain of SphS (Figure 1.6; Juntarajumnong *et al.*, 2007). The inactivation of *sphU* in *Synechocystis* 6803 resulted in accumulation of polyPs and constitutive expression of alkaline phosphatase activity regardless of $P_i$ bioavailability (Morohoshi *et al.*, 2002; Juntarajumnong *et al.*, 2007). Under $P_i$-limiting conditions, the inhibitory effect of SphU is alleviated so that the sensory relay is activated for the upregulation of the pho regulon (Juntarajumnong *et al.*, 2007). *sphU* lies outside of the pho regulon and, therefore, expression does not increase under $P_i$-limiting conditions (Juntarajumnong *et al.*, 2007; Scanlan *et al.*, 2009). This could be an environmental adaptation for *Synechocystis* 6803 to low $P_i$ environments and essential for strict regulation and may involve conformational states of a complex comprised of SphU (Juntarajumnong *et al.*, 2007; Burut-Archanai, 2009). The biotechnological implications utilizing strains lacking functional SphU to promote constitutive expression of the pho regulon in *Synechocystis* 6803 and, by extension, the continuous uptake of phosphate from the environment are currently being explored (Burut-Archanai *et al.*, 2013; Krasaesueb *et al.*, 2019). As for a research perspective, the system may be interrelated to downstream cytosolic processes which ultimately require $P_i$, such as photosynthesis.

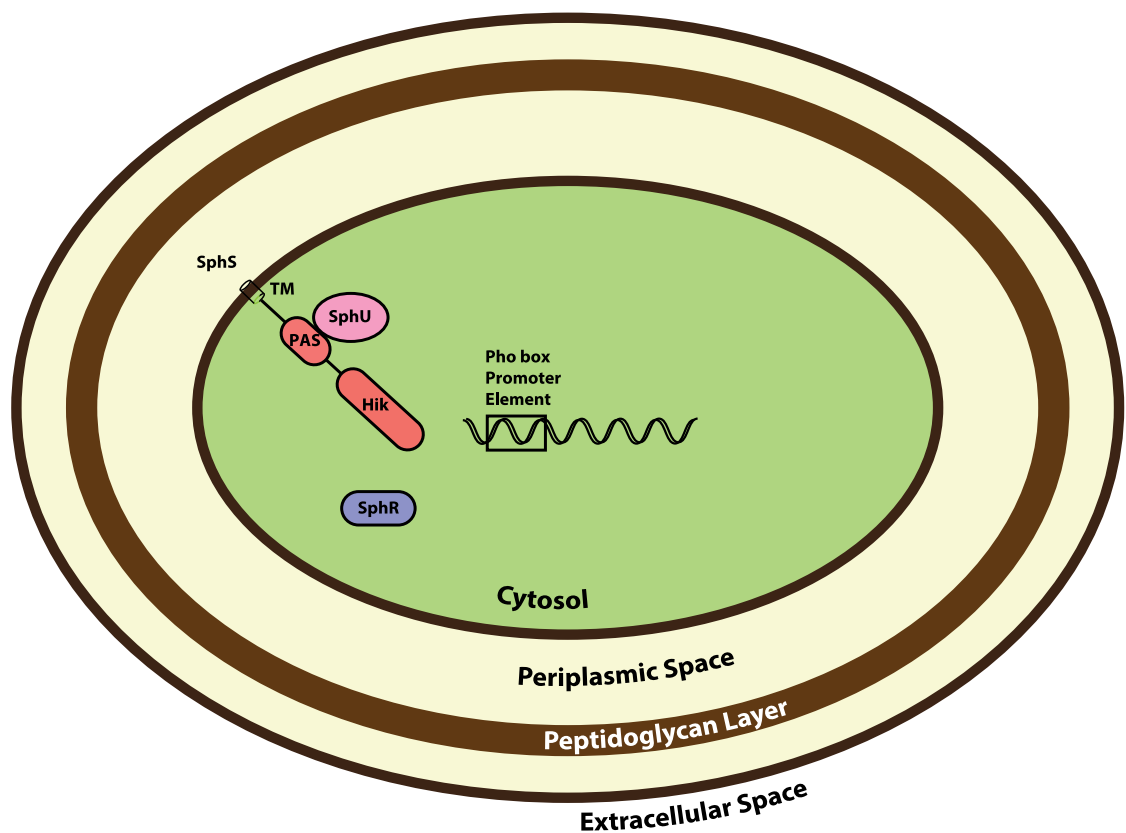

**Figure 1.6 SphU Regulation of SphS-SphR Signal-Transduction System in *Synechocystis* 6803.** The SphS-SphR signal transduction system (red and purple, respectively) is negatively regulated via the association of SphU (pink) with the Per-Arnt-Sim (PAS) domain of SphS.

## 1.6 Photosynthesis

Through evolutionary stability, a plethora of photosynthetic species have arisen with varying approaches to photonic capture as well as the compounds which are able to act as electron donors. The general photosynthetic equation may be summarized as:

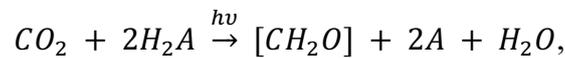

$$CO_2 + 2H_2A \xrightarrow{h\nu} [CH_2O] + 2A + H_2O,$$

which was originally proposed by Cornelis van Niel in 1931. Photosynthesis is comprised of two processes, namely the light-dependent and light-independent reactions which obtain their name, as implied, based solely off of the dependence and independence of photonic capture to complete the reaction. The light-independent reaction utilizes ATP and NADPH for the Calvin-Benson cycle in order to fix atmospheric $CO_2$ into carbohydrates which is then utilized as a source of energy from higher trophic levels. The light-dependent reaction utilizes compartmentalization formed by the thylakoid membranes in order to produce ATP and NADPH via an electron transport chain.

### 1.6.1 Photosynthetic Apparatus

The photosynthetic apparatus of *Synechocystis* 6803 is comprised of an electron transport chain (ETC) – a series of protein complexes which utilize redox reactions, simultaneous reduction and oxidation processes, to produce a trans-membrane electrochemical potential of protons, or proton motive force (Figure 1.7). The photosynthetic ETC requires three main protein complexes: Photosystem II (PS II), a water-plastoquinone oxidoreductase; the cytochrome *b6f* complex (Cyt *b6f*), a plastoquinol-plastocyanin oxidoreductase; and Photosystem I (PS I), a plastocyanin-ferredoxin oxidoreductase. Electrons (e⁻) are shuttled down the photosynthetic ETC – this ultimately requires light (*hv*) which aids in separating molecular charges, creating a path that e⁻ can utilize and increasing their energetic states through the Z-scheme (Govindjee *et al.*, 2017). Protons are pumped (H⁺) into the lumen, building a gradient which is used in order to store energy in a chemical form which is readily available for metabolic processes, namely the oxidation of adenosine diphosphate (ADP) to adenosine triphosphate (ATP) and the reduction of nicotinamide adenine dinucleotide phosphate (NADP⁺) to NADPH via the ferredoxin-NADP⁺ oxidoreductase (FNR). Photosynthesis is also able to implement a strategy by which FNR is able to recycle e⁻ back into the photosynthetic ETC through donation back to Cyt $b_6f$ – this is referred to as cyclic electron transport and alleviates oxidative stress. PSII is typically considered the initial starting point for the photosynthetic ETC, as it is essential in the extraction of e⁻ from water, yet there is an earlier step which aids in photonic capture – the phycobilisome.

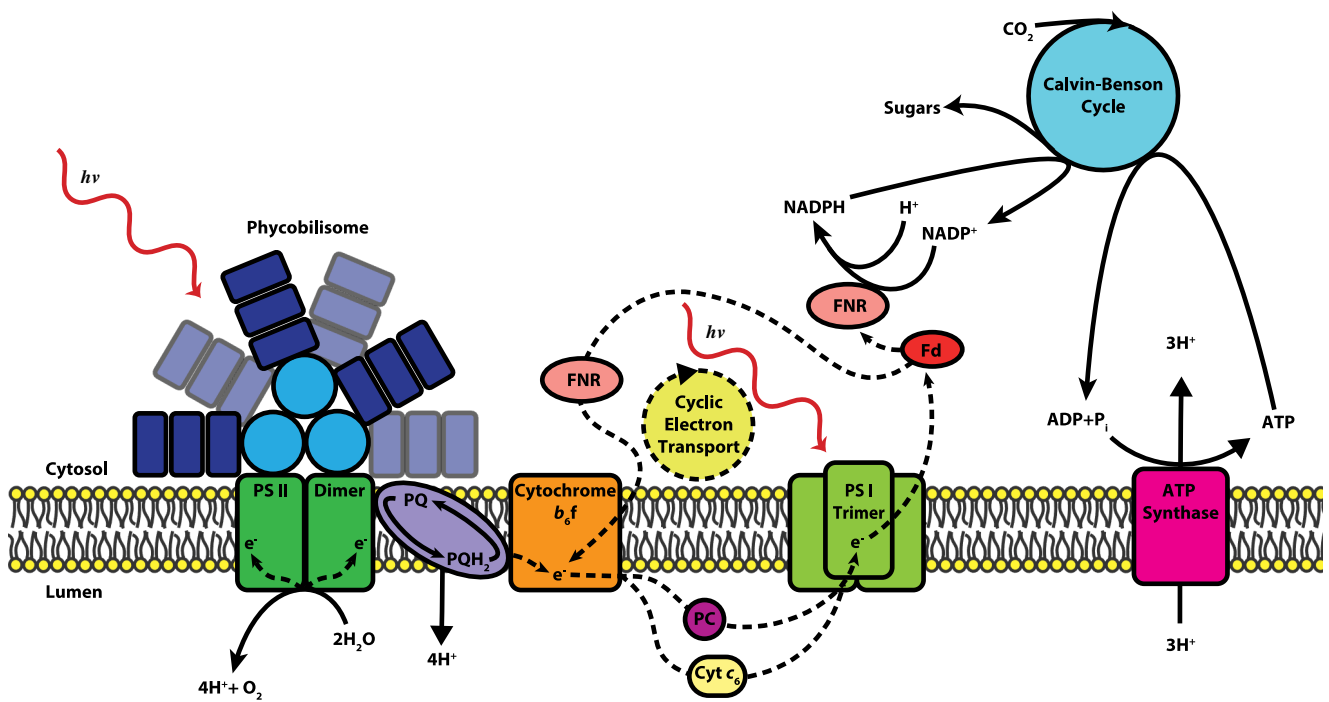

**Figure 1.7 The Photosynthetic Electron Transport Chain of *Synechocystis* 6803.** The Photosynthetic ETC is a sequential process of electron (e⁻) flow through complexes oscillating between oxidized and reduced states. Light ($h\nu$) harvested from the phycobilisome or directly from the network of chlorophyll *a* associations is utilized via PS II and PS I to convert photonic energy into redox potential. The plastoquinone (PQ) and plastocyanin (PC) are the mobile carriers, which translocate electrons from PS II and cytochrome $b_6f$, respectively, to cytochrome $b_6f$ and PS I, respectively. Cytochrome $c_6$ (Cyt $c_6$) is also able to donate electrons to PS I. As this happens, a trans-membrane electrochemical potential of protons is formed – this is utilized for adenosine triphosphate (ATP) conversion via ATP Synthase. Moreover, the ferredoxin (Fd) terminal acceptor of PS I, generates NADPH via the incorporation into a protein complex association referred to as Fd-NADP⁺ oxidoreductase (FNR). FNR is also able to donate e⁻ to cytochrome $b_6f$, creating a cyclic flow of e⁻. The primary by-products, namely ATP and NADPH, are utilized to fix atmospheric carbon into carbohydrates. Adapted from Liu (2016).

## 1.6.2 The Phycobilisome

The majority of light-harvesting capacity of the photosynthetic apparatus stems from the association of large antennae complexes referred to as the phycobilisomes (Figure 1.8; Glazer *et al.*, 1983; Bryant, 1991). The phycobilisome (PBS) is a hemi-discoidal pigment-protein complex, comprised of a core and rod protein structures with energetically coupled open-chain tetrapyrroles, referred to as phycobilins, that transfer light energy to the photosynthetic reaction centers via the photosystem network of Chlorophyll *a* (Bryant *et al.*, 1979). There are two main classes of proteins associated with the phycobilins (phycobiliproteins) within *Synechocystis* 6803, namely phycocyanin (PC) and allophycocyanin (APC), whereby a covalent thioether bond and cysteine residue stabilize the phycobilin-phycobiliprotein association (Oi *et al.*, 1982; Yu and Glazer, 1982; Beale, 1993). The rod structures are aggregated trimeric disks of PC (MacColl, 1998). The core is predominantly comprised of trimeric disks of APC, which is mainly composed of α and β polypeptides associated with high-energy phycobilins (Biggins and Bruce, 1989; MacColl, 1998). However, a bridge between the PBS and photosystem network of Chlorophyll *a* is formed via two polypeptides associated with low-energy phycobilins – these are referred to as the terminal emitters (TE). The phycobiliproteins within the PBS are arranged from shortest to longest wavelengths of maximum absorption from the rods to the core, respectively, optimizing energy transference. This is analyzed through low-temperature fluorescence emission spectroscopy with an excitation wavelength of 580 nm and maximum fluorescence for PC emitted at ~650 nm, APC emitted at ~665 nm, and TE emitted at ~685 nm. PBS associations within the photosynthetic apparatus are providing more insight into the dynamic process, yet it is undoubted that the PBS plays an essential role with Photosystem II (Kirilovsky *et al.*, 2014).

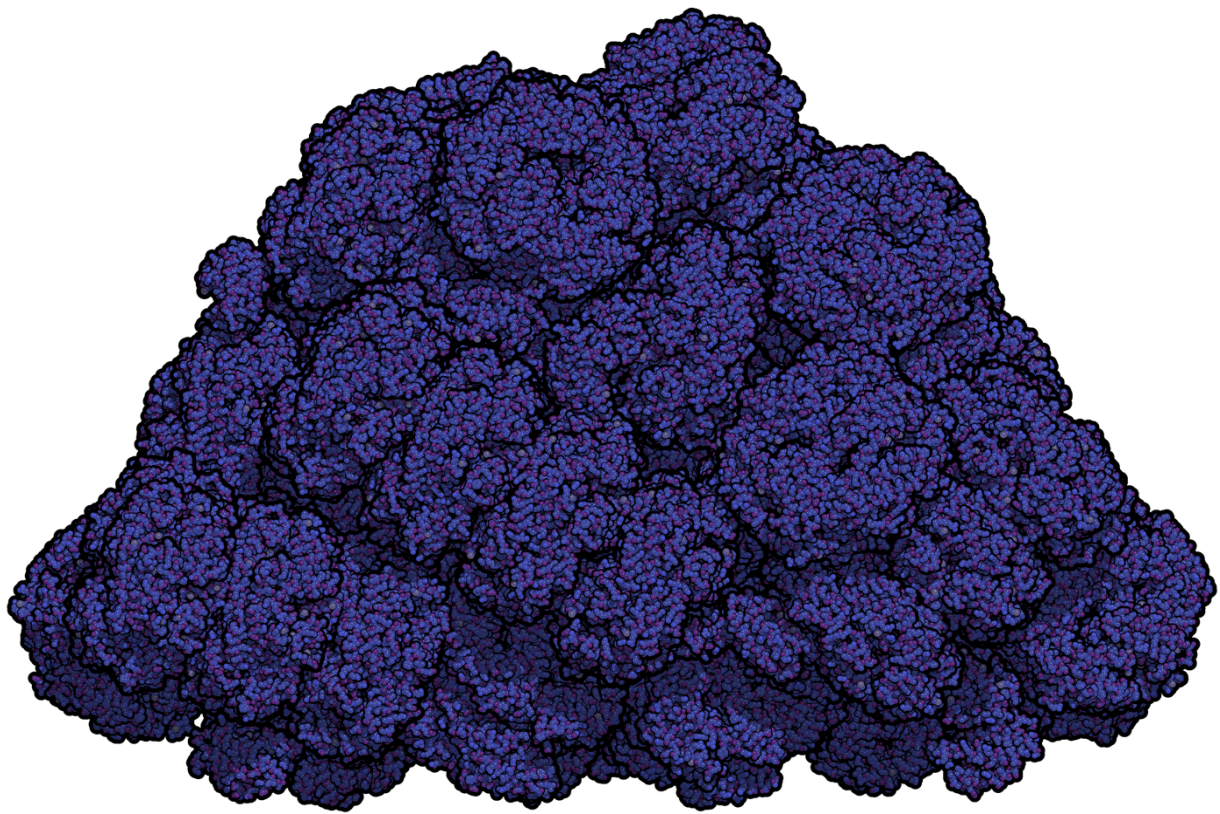

**Figure 1.8 PBS Structure from Red Alga *Porphyridium purpureum*.** An example of the PBS structure obtained by X-ray crystallography (PDB: 6KGX; Ma *et al.*, 2020).

### 1.6.3 The Photosystem II Complex

PS II, a homodimer, is comprised of more than 30 transmembrane and peripheral protein associations (Figure 1.9). PS II core complexes are composed of a reaction center (RC) and chlorophyll-binding, inner antenna proteins (Vermaas *et al.*, 1988). The RC, formed via an isoform of D1 and the D2 protein interface, can be encoded by *psbA1-3* and *psbD*, respectively (Summerfield *et al.*, 2008; Sheridan *et al.*, 2020). The chlorophyll-binding proteins, namely

CP43 and CP47, which transfer energy to the RC, are encoded by *psbC* and *psbB*, respectively (Figure 1.10; Komenda *et al.*, 2004).

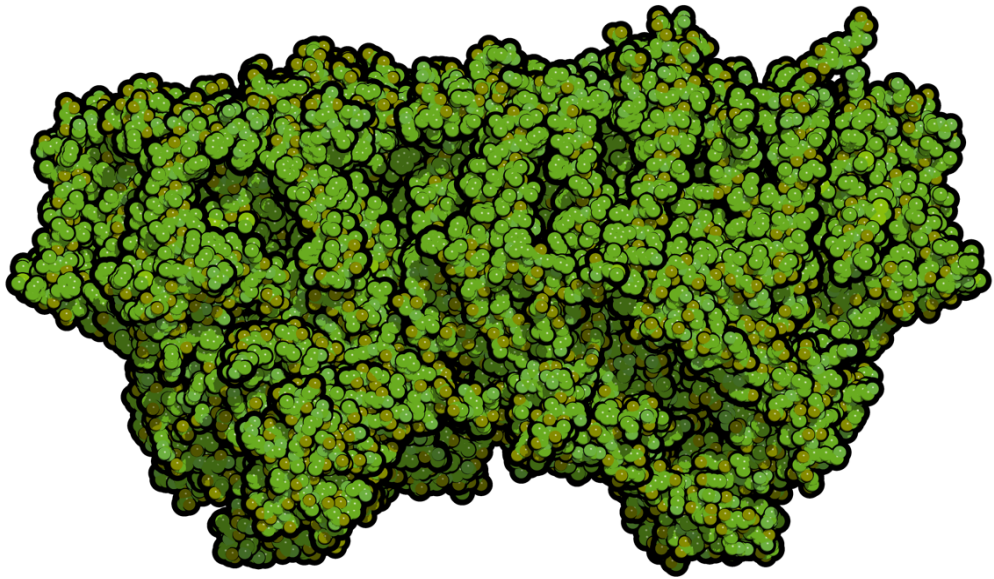

**Figure 1.9 PS II Structure from *Thermosynechococcus vulcanus*.** An example of the PS II structure obtained by X-ray crystallography (PDB: 5V2C; Wiwczar *et al.*, 2017).

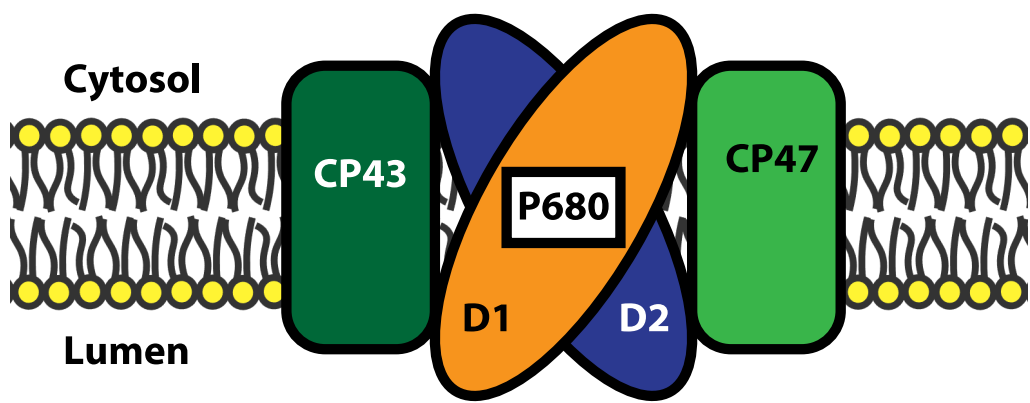

**Figure 1.10 PS II Core Complex.** The RC, comprised of D1 and D2 proteins, harbor two chlorophyll *a* molecules (P680) which form the charge separation along with pheophytin from D1. Chlorophyll-binding antenna proteins, namely CP43 and CP47, utilize a network of chlorophyll *a* molecules in order to transfer energy to the RC.

Oxygenic photosynthesis derives its name from the ability of PS II to oxidize water at ambient temperature, creating molecular oxygen as a by-product. This process is referred to as the primary charge separation, followed by subsequent electron transfer within PS II and throughout the photosynthetic apparatus, and occurs via a $Mn_4O_5Ca$ cluster referred to as the oxygen-evolving complex (OEC), stabilized from binding to CP43 and D1 (Nanba and Satoh, 1987; Ferreira *et al.*, 2004; Loll *et al.*, 2005; Umena *et al.*, 2011). The water-extracted electron is utilized in the conversion of photon-generated resonance energy at the RC to form a charge separation between two chlorophyll molecules, denoted P680, and a pheophytin from the D1 protein – this process is called the primary photochemical reaction (Parson and Cogdell, 1975; Werner *et al.*, 1978). Due to various charge separated states of molecules utilized in these processes of light conversion, photodamage, maintenance of PS II integrity is required, particularly for the D1 protein (Aro *et al.*, 1992; Lindahl *et al.*, 2000; Lupínková and Komenda, 2004). This has led to an entire area of research essential for understanding the biogenesis pathways by which functional PS II is recycled and repaired (Mabbitt *et al.*, 2014; Fagerlund *et al.*, 2020). The PS II biogenesis is analyzed through low-temperature fluorescence emission spectroscopy with an excitation wavelength of 440 nm and peaks form for CP43 and, by extension, PS II monomers with maximum fluorescence emitted at ~685 nm and CP47 and, by extension, functional PS II dimers with maximum fluorescence emitted at ~695 nm. Importantly, the spectra provide a ratio of PS II:PS I.

**1.6.4 Photosystem I**

The Photosystem I complex (PS I) is comprised of monomeric associations of at least eleven polypeptides with ~100 chlorophyll *a* molecules and three 4Fe-4S clusters (Figure 1.11; Li *et al.*, 1991; Golbeck, 1992). The reaction center of PS I is formed by two chlorophyll-binding proteins, encoded by *psaA* and *psaB*, and harbor the electron transfer centers, P700, and the 4Fe-4S center $F_X$ (Parrett *et al.*, 1989; Golbeck, 1992). Photonic capture via the peripheral pigment-protein associations is utilized to reduce ferrodoxin, subsequently generating NADPH or utilized for cyclic electron flow (Chitnis, 1996; Munekage *et al.*, 2004). PS I is analyzed via low-temperature fluorescence emission spectroscopy with an excitation wavelength of 440 nm and maximum fluorescence emitted at ~ 725 nm.

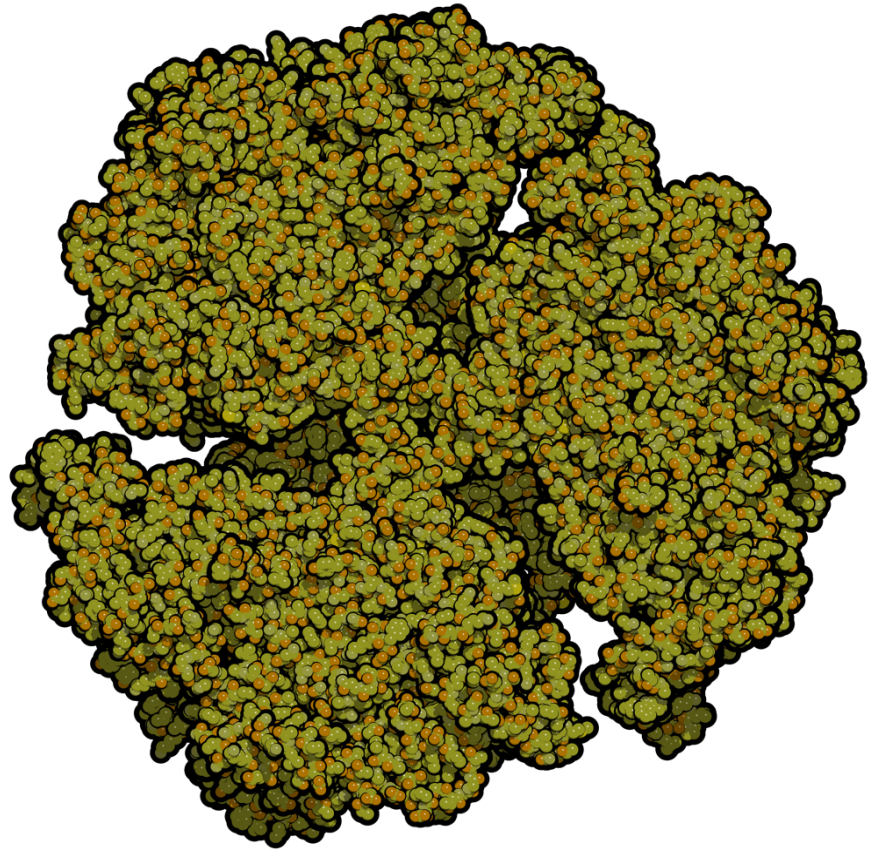

**Figure 1.11 PS I Structure from *Synechocystis* 6803.** An example of PS I structure obtained by X-ray crystallography (PDB: 5OY0; Malavath *et al.*, 2018).

Iron-limiting conditions have indicated an association between PS I and a chlorophyll-binding protein encoded by *isiA* and homologous to CP43 from PS II; therefore, it is referred to as IsiA or CP43' (Figure 1.12; Laudenbach and Straus, 1988; Burnap *et al.*, 1993). Observations and crystallographic structures have resolved the formation of a ring of CP43' proteins around a centralized PS I trimer, aiding in the dissipation of energy (Figure 1.12; Kouřil *et al.*, 2005; Schoffman and Keren, 2019; Toporik *et al.*, 2019; Cao *et al.*, 2020). It was therefore proposed that CP43' may provide a more generalized acclimation response for oxidative stress due to redox imbalance (Havaux *et al.*, 2005; Ihalainen *et al.*, 2005; Wilson *et al.*, 2007). Absorption and fluorescence emission spectra stemming from CP43' are characterized by blue-shifts of peaks (Dühring *et al.*, 2006; Rakhimberdieva *et al.*, 2007; van der Weij-de *et al.*, 2007; Ma *et al.*, 2017).

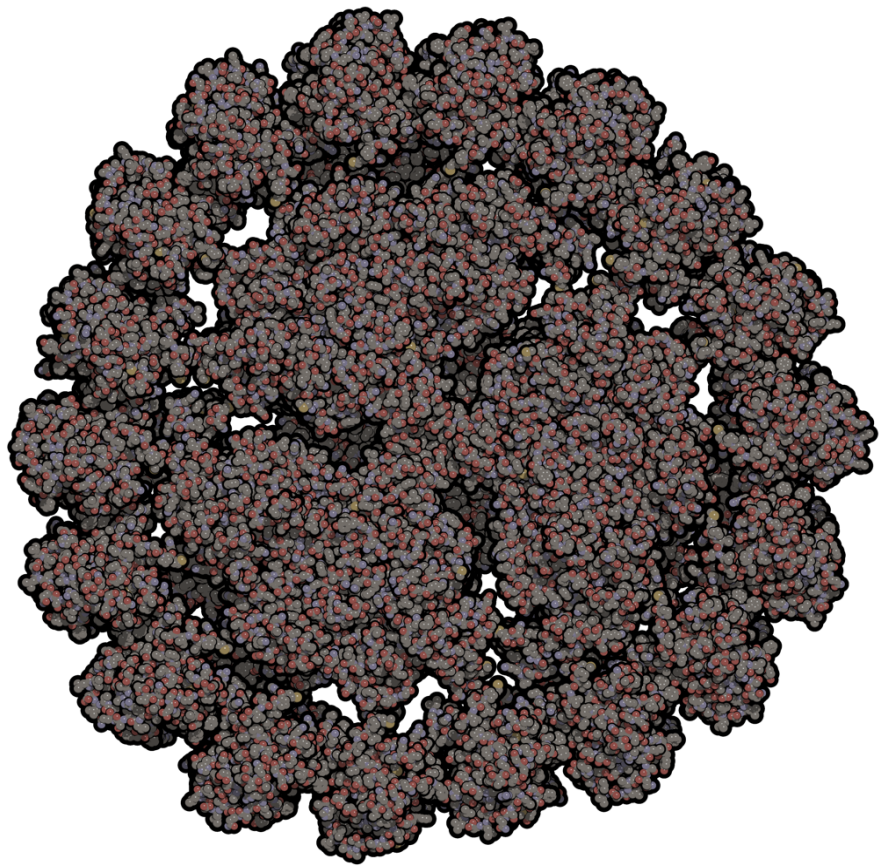

**Figure 1.12 PS I-IsiA Antenna Supercomplex Structure from *Synechocystis* 6803.** An example of PS I-IsiA supercomplex structure obtained by X-ray crystallography (PDB: 6NWA; Toporik *et al.*, 2019).

### 1.6.5 Phosphatidylglycerol

Phosphatidylglycerol (PG) is the only phospholipid in thylakoid membranes. Photosynthesis was directly impeded by PG digestion utilizing phospholipase $A_2$ and inhibited PS II activity and whole-chain electron transport as well as the removal of a PG phosphate synthase encoded by the *pgsA* gene (Jordan *et al.*, 1983; Siegenthaler *et al.*, 1987; Hagio *et al.*, 2000). Phosphatidylglycerol is integral for PS II dimerization stability and facilitates the binding of CP43 with the PS II RC via two PG molecules (Kruse and Schmid, 1995; Laczkó-Dobos *et al.*, 2008; Guskov *et al.*, 2009). *Synechocystis* 6803 lacking PG also degraded trimeric PS I with accompanied depletion of chlorophyll due to inhibition of the chlorophyll biosynthetic pathway (Domonkos *et al.*, 2004; Sato *et al.*, 2004; Kopečná *et al.*, 2015).

**1.7 Aims**

The functions of two periplasmic PBPs, SphX and SphZ, from *Synechocystis* 6803 have remained elusive, yet each contributes a role within a network of proteins and, by extension, metabolic processes from an evolutionary and biochemistry standpoint. This study was designed with a "wide-net" approach so that the collected data could indicate any differences between SphX and SphZ as well as from that of WT. As information was uncovered, experiments were then designed for exploring responses of the photosynthetic apparatus during nutrient limitation. Ultimately, this study aimed to contribute evidence for the functional roles of SphX and SphZ within *Synechocystis* 6803 for the acquisition of $P_i$ during periods of nutrient deficiency and present an acclimation response of the photosynthetic apparatus.

*Chapter 1*

# Chapter Two: Materials and Methodology

## 2.1 Bioinformatics

The "PstS" domain family was studied through the National Center for Biotechnology Information (NCBI; https://www.ncbi.nlm.nih.gov/). Amino acid sequences of proteins classified within the "COG0226" domain from *Synechocystis* 6803 and *E. coli* were retrieved from NCBI.

### 2.1.1 Phylogenetic Analysis of Cyanobacterial Phosphate-binding Protein Homologues

A customized BLAST search was performed in order to find cyanobacterial homologues for each of the PBPs from *Synechocystis* 6803 utilizing 206 cyanobacterial genomes retrieved from Joint Genome Institute (JGI; https://jgi.doe.gov/) and NCBI, which were exported into Geneious (Sheridan *et al.*, 2020; Biomatters Ltd., NZ). Signal peptides were cleaved using SignalP-5.0 (http://www.cbs.dtu.dk/services/SignalP/). The amino acid sequences of these cyanobacterial homologues were then aligned via the Clustal Omega multiple sequence alignment program within Geneious, parameterized by global alignment with free end gaps and a cost matrix of Blosum62. A phylogenetic tree was constructed via PhyML software utilizing the WAG amino acid substitution model, supported by bootstrap values, and rooted to the PstS amino acid sequence from *E. coli* in Geneious.

### 2.1.2 Structural Prediction

Amino acid sequences of SphX and SphZ from *Synechocystis* 6803 were uploaded to RaptorX to ascertain their predicted protein secondary structures (http://raptorx.uchicago.edu/). PDB2PQR was then used in order to obtain the Poisson-Boltzmann electrostatics of their predicted structures (http://server.poissonboltzmann.org/; Dolinsky *et al.*, 2004). MolProbity was used to analyze the Ramachandran plots of the predicted structures (http://molprobity.biochem.duke.edu/). The PyMOL molecular graphics system (Schrödinger, Inc., USA) was used to visualize the predicted protein structures. InterPro was used in order to indicate the predicted phosphate-binding ligands for both of the proteins (https://www.ebi.ac.uk/interpro/).

## 2.2 General Techniques and Solutions

Standard microbiological aseptic techniques were implemented throughout this study with the usage of: a Class II Biological Safety Cabinet (BTR Environmental, USA); a MilliporeSigma™ Milli-Q™ Direct Water Purification System (Millipore, USA); and a Hawkins® BIGBOY™ pressure cooker or a SQUARE autoclave (Astell Scientific Ltd., UK).

## 2.3 Strains and Growth Conditions

*Synechocystis* sp. PCC 6803, subtype GT-O1, was derived from the glucose-tolerant Williams strain (Williams, 1988) and will be considered wild type (WT) of this particular strain moving forward (Figure 2.1; Appendix 1). Site-directed mutagenesis was used to construct the Δ*sphX* plasmid, by which the start codon was mutated along with the insertion of a 2.0 kb kanamycin-resistance cassette upstream of the *pst-1* operon (Cabout, 2013; Appendix 2). The Δ*sphZ* plasmid was constructed via interruption of *sphZ* with a 2.0 kb chloramphenicol-resistance cassette (Cabout, 2013; Appendix 2). These plasmids were incorporated into the WT background via homologous recombination (Cabout, 2013; Appendix 2). The Δ*sphZ* plasmid was also transformed into the ΔSphX background to create the ΔSphX:SphZ strain (Cabout, 2013; Appendix 2). Strains were maintained at 30°C under constant ~40 µE·m$^{-2}$·s$^{-1}$ illumination of warm, white light on solid BG-11 agar plates, in which BG-11 was supplemented with 1.5% (w/v) agar, 10 mM TES-NaOH (pH 8.2), and 0.3% (v/v) sodium thiosulfate. Where appropriate, 46.4 µM of chloramphenicol and/or 51.6 µM of kanamycin were/was added. BG-11 liquid media was used for cultures as an optimal environmental condition and is denoted as BG-11 (Rippka *et al*., 1979). BG-11 liquid media lacking phosphate yet containing equimolar potassium chloride (KCl) is denoted as BG-11 − $P_i$ (Table 2.1; Hirani *et al*., 2001). BG-11 liquid media lacking dipotassium phosphate is denoted as BG-11 − $K_2HPO_4$ (Table 2.1). BG-11 liquid media supplemented with $\frac{1}{10^{th}}$ or $\frac{1}{100^{th}}$ of iron in the form of ferric ammonium citrate is denoted as BG-11 + $\frac{1}{10^{th}}$ $Fe^{3+}$ or BG-11 + $\frac{1}{100^{th}}$ $Fe^{3+}$, respectively (Table 2.1).

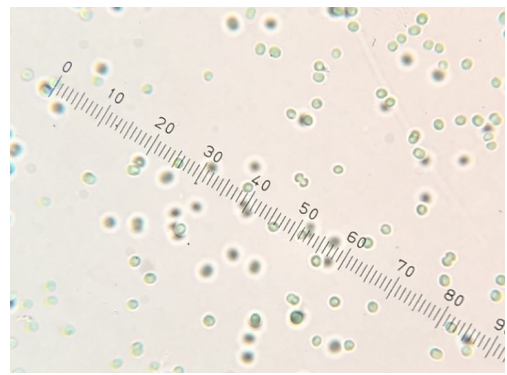

**Figure 2.1 Organism of Study.** 400x magnification of *Synechocystis* sp. PCC 6803 subtype GT-O1 (WT) using a compound light microscope.

*Table 2.1 Supplemental Nutrients for Each BG-11 Environmental Condition*

| Media: | Ferric ammonium citrate (µM): | Sodium carbonate (µM): | Dipotassium phosphate (µM): | Potassium chloride (µM): |
|---|---|---|---|---|
| **BG-11** | 22.9 | 188.7 | 175.1 | - |
| **BG-11 − $P_i$** | 22.9 | 188.7 | - | 175.0 |
| **BG-11 − $K_2HPO_4$** | 22.9 | 188.7 | - | - |
| **BG-11 + $\frac{1}{10^{th}}$ $Fe^{3+}$** | 2.29 | 188.7 | 175.1 | - |
| **BG-11 + $\frac{1}{100^{th}}$ $Fe^{3+}$** | 0.229 | 188.7 | 175.1 | - |

***100X BG-11 without iron, carbonate, or phosphate:*** 1.76 M $NaNO_3$; 30.4 mM $MgSO_4 \cdot 7H_2O$; 2.86 mM citric acid ($C_6H_8O_7$); 0.22 mM NaEDTA (pH 8.0); 10% (v/v) trace minerals

***Trace Minerals:*** 42.26 mM $H_3BO_3$; 8.9 mM $MnCl_2 \cdot 4H_2O$; 0.77 mM $ZnSO_4 \cdot 7H_2O$; 0.32 mM $CuSO_4 \cdot 5H_2O$; 0.17 mM $Co(NO_3)_2 \cdot 6H_2O$

***BG-11:*** 1X BG-11 (without iron, carbonate, or phosphate); 22.9 µM ferric ammonium citrate $\{(NH_4)_5[Fe(C_6H_4O_7)_2]\}$; 188.7 µM sodium carbonate ($Na_2CO_3$); 175.1 µM dipotassium phosphate ($K_2HPO_4$)

***BG-11 − P$_i$:*** 1X BG-11 (without iron, carbonate, or phosphate); 22.9 µM ferric ammonium citrate $\{(NH_4)_5[Fe(C_6H_4O_7)_2]\}$; 188.7 µM sodium carbonate ($Na_2CO_3$); 175.0 µM potassium chloride (KCl)

***BG-11 − K$_2$HPO$_4$:*** 1X BG-11 (without iron, carbonate, or phosphate); 22.9 µM ferric ammonium citrate $\{(NH_4)_5[Fe(C_6H_4O_7)_2]\}$; 188.7 µM sodium carbonate ($Na_2CO_3$)

***BG-11 + $\frac{1}{10^{th}}$ Fe$^{3+}$:*** 1X BG-11 (without iron, carbonate, or phosphate); 2.29 µM ferric ammonium citrate $\{(NH_4)_5[Fe(C_6H_4O_7)_2]\}$; 188.7 µM sodium carbonate ($Na_2CO_3$); 175.1 µM dipotassium phosphate ($K_2HPO_4$)

***BG-11 + $\frac{1}{100^{th}}$ Fe$^{3+}$:*** 1X BG-11 (without iron, carbonate, or phosphate); 0.229 µM ferric ammonium citrate $\{(NH_4)_5[Fe(C_6H_4O_7)_2]\}$; 188.7 µM sodium carbonate ($Na_2CO_3$); 175.1 µM dipotassium phosphate ($K_2HPO_4$)

## 2.4 Molecular Biology

### 2.4.1 Polymerase Chain Reaction (PCR) Primers

PCR primers were ordered from Sigma Aldrich, Australia (Table 2.1).

*TABLE 2.2 PRIMERS USED IN STUDY.*

| Primer | Sequence | Reference |
|---|---|---|
| Sll0540C-F | 5'-GTACCGCAATATGAACTAATTAGACC-3' | (Cabout, 2013) |
| Sll0540C-R | 5'-CTAAATAAGCACTATTTGCTCACGGAGCGG-3' | (Cabout, 2013) |
| Sll0540S-F | 5'-GGAACTAGCAGATAGATCAACTTTGC-3' | (Cabout, 2013) |
| Sll0540S-R | 5'-CGAACCAGGGCAGTTACCTC-3' | (Cabout, 2013) |
| PstS1/SphX Del For | 5'-GACAATGAAAGTAAGAACAATCAGC-3' | (Cabout, 2013) |
| SphX Del Rev | 5'-GGTATTACTAAATCGCTGTCGG-3' | (Cabout, 2013) |

### 2.4.2 Colony PCR

A KAPA3G Plant PCR Kit (Kapa Biosystems, Inc., USA) was used for the PCR reactions whereby the final concentration of the 25 µL reaction was 0.2 mM of each of the dNTPs, 1.5 mM of $MgCl_2$, 0.3 µM of each primer, and 0.5 U of KAPA3G Plant DNA Polymerase. Cells were scraped from BG-11 agar plates containing WT or mutant strains, suspended in 20 µL of Milli-Q and, dependent on turbidity, either 0.5 or 1.0 µL was used for the PCR reaction (Table 2.2).

*TABLE 2.3 THERMAL CYCLING PROTOCOL.*

| Step | | Temperature | Time |
|---|---|---|---|
| **INITIAL DENATURATION** | | 95°C | 10:00 |
| **10X** | Denaturing | 95°C | 00:20 |
| | Annealing | 68°C | 00:15 |
| | Elongation | 72°C | 04:00 |
| **20X** | Denaturing | 95°C | 00:20 |
| | Annealing | 63°C | 00:15 |
| | Elongation | 72°C | 04:00 |
| **FINAL EXTENSION** | | 72°C | 10:00 |
| **HOLD** | | 10°C | --:-- |

### 2.4.3 Gel Electrophoresis

A 1:2 dilution of PCR products and Milli-Q combined with loading dye, which comprised 16.7% (v/v), were loaded onto a 1.0% (w/v) agarose gel made with 1X TBE buffer. A constant electric potential difference of 100 V was applied for 50 min.

***1X TBE Buffer:*** 89 mM Tris, 89 mM boric acid, 2 mM EDTA
***Loading Dye:*** 0.25% (w/v) bromophenol blue, 0.25% (w/v) xylene cyanol FF, 30% (v/v) glycerol

### 2.4.4 PCR Cleanup

PCR products were purified following the protocol established in the GenepHlow™ Gel/PCR Kit (Geneaid Biotech Ltd., TW).

### 2.4.5 Spectroscopic Quantification

A NanoDrop™ 2000 UV-Vis spectrophotometer (Thermo Scientific, USA) was used to measure the concentration of nucleic acids with 2.0 µL of sample.

### 2.4.6 DNA Sequencing

200 pg · µL$^{-1}$ per 100 bp of purified PCR products, a final primer concentration of 0.64 pmol · µL$^{-1}$, and a final volume of 5 µL were submitted to the Genetic Analysis Services (Department of Anatomy, University of Otago, NZ) for sanger sequencing. Results were analyzed via Geneious (Biomatters Ltd., NZ).

### 2.5 Physiological Analyses

Liquid cultures were grown in a modified 150 mL Erlenmeyer flask with a 20 mm diameter glass sidearm and a glass capillary tube, which has an outer diameter of 6.0 mm, an inner diameter of 1.7 mm, and extends 30 mm above self-intersection and stops 10 mm from the bottom of the flask (Eaton-Rye, 2004). Aeration was provided by an aquarium pump through a 0.2 µm Millipore filter and connected to the glass capillary tube via Cole Palmer Masterflex tubing (Eaton-Rye, 2004).

Prior to physiological experiments, BG-11 was inoculated with cells scraped from BG-11 agar plates, incubated for at least 3 hours without aeration, and then grown to mid-exponential phase ($OD_{730nm} = 0.8 - 1.0$) as previously described with aeration – this is referred to as a starter culture. Cells were then harvested from these starter cultures through centrifugation at 2760 $g$ for 10 min at room temperature. The supernatant was discarded, cells were resuspended in 50 mL of BG-11 – $P_i$ or BG-11 – $K_2HPO_4$, and then the suspension was centrifuged at 2760 $g$ for 10 min at room temperature – this is referred to as "washing" the cells and was performed twice. After the second wash, the cells were resuspended in 2 mL of BG-11 – $P_i$ or BG-11 – $K_2HPO_4$ and the $OD_{730nm}$ was determined via spectrophotometry. Fresh media was inoculated with washed cells to an $OD_{730nm} = 0.05$ with appropriate antibiotics and grown under environmental conditions previously described for 36 h or one week, depending on the physiological experiment being performed.

Phosphate deprivation has previously been analyzed after 36 h due to the upregulation of the pho regulon; however, this approach only provides a snapshot of the molecular response for acclimatization of *Synechocystis* 6803 when subjected to inadequate phosphate concentrations (Hirani *et al.*, 2001; Suzuki *et al.*, 2004; Juntarajumnong *et al.*, 2007; Burut-Archanai *et al.*, 2009; Pitt *et al.*, 2010). Moreover, short term approaches for analyzing $P_i$ uptake rates across varying concentrations of supplied phosphate after the removal of either Pst system provide insight into $P_i$ acquisition via the Pst systems; however, this approach does not provide a generalized acclimatization strategy utilizing the network of proteins (Burut-Archanai *et al.*, 2011). Since there are two innate Pst systems within *Synechocystis* 6803 and mutagenic strains were designed to target periplasmic PBPs, namely SphX and SphZ which lack an elucidated role in facilitating the transport of said anions, a longer duration of data collection was used for comparison with that of the exhibited growth for each of the strains in comparison to WT due to $P_i$ being necessary for various metabolic processes over the phases

**35**

of growth. If metabolic processes were altered due to the deprivation of $P_i$, then these differences would be showcased over the fourth dimension of time.

### 2.5.1 Cell Counting

After 36 h, cells were harvested, washed, and diluted to an $OD_{730nm} = 0.01$, which has previously been determined to lie within the linear trend (Morris *et al*., 2017). A Bright-Line™ hemocytometer (Hausser Scientific, USA) with a chamber depth of 0.1 mm and total volume of $2.50_E$-4 mm$^3$ was used in order to count the cells. One µL was loaded onto the chamber and cells were counted with 400x magnification under a compound light microscope. Counting entailed two technical replicates per biological replicate following the improved Neubauer ruling protocol (Judson, 1925). A one-way analysis of variance (ANOVA) was performed within the R environment (R Core Team, 2013).

### 2.5.2 Photoautotrophic Growth

*Synechocystis* 6803 strains were grown in BG-11 liquid media or phosphate-limiting BG-11 in the form of BG-11 – $P_i$ or BG-11 – $K_2HPO_4$. The culture density was monitored via measuring $OD_{730nm}$ in a 1 cm cuvette using a Jasco V-550 UV/VIS spectrophotometer (Jasco Inc., USA) and maintained under a value of 0.4 through dilution in order to ensure an accurate reading of cell growth. A linear transformation was then applied utilizing the cell counts to ascertain the number of cells. Each environmental condition was analyzed with a one-way ANOVA, performed within the R environment.

Mathematical models were then generated for each strain under each environmental condition according to best fit. Strains grown in BG-11 were modeled based off of the logistic growth differential equation, more commonly referred to as the Verhulst equation (Bacaër, 2011):

$$\frac{dP}{dt} = rP \cdot \left(1 - \frac{P}{K}\right),$$

where the rate of change of the population with respect to time $\left(\frac{dP}{dt}\right)$ is constrained by a growth rate ($r$) and a carrying capacity ($K$).

Strains grown in BG-11 - $P_i$ were modeled utilizing a Fourier series:

$$y = a_0 + \sum_{i=1}^{n} a_i \cos(iwx) + b_i \sin(iwx),$$

where $a_0$ is a constant and intercept for the function, $w$ dictates the periodicity, and $n$ represents the harmonics, in this case $n = 2$.

Strains grown in BG-11 – $K_2HPO_4$ were modeled via a rational polynomial:

$$y = \frac{\sum_{i=1}^{n+1} p_i x^{n+1-i}}{x^m + \sum_{i=1}^{m} q_i x^{m-1}},$$

where $p$ and $q$ are constants which change with respect to each iterative step, $n$ is the degree of the numerator polynomial and in this case $n = 4$, and $m$ is the degree of the denominator polynomial, in this case $m = 1$.

The modelled curves were generated in MATLAB® (The MathWorks Inc., USA) and confirmed in the R environment through nonlinear least-squares estimates for the parameters. The first and second derivatives were then calculated to further analyze the growth kinetics.

### 2.5.3 Ascorbic Acid Method of Detection for Total Phosphorus

All glassware used to store reagents was washed with 10% (v/v) HCl to remove residual concentrations of phosphorus adsorbed to the glass surface and was rinsed thoroughly with distilled water. The reagents were stored at room temperature in glass-stoppered bottles where appropriate, except the ascorbic acid solution which was stored at 4°C and wrapped in tin foil to limit light exposure. The combined reagent was used for measurements, and all reagents reached room temperature prior to their combination and were mixed after addition of each reagent. A stock solution of 0.526 M phosphate was prepared using previously dried (1 h at 110°C) potassium dihydrogen phosphate, $KH_2PO_4$, and a standard solution containing 2.5 μg $\cdot$ mL$^{-1}$ was prepared through dilution of the stock phosphate solution. Calibration curves were prepared by varying phosphate concentrations through dilution and the combined reagent to determine detection limits (Appendix 3). All spectrophotometric measurements at 880 nm were carried out with a Jasco V-550 UV/VIS spectrophotometer (Jasco Inc., USA) in a 1 cm cuvette.

***Combined Reagent:*** 2.5 $N$ sulfuric acid ($H_2SO_4$), 0.41 mM Sb from potassium antimonyl tartrate solution [$K(SbO) \cdot C_4H_4O_6 \cdot \frac{1}{2}H_2O$], 4.85 mM ammonium molybdate solution [$(NH_4)_6Mo_7O_{24} \cdot 4H_2O$], 30 mM ascorbic acid solution ($C_6H_8O_6$)

The concentration of total phosphorus was measured every 24 h during photoautotrophic growth of liquid cultures. Three 1 mL samples were collected, centrifuged at 16,000 $g$ for 10 min and then passed through a 0.45 µm ReliaPrep™ syringe filter (Ahlstrom-Munksjö, DE). A 13.8% (v/v) of combined reagent and filtrate were mixed thoroughly. After a 20-min incubation period at room temperature, the absorbance of each sample was measured at 880 nm.

A linear transformation utilizing a standard curve for the concentration of phosphorus was applied to ascertain the concentration of extracellular phosphorus. The concentrations were normalized to an initial concentration of 2500 µg $\cdot$ L$^{-1}$. Vectors were then generated with the following form:

$$< t, c, p >,$$

where $t$ represents time (h), $c$ represents the number of cells (n), and $p$ represents the concentration of extracellular phosphorus (µg $\cdot$ L$^{-1}$).

A biharmonic spline interpolant surface could then be generated for each strain by which a surface was constructed from measurements taken during photoautotrophic growth and phosphorus depletion from BG-11 and time parameterized in $\mathbb{R}^3$ within MATLAB®. A contour plot was subsequently constructed with growth rates and time as the predictors and extracellular phosphorus depicted with the contour lines in MATLAB®. Based upon the contour lines, phosphorus depletion was directly plotted against photoautotrophic growth.

## 2.5.4 Alkaline Phosphatase Activity

Liquid cultures were grown in either BG-11, BG-11 – $P_i$ , or BG-11 – $K_2HPO_4$ for 36 hours, harvested, washed twice, and resuspended in either BG-11 – $P_i$ or BG-11 – $K_2HPO_4$. The $OD_{730nm}$ was measured for each sample using a Jasco V-550 UV/VIS spectrophotometer (Jasco Inc., USA). A constant volume (30 μL) of each sample was incubated in 0.2 M Tris-HCl (pH 8.5), containing 2 mM $MgCl_2$ and 3.6 mM *p*-nitrophenyl phosphate, in a reaction volume of 1 mL at 37°C for 20 min. The reaction was stopped with the addition of 600 mM NaOH, and then the samples were centrifuged at 12,000 *g* for 5 minutes to remove the cells. The $OD_{400nm}$ values of the supernatants were measured directly for strains grown in BG-11 and after a ten-fold dilution for strains grown in phosphate-limiting BG-11. Alkaline phosphatase activity was calculated by the following formula:

$$1000 \times \frac{OD_{400nm}}{t \times v \times OD_{730nm} \times c},$$

where *t* is the reaction time (min), *v* is the sample volume used (mL), and *c* is the number of cells per $OD_{730nm}$ (n · $OD_{730nm}^{-1}$).

## 2.5.5 Whole-Cell Absorption Spectra

Liquid cultures were grown photoautotrophically in either BG-11, BG-11 – $P_i$, BG-11 – $K_2HPO_4$, or BG-11 $+ \frac{1}{10^{th}} Fe^{3+}$ for one week, harvested, washed twice, and resuspended in respective, fresh media supplemented with 25 mM HEPES/NaOH (pH 7.5), except for when grown in iron-limiting media as the cells were washed a resuspended in BG-11 lacking any ferric ammonium citrate. The $OD_{800}$ was measured for each sample using a Jasco V-550 UV/VIS spectrophotometer (Jasco Inc., USA). Samples were diluted to an $OD_{800} = 0.3$ in a total volume of 3 mL in a 10 mm path length quartz glass cuvette (Starna Scientific, UK). A Jasco V-550 UV/VIS spectrophotometer (Jasco Inc., USA) was used to measure the cell absorption spectra between 400 and 800 nm. Cellulose tape was placed on the exit side of the sample holder to account for the light scattering effect caused by cells. The spectra were plotted and absorbance peaks were found within MATLAB®.

### 2.5.6 Chlorophyll *a* (Chl *a*) Measurements

1:100 or 1:10 dilutions of cells and 100% methanol were prepared in Eppendorf tubes

and mixed thoroughly to extract pigments. An Eppendorf 5415D centrifuge was used to isolate

chlorophyll *a* pigment (Chl *a*) with a centrifugal force of 16000 *g*. The supernatant was

aspirated into a 1 mL glass cuvette, and the absorbance at 663 nm was measured with a Jasco

V-550 UV/VIS spectrophotometer (Jasco Inc., USA). The concentration of Chl *a* was

calculated, utilizing the absorption coefficient:

$$\text{Chl } a \; (\mu g \cdot mL^{-1}) = 12.2 \cdot A_{663nm},$$

where $A_{663nm}$ must take into account the original dilution and any subsequent dilution. Chl *a*

was monitored over the course of a week as WT acclimated to $BG\text{-}11 - P_i$.

### 2.5.7 Low-Temperature (77 K) Fluorescence Emission Spectroscopy of Cells

Liquid cultures were grown photoautotrophically in either BG-11, $BG\text{-}11 - P_i$, BG-11

$- K_2HPO_4$, or $BG\text{-}11 + \frac{1}{10^{th}} Fe^{3+}$ for one week, harvested, washed twice, and resuspended in

respective, fresh media supplemented with 25 mM HEPES/NaOH (pH 7.5), except for when

grown in iron-limiting media as the cells were washed a resuspended in BG-11 lacking any

ferric ammonium citrate. Cells were transferred to an Erlenmeyer flask with a concentration of

Chl $a = 5 \mu g \cdot mL^{-1}$ in respective BG-11 media (pH 7.5) and shaken for 30 min in growth room.

Once again, the iron-limited WT strain was resuspended in BG-11 lacking any ferric ammonium citrate. After the incubation period, [Chl *a*] was calculated again to ensure an accurate reading for the 77 K measurements. A glass tube (4 mm internal diameter, 6 mm external diameter) was supplied with 2.5 µg of Chl *a* and frozen with liquid nitrogen.

Spectra were measured via a modified MPF-3L fluorescence spectrometer (Jackson,

2012) equipped with a customized silver-lined Dewar flask. Emission spectra were measured from 600 to 780 nm with a constant emission slit of 2 nm. An excitation wavelength of 440 or 580 nm with an excitation slit of 12 or 8 nm, respectively. Two technical replicates were measured for each of the three biological replicates. R scripts were used to process spectral data through baseline subtraction by which the baseline fitting was optimized via the incorporation of emission relative maxima specific to *Synechocystis* 6803 at various wavelengths, written as a Gaussian (Jackson, 2012). The spectra were normalized with equivalent total area under their traces. Normalization in this regard allows for the functions to maintain equivalent ratios of the emission from each point yet alleviates the assumption that there is an equivalence of fluorescence from a specific wavelength – ultimately due to an equivalence of only one photosystem, complex, or subunits – and, therefore, provides a generalized representation for changes within the photosynthetic apparatus.

In addition, 77 K measurements were taken every 24 h for WT cultures grown in BG-11 – $P_i$ over the course of one week to analyze the acclimation of the photosynthetic apparatus to phosphate limitation. In order to pursue this analysis, Chl *a* was <u>not</u> held constant. Instead, a constant volume of 500 μL was taken from each culture over the time course. The spectra were normalized to an artifact normally seen in spectra collected specifically from the machine, namely at 620 nm for spectra obtained from an excitation wavelength of 440 nm and at 615 nm for spectra obtained from an excitation wavelength of 580 nm. A biharmonic spline interpolant surface was generated from the parametric form:

$$< t, \lambda, F >,$$

where *t* represents time (h), $\lambda$ represents wavelength (nm), and *F* represents average fluorescence emission values (arbitrary units, a.u.). Contour plots were constructed from the generated surfaces with indicated heights corresponding to the fluorescence.

Moreover, a direct comparison of spectral measurements obtained from an excitation wavelength of 440 nm with those obtained from the 580 nm excitation wavelength is needed. Vectors were once again parameterized in $\mathbb{R}^3$ with the following form:

$$< \lambda, F_{440}, F_{580} >,$$

where $\lambda$ represents wavelength (nm), $F_{440}$ represents fluorescence (a.u.) with an excitation wavelength of 440 nm, and $F_{580}$ represents fluorescence (a.u.) with an excitation wavelength of 580 nm. This ultimately led to the two dependent spectra being plotted against each other, to produce an actual parametric curve:

$$< F_{440}, F_{580} >.$$

To prove that this parametric form is an accurate depiction of differences that innately arise from whole cells due to the plotting of these dependent variables against each other, an exhaustive combination of spectra was used by which the Cartesian product $F_{440} \times F_{580}$ produces the set of all possible ordered pairs:

$$F_{440} \times F_{580} = \{(f_{440}, f_{580}): f_{440} \in F_{440} \text{ and } f_{580} \in F_{580}\}.$$

All 36 possible combinations of spectra obtained from an excitation wavelength of 440 nm along with spectra obtained from an excitation wavelength of 580 nm were used to generate the parametric curves. K-means clustering analysis was then implemented in the R environment to determine differences of fluorescence emission between 683-687 nm for each environmental condition.

## 2.6 Protein Analyses

Analyses were carried out in a 4ºC dark room, illuminated with green light in order to limit photodamage.

### 2.6.1 Thylakoid Membrane Isolation

A WT culture was grown in BG-11 to an $OD_{730nm} = 0.8$, harvested, and washed twice in BG-11 – $P_i$. The entire resuspension was transferred to 1 L of BG-11 – $P_i$ for 186 h. The 1 L culture was harvested into 3x 15 mL conical tubes and washed in BG-11 – $P_i$ or BG-11 lacking ferric ammonium citrate. The resuspensions were transferred to three fresh 2 L flasks containing 1 L of BG-11, BG-11 – $P_i$, or BG-11 + $\frac{1}{100^{th}}$ $Fe^{3+}$. After 126 h, cells were harvested, washed with cell-washing buffer, frozen in liquid nitrogen, and stored at -80ºC for thylakoid isolation (Figure 2.2).

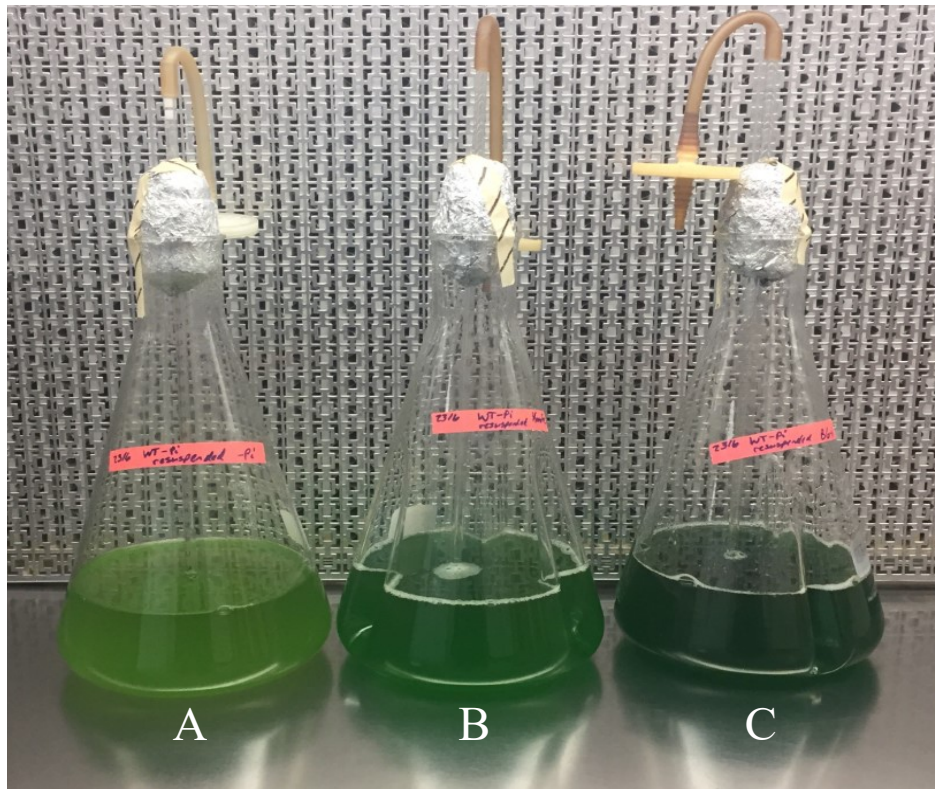

**Figure 2.2 Cultures Prepared for Thylakoid Extraction.** 1 L cultures grown in BG-11 – $P_i$ (**A**), BG-11 + $\frac{1}{100^{th}}$ $Fe^{3+}$ (**B**), and BG-11 (**C**) for 126 h after having been phosphate-stressed for 186 h.

Cell pellets were thawed, transferred to a new 15 mL conical tube, and washed with 40 mL of cell-washing buffer. Cells were then resuspended in 1.6 mL of disruption buffer per liter of initial culture and incubated on ice for 30 min. Zirconia beads with a diameter of 0.1 mm (BioSpec Products, USA) with an equivalent volume of 700 µL of Milli-Q in a 2-mL bead beating tube were then used to mechanically disrupt cells in a bead-beater over 5 cycles, each of which consisted of: 20 sec at 4800 rpm and a 5-min incubation on ice. Afterwards, the cell and bead contents were transferred to 15 mL conical tubes. Zirconia beads (BioSpec Products, USA) were removed using centrifugation at 2000 $g$ for 2 min at 4°C. Five mL of the supernatant was transferred to a new 15 mL conical tube. The zirconia beads were washed with 5 mL of disruption buffer and centrifuged again. 5 mL of the supernatant was also added to the new 15 mL conical tube. The conical tube containing the combined 10 mL of supernatant was centrifuged at 8000 $g$ for 5 min at 4°C to remove unbroken cells. Nine mL of the supernatant were transferred to a 10 mL ultracentrifuge bottle, and thylakoid membranes were isolated via ultracentrifugation at 60000 $g$ for 1 h using a 75 Ti rotor (Beckman-Coulter, USA). The supernatant was removed, and the pellet was resuspended in 100 µL of solubilizing buffer. 25 µL aliquots were prepared and stored at -80°C.

***Cell Wash Buffer:*** 50 mM HEPES-NaOH (pH 7.5), 20 mM $CaCl_2$, 10 mM $MgCl_2$, 1 mM 6 amino-caproic acid, 1 mM PMSF, 2 mM benzamidine
***Solubilization Buffer:*** 25 mM BisTris-HCl (pH 7.0), 20% (w/v) glycerol, 1 mM Pefabloc

## 2.6.2 Sodium Dodecyl Sulfate-Polyacrylamide Gel Electrophoresis (SDS-PAGE)

Isolated thylakoid membranes were analyzed by size fractionation, utilizing Bolt™ Bis-Tris Plus 4-12% pre-cast gels (Life Technologies, USA) ran within a Mini-PROTEAN 3 system (Bio-Rad, USA). 5 µg of chlorophyll $a$ for each sample per lane were prepared with LDS Sample Buffer and heated at 70ºC for 10 min. An 1X MES SDS Running Buffer was used to run the samples for 45 min at 200 V and 4ºC.

***4X LDS Sample Buffer:*** 141mM Tris base, 106 mM Tris HCl, 2% (w/v) LDS, 0.51 mM EDTA, 0.22 mM SERVA Blue G-250, 0.175 mM phenol red (pH 8.5)
***20X MES SDS Running Buffer:*** 50 mM MES, 50 mM Tris Base, 0.1% (w/v) SDS, 1 mM EDTA (pH 7.3)

## 2.6.3 Blue-native PAGE

Isolated thylakoid membranes were diluted with solubilization buffer with a concentration of 250 µg mL$^{-1}$ of Chl *a*. Native protein complexes were disassociated from thylakoid membranes with the addition of 3% (w/v) β -D-dodecylmaltoside (DDM; Affymetrix, USA) to a final percentage of 1.125% (v/v) over the course of 3 additions with gentle mixing and incubation on ice for 2 min. Thylakoid membranes were removed by centrifugation at 16000 *g* for 15 min at 4ºC. Solubilized protein complexes were transferred to a new Eppendorf tube (Figure 2.3). 0.7 µg of Chl *a* with 1X sample buffer and 0.37% (v/v) Serva G-250 (Serva, DE) were loaded onto a 3-12% Bis-Tris 1.0 mM NativePAGE precast gel (Life Technologies, USA). Gel electrophoresis was carried out in a Mini Gel Tank (Invitrogen, USA) with 1X anode buffer at 4ºC in the dark at 80 V for 1 h with 1X dark blue cathode buffer, and then 100 V for 30 min and 150 V for 2.5 h with 1X light blue cathode buffer.

Following electrophoresis, BN-PAGE gels were removed from the cassette and washed with Milli-Q. Gels were imaged using an ImageScanner III color scanner (Epson, USA).

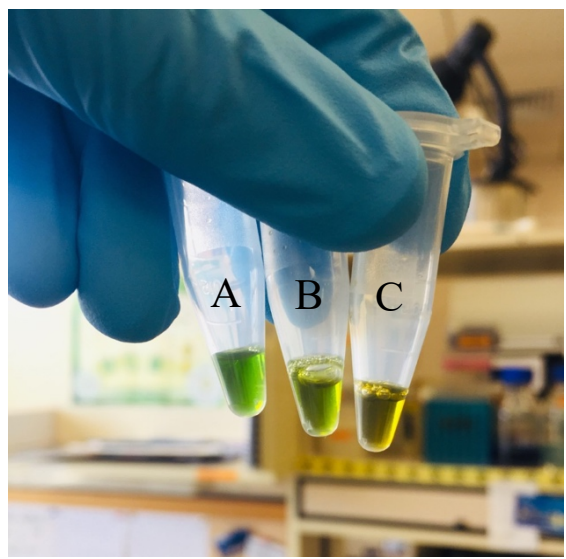

**Figure 2.3 Solubilized Thylakoid Membranes.** Isolated thylakoid membranes after being grown in BG-11 (**A**), BG-11 $+ \frac{1}{100^{th}}$ Fe$^{3+}$ (**B**), and BG-11 $-$ P$_i$ (**C**) resuspended in solubilization buffer.

*4X BN-PAGE Sample Buffer:* 50 mM Bis-Tris, 6 M HCl, 10% (v/v) glycerol, 50 mM NaCl, 0.004% (w/v) Ponceau S
*10X BN-PAGE Running Buffer:* 50 mM Bis-Tris, 50 mM Tricine (pH 6.8)
*10X Cathode Buffer:* 0.2% (w/v) Coomassie G-250 in Milli-Q
*1X Anode Buffer:* 5 mM Bis-Tris, 5 mM Tricine (pH 6.8)
*1X Dark Blue Cathode Buffer:* 5 mM Bis-Tris, 5 mM Tricine (pH 6.8), 0.02% (w/v) Coomassie G-250 in Milli-Q
*1X Light Blue Cathode Buffer:* 5 mM Bis-Tris, 5 mM Tricine (pH 6.8), 0.0002% (w/v) Coomassie G-250 in Milli-Q

## 2.6.4 Western Blotting

SDS-PAGE and BN-PAGE gels, scotch pads, 3MM Whatmann filter paper, and a polyvinylidene fluoride (PVDF) membrane (Bio-Rad, USA) were submerged in pre-chilled Nu-PAGE Transfer Buffer or Electroblot Buffer, respectively, for 10 min. The PVDF membrane was also immersed in 100% methanol for 1 min immediately prior to constructing the transfer sandwich. Once assembled, the cassette was transferred to a Mini Trans-Blot Electrophoretic Transfer Cell (Bio-Rad, USA) with a magnetic stirrer bar and placed on a stirrer plate at 4ºC. Transfer of proteins from SDS-PAGE gels to PVDF membrane was completed with 30 V for 1 h, and proteins from BN-PAGE gels were transferred to PVDF membrane with 55 V for 20 h.

After transferring the proteins, the PVDF membrane was washed with Milli-Q or 100% methanol, respectively. Membranes were then shaken in Blot-O for 1 h at room temperature, washed twice with Milli-Q for 10 min, and shaken with a primary antibody for 14 h at 4ºC. Primary antibodies used in this study include: $\alpha$-CP43 (AS11 1787; Agrisera) which is polyclonal, rabbit serum extract, raised against a synthetic peptide of a highly conserved region of PsbC (CP43) from *Arabidopsis thaliana* conjugated with keyhole limpet hemocyanin (KLH), a copper-containing protein often as an adjuvant used for antibody production in mammalian hosts; $\alpha$-CP43' (AS06 111; Agrisera) which is polyclonal, rabbit serum extract, raised against a synthetic peptide of a "near perfectly conserved" region of IsiA (CP43') from *Synechocystis* 6803 conjugated with KLH; and $\alpha$-PsaA (AS06 172; Agrisera) which is polyclonal, rabbit serum extract, raised against the N-terminus of PsaA from *Chlamydomonas reinhardtii* conjugated with KLH.

After incubation, primary antibodies were decanted into a conical tube and PVDF membranes were washed thrice with 25 mL of 0.1% (v/v) Tween-20 in TBS (pH 7.4) for 15 min and then twice for 5 min at ambient temperature. The PVDF membranes were then incubated in a secondary antibody for 1 h at room temperature, and then washed thrice for 5 min each with 0.1% (v/v) Tween-20 in TBS (pH 7.4). Immunoblotting visualization was performed with enhanced chemiluminescence (ECL) via reagents (Abcam, USA). Immunoblots were detected with an Odyssey® Fc imaging system (LI-COR Biotechnology, USA).

***Electroblot Buffer:*** 25 mM Tris, 192 mM glycine, 20% (v/v) methanol
***NuPAGE Transfer Buffer:*** 25 mM Tricine, 25 mM Bis-Tris, 1 mM EDTA, 10% (v/v) methanol
***Blot-O:*** 4% (w/v) Bovine serum albumin (BSA), 0.02% (w/v) $NaN_3$ in TBS (pH 7.4)
***TBS (pH 7.4):*** 0.137 M NaCl, 2.7 mM KCl, 25 mM Tris, ~37% (v/v) HCl
***Secondary Antibody:*** 1:20,000 dilution of anti-Rabbit IgG-peroxidase antibody, produced via goat (Sigma), in 50 mM Tris-HCl (pH 7.5), 150 mM KCl, 3% (w/v) bovine serum albumin (BSA)

## 2.5.7 Low-Temperature (77 K) Fluorescence Emission Spectroscopy of Thylakoid Membranes and Native Proteins

2.5 μg of Chl *a* from isolated thylakoid membranes and solubilized proteins used for BN-PAGE from WT cultures grown in BG-11, BG-11 − $P_i$ , or BG-11 + $\frac{1}{100th}$ $Fe^{3+}$ were transferred to a glass tube (4 mm internal diameter, 6 mm external diameter) and frozen with liquid nitrogen. The data was processed as aforementioned (Jackson, 2012). After baseline subtraction, deconvolution of the spectra was carried out in MATLAB® (Maxim, 2020).

### 2.5.8 Rationale of Experiments

Photoautotrophic growth is analyzed for any research when altering a gene within *Synechocystis* 6803. Since these mutagenic strains targeted periplasmic PBPs, potentially involved in the regulation of the pho regulon, alkaline phosphatase activity was measured. Whole-cell absorption spectra were utilized to indicate that each of the strains were, in fact, phosphate deficient when grown with BG-11 − $P_i$ and BG-11 − $K_2HPO_4$ liquid media. Ascorbic acid method of detection for total phosphorus was used for analyzing the depletion of phosphorus from BG-11 media over the course of each strain's respective phases of growth as well as confirming that phosphorus was removed from BG-11 − $P_i$ and BG-11 − $K_2HPO_4$ liquid media. However, some of the presented experiments were arranged differently across the temporal span of this study. Cells were actually counted for each of the strains due to differences of whole-cell absorption spectra, indicating that cell sizes varied since measurements were taken with a constant $OD_{800nm}$. Low-temperature (77 K) fluorescence emission spectroscopy was performed on each of the strains since phosphate is essential for the production of ATP, which stems mostly from the PETC since the cells were grown photoautotrophically. Afterwards, the study shifted focus toward the acclimatization strategy of the photosynthetic apparatus. A comparison with iron deficiency stemmed from an increased ratio of 685:695 nm fluorescence emission obtained with an excitation wavelength of 440 nm when cells were grown with BG-11 – $P_i$. Thylakoids had to be removed from phosphate deficient cells, so an experiment was designed utilizing 77 K fluorescence emission spectroscopy to ensure that the initial fluorescence emission response was present. SDS-PAGE and BN-PAGE were performed to indicate if PS I-IsiA antenna supercomplexes were present. 77 K fluorescence emission spectroscopy was also used on thylakoid membranes due to differences of color after extraction.

# Chapter Three: Characterization of ΔSphX, ΔSphZ, and ΔSphX:SphZ Strains

## 3.1 Bioinformatics

### 3.1.1 Phylogenetic Analysis of PBPs from *Synechocystis* 6803

After several rounds of trimming, 312 cyanobacterial amino acid sequences were analyzed. When rooted with PstS from *E. coli*, the two primary clades – PstS and SphX as previously described – were apparent (Figure 3.1; Appendix 4; Appendix 5; Pitt *et al.*, 2010). Moreover, the two clades diverged, indicating the PstS1 and PstS2 subclades as well as the SphX and SphZ subclades (Figure 3.1; Appendix 4; Appendix 5). There is notable similarity in the number of PstS1 and SphX homologues with 118 and 116, respectively, as well as the number of PstS2 and SphZ homologues with 38 and 40, respectively (Figure 3.1; Appendix 4; Appendix 5). All four subclades contained amino acid sequences from cyanobacteria isolated from freshwater, marine, and soil samples (Appendix 5). However, PstS1 incorporated a large fraction of marine *Synechococcus* and *Prochlorococcus* species: these species lacked PstS2 and *Prochlorococcus* species also lacked SphX and SphZ homologues (Appendix 5). The finished phylogenetic tree incorporated an uncharacterized protein amino acid sequence with multiple predicted transmembrane domains from *E. coli* that was most similar to SphZ (Figure 3.1; Appendix 5).

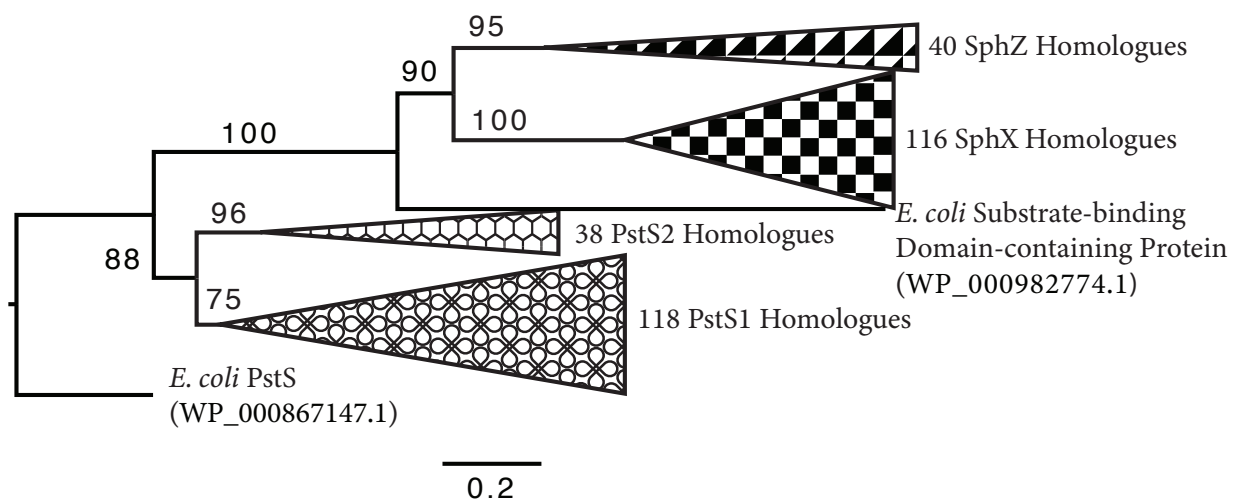

**Figure 3.1 Phylogenetics of Cyanobacterial PBP Homologues.** Phylogenetic tree constructed via PhyML software in Geneious on a Clustal Omega alignment of cyanobacterial periplasmic phosphate-binding proteins, rooted with *E. coli* PstS sequence. Indicated values are bootstrap support above 70. Parentheses indicate NCBI accession numbers.

## 3.1.2 Predicted Secondary Structures of SphX and SphZ

The pdb files generated from RaptorX were assessed on their Ramachandran plots and viewed in PyMOL (Figure 3.2; Appendix 6). The predicted structure of both proteins represented the typical "venus fly trap" structure known for the protein family (Figure 3.2.A; Figure 3.2.B). In addition, SphZ contained a transmembrane domain uncharacteristic of other PBPs (Figure 3.2.B). The protein structures were aligned in PyMOL to show their similarity concentrating around the phosphate-binding site (Figure 3.2.C). The structural alignment indicates that SphZ is characteristic of the phosphate-binding domain – as seen with the overlay of SphX – and also contains a hinge, a transmembrane domain, and a cytosolic motif (Figure 3.2.C).

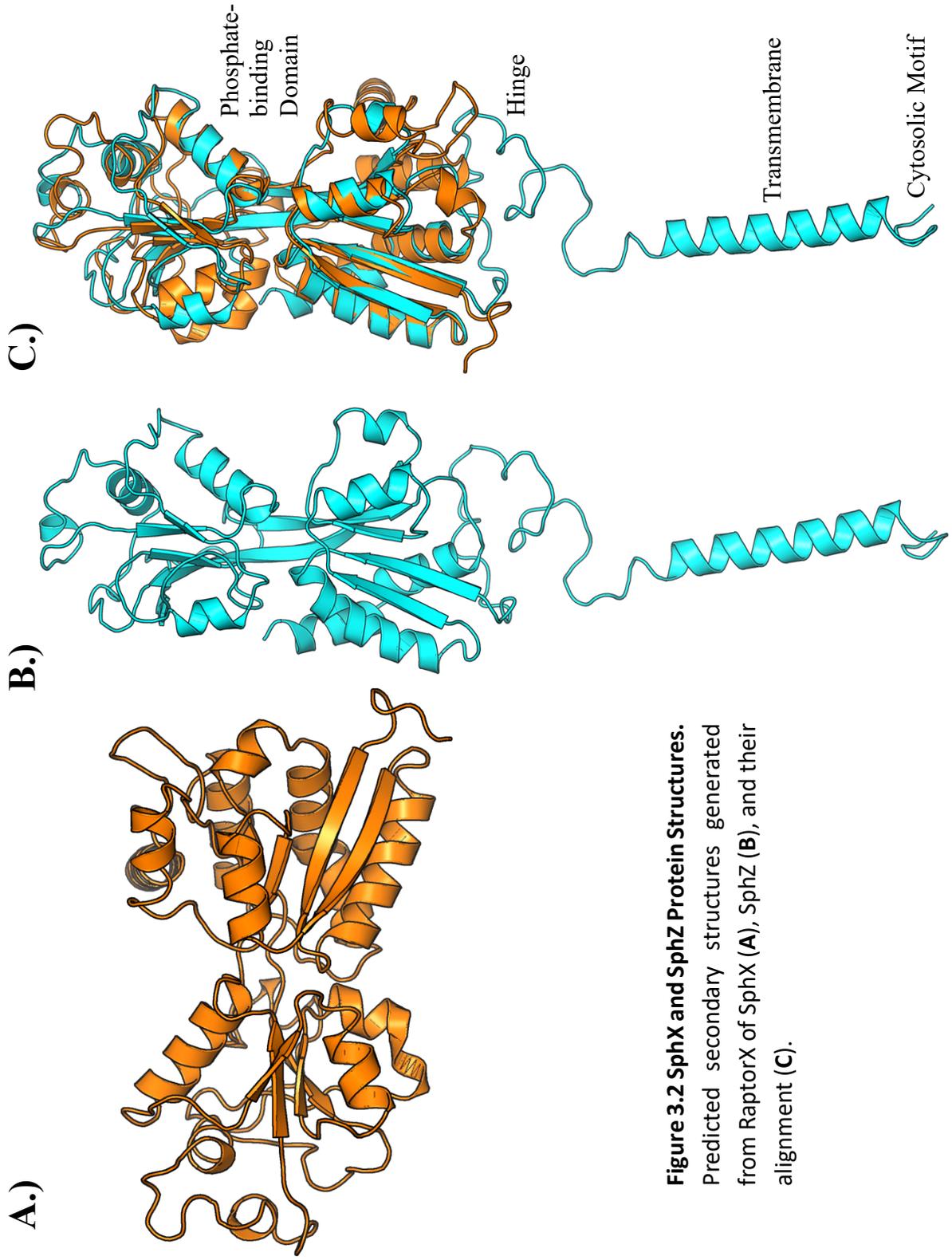

**Figure 3.2 SphX and SphZ Protein Structures.** Predicted secondary structures generated from RaptorX of SphX (**A**), SphZ (**B**), and their alignment (**C**).

### 3.1.3 Electrostatic Surfaces of SphX and SphZ

The pdb files generated from RaptorX were uploaded to the PDB2PQR server in order to obtain Poisson-Boltzmann biomolecular electrostatics for each of their secondary protein structures with the default parameters for solvation energy (PARSE) force field and PROPKA to predict protonation states (Sitkoff *et al*., 1994; Li *et al*., 2005; Tang *et al.*, 2007; Søndergaard *et al*., 2011). The pqr files were then imported into PyMOL and the electrostatic potential was displayed through the adaptive Poisson-Boltzmann solver (APBS) plug-in (Figure 3.3).

**A.)**

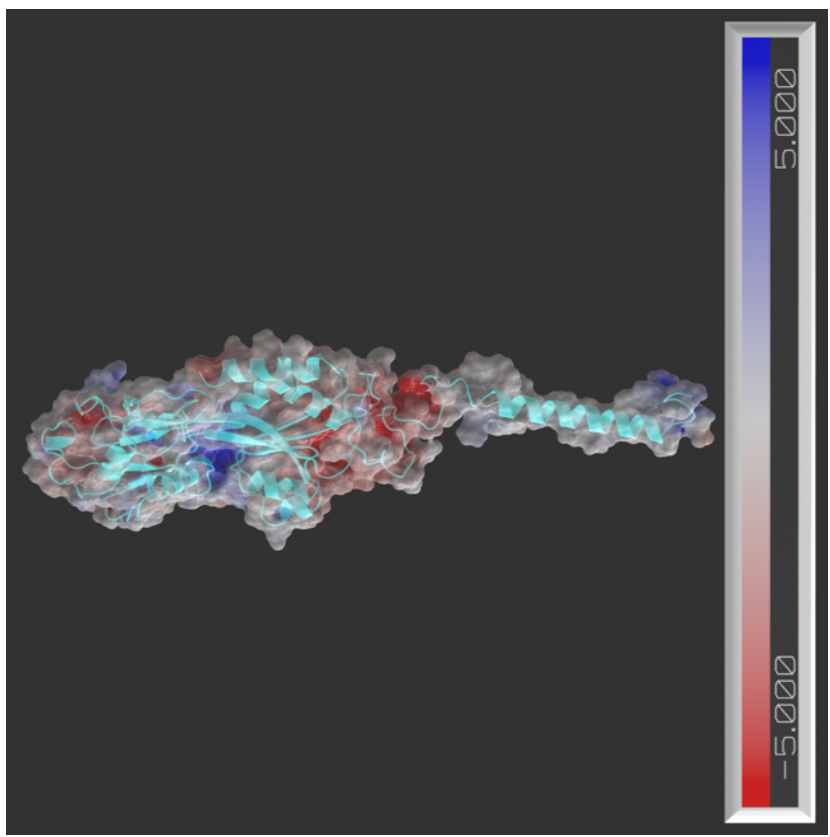

**B.)**

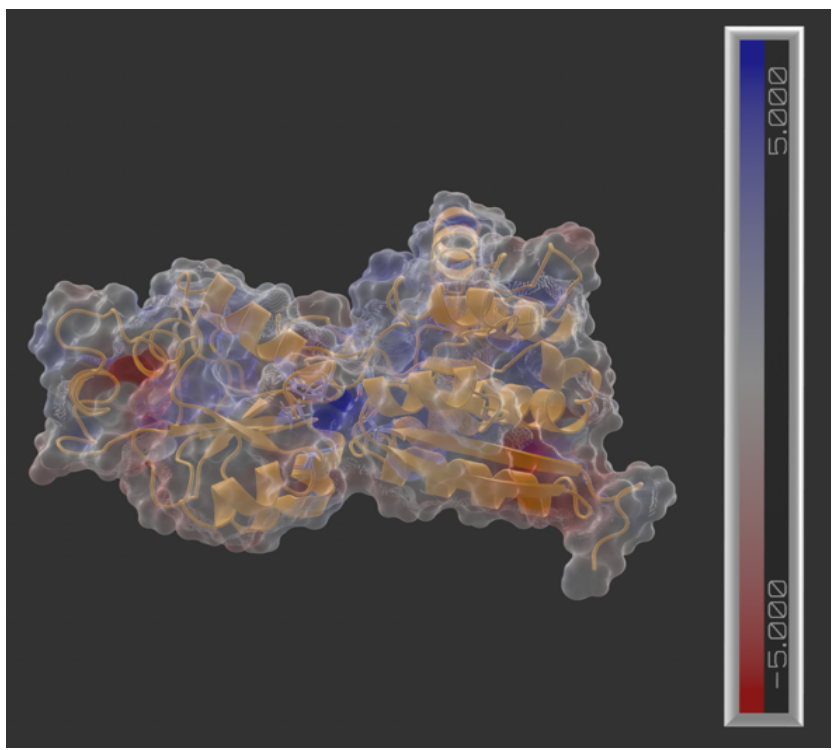

**Figure 3.3 Protein Surface Charges.** Electrostatic surface potentials for SphX (**A**) and SphZ (**B**) are shown (red, electronegative; white, neutral; blue, electropositive; ± 5 kJ · mol$^{-1}$ · e$^{-1}$). Calculations were performed with a protein dielectric of 2, solvent dielectric of 78, temperature of 310 K, and 0.15 M ionic strength.

**3.1.4 Phosphate-Binding Pocket of SphX and SphZ**

The putative phosphate-binding ligands were retrieved from InterPro and were annotated with another color, namely purple for SphX and pink for SphZ. The predicted phosphate-binding ligands of SphX were Ser-38, Ser-86, Pro-165, Ser-169, Gly-170, Thr-171, Phe-172, and Asp-196. The predicted phosphate-binding ligands of SphZ were Ser-78, Ser-126, Arg-193, Ser-197, Gly-198, Thr-199, Gln-200, and Asp-225.

The electrostatic surfaces were analyzed to see if a positively charged pocket formed near the predicted phosphate-binding ligands through altering the transparency of the generated surface (Figure 3.4; Figure 3.5). The SphX electrostatic surface generated a positive pocket of 5 kJ $\cdot$ mol$^{-1}$ $\cdot$ e$^{-1}$ near its center (Figure 3.4.A). The binding-pocket was in close proximity to the putative phosphate-binding ligands (Figure 3.4.B-C).

The SphZ electrostatic surface also generated a positive pocket of 5 kJ $\cdot$ mol$^{-1}$ $\cdot$ e$^{-1}$ near the center of the phosphate-binding domain (Figure 3.5.A). The binding pocket was also in close proximity to the its putative phosphate-binding ligands (Figure 3.5.B-C). It was also noted that the hinge structure generated a negative electrostatic surface around the entire protein (Figure 3.5.A).

The potential phosphate-binding ligands were then overlaid (Figure 3.6). The position of aspartic acid residues, namely Asp-196 from SphX and Asp-225 from SphZ, varied in position within $\mathbb{R}^3$ (Figure 3.6.A-C). Only two differences between phosphate-binding ligands were noted, namely Pro-165 from SphX and Arg-193 from SphZ as well Phe-172 from SphX and Gln-200 from SphZ (Figure 3.6.A-C).

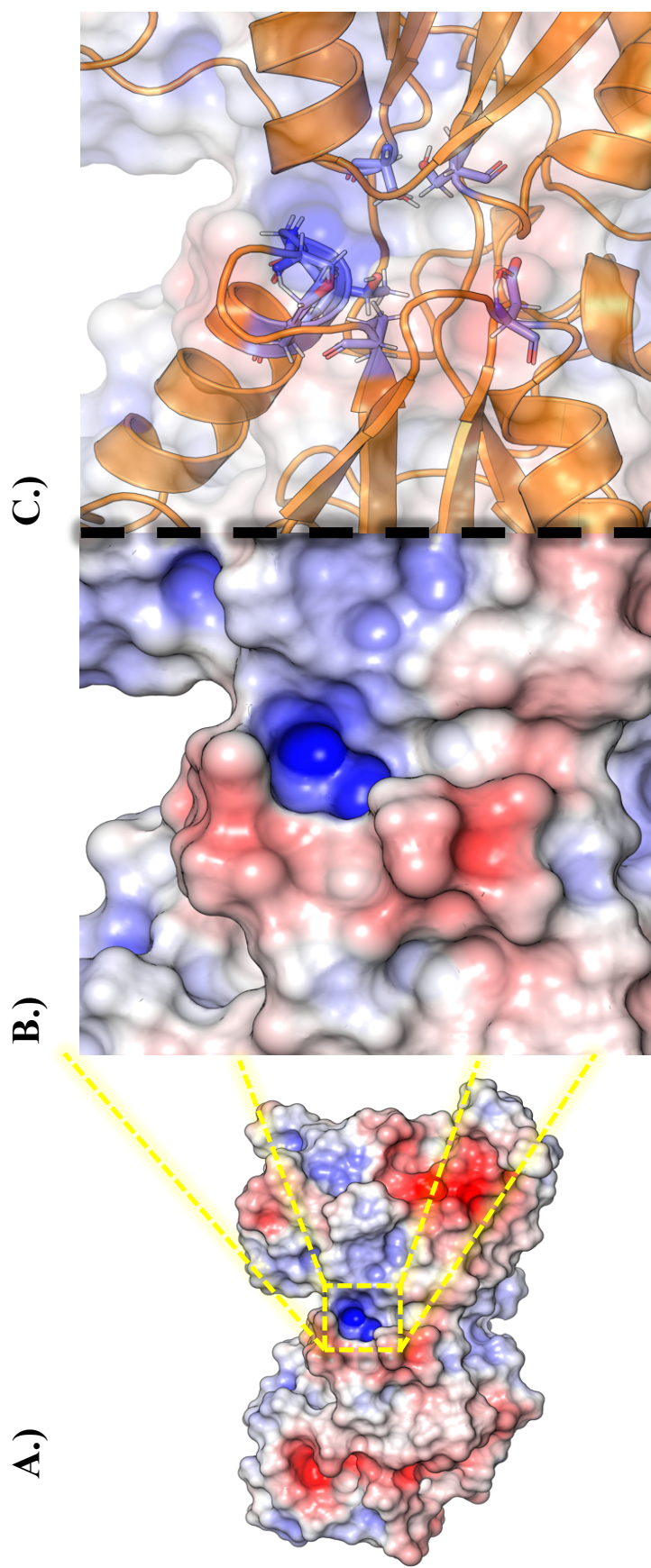

**Figure 3.4 SphX Phosphate-Binding Site.** The electrostatic surface of SphX (**A**), predicted phosphate-binding site (**B**), and the phosphate-binding ligands, indicated in purple (**C**). Electrostatic surface potentials for SphX are shown (red, electronegative; white, neutral; blue, electropositive; $\pm$ 5 kJ $\cdot$ mol$^{-1}$ $\cdot$ e$^{-1}$). Calculations were performed with a protein dielectric of 2, solvent dielectric of 78, temperature of 310 K, and 0.15 M ionic strength.

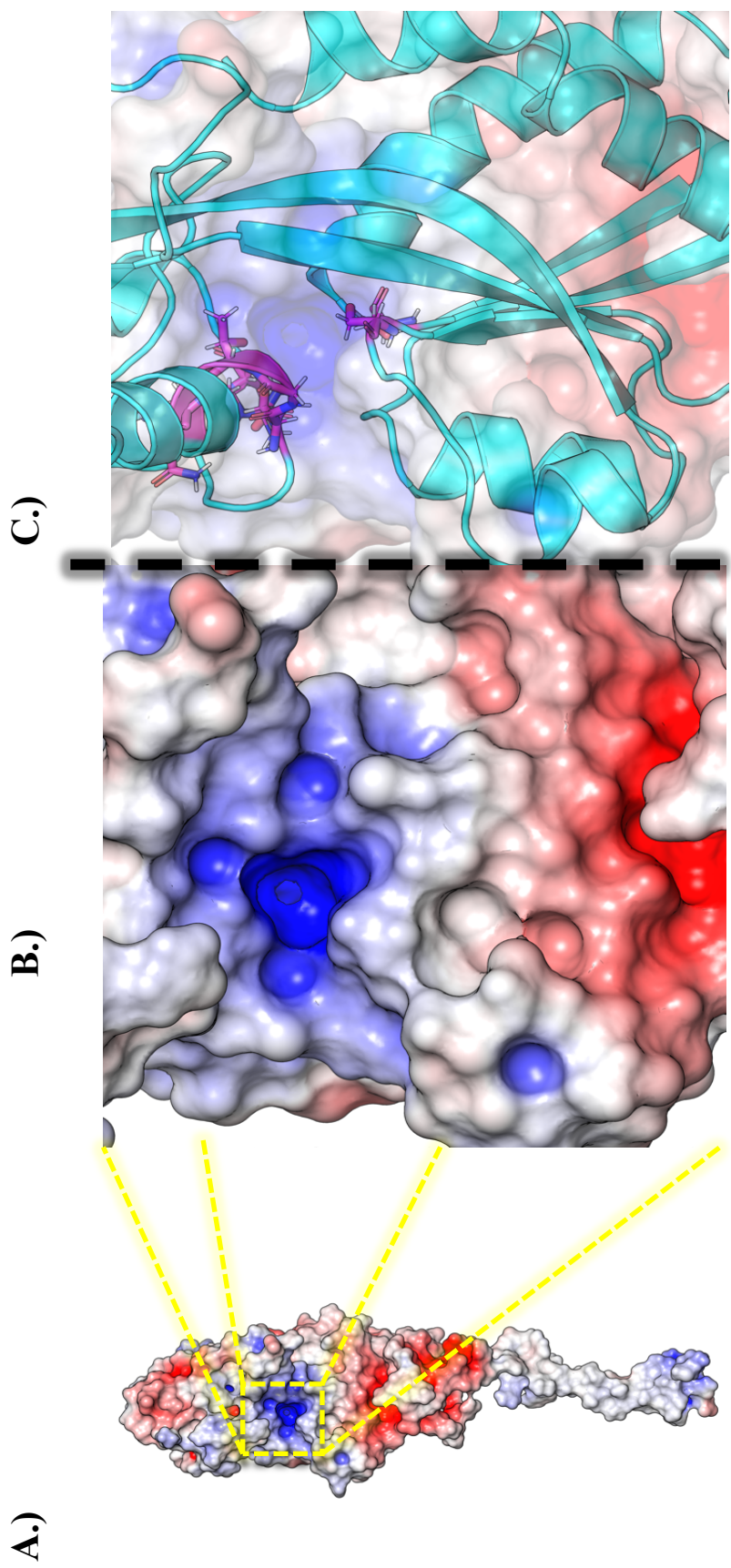

**Figure 3.5 SphZ Phosphate-Binding Site.** The electrostatic surface of SphZ (**A**), predicted phosphate-binding site (**B**), and the phosphate-binding ligands, indicated in pink (**C**). Electrostatic surface potentials for SphZ are shown (red, electronegative; white, neutral; blue, electropositive; ± 5 kJ · mol⁻¹ · e⁻¹). Calculations were performed with a protein dielectric of 2, solvent dielectric of 78, temperature of 310 K, and 0.15 M ionic strength.

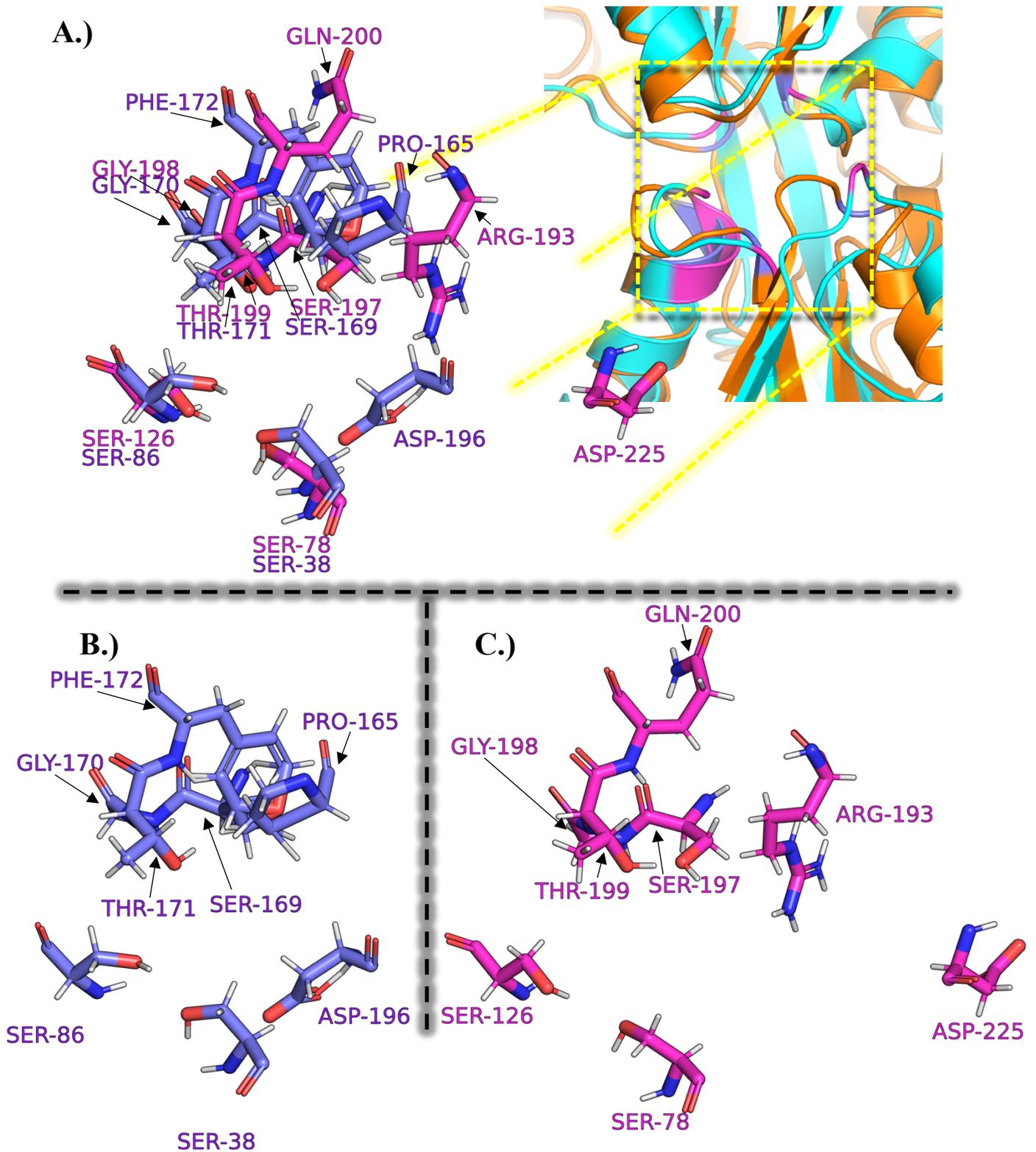

**Figure 3.6 Comparison of SphX and SphZ Phosphate-Binding Site Residues.** The phosphate-binding site and ligands of SphX overlaid on SphZ (**A**). The phosphate-binding ligands of SphX: Ser-38, Ser-86, Pro-165, Ser-169, Gly-170, Thr-171, Phe-172, Asp-196 (**B**). The phosphate-binding ligands of SphZ: Ser-78, Ser-126, Arg-193, Ser-197, Gly-198, Thr-199, Gln-200, Asp-225 (**C**).

## 3.2 Physiological Characterization

### 3.2.1 Cell Counts per OD$_{730nm}$

After being grown for 36 h in BG-11, strains were analyzed with the same culture density of an OD$_{730nm}$ = 0.01 (Figure 3.7). There were statistically significant differences between group means, determined by one-way ANOVA ($F$(3,8) = 10.49, $MSE$ = 5.36$_E$14, $p <$ 0.005). Post hoc analyses were conducted using Tukey's HSD test: there were statistically significant differences between $\Delta$SphX and WT ($p <$ 0.05), $\Delta$SphZ ($p <$ 0.005), and $\Delta$SphX:SphZ ($p <$ 0.05).

### 3.2.2 Growth Curves

Growth was analyzed over the course of one week by measuring the turbidity of the culture, namely OD$_{730nm}$. These measurements were converted to the number of cells through linear transformation, utilizing the cell counts and normalized with an initial number of cells set to $10^7$ (Figure 3.8).

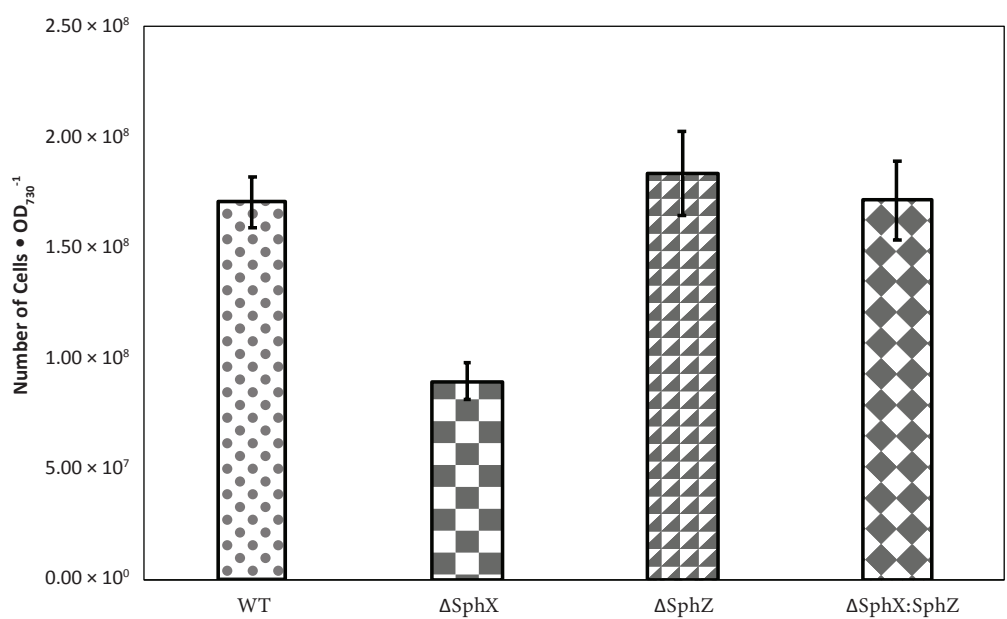

**Figure 3.7 Number of Cells per OD$_{730nm}$.** Number of cells counted at an OD$_{730nm}$ of 0.01 after WT, $\Delta$SphX, $\Delta$SphZ, and $\Delta$SphX:SphZ strains were grown in BG-11 for 36 h. Values are means ± S.E. (n = 3).

There were statistically significant differences between group means for strains grown in BG-11 after 168 h, determined by one-way ANOVA ($F(3,8) = 17.16$, $MSE = 1.75_E15$, $p < 0.001$; Figure 3.8). Post hoc analyses were conducted using Tukey's HSD test. There were statistically significant differences between ΔSphX and: WT ($p < 0.005$), ΔSphZ ($p < 0.005$), and ΔSphX:SphZ ($p < 0.005$).

There were statistically significant differences between group means for strains grown in BG-11 – $P_i$ after 168 h, determined by one-way ANOVA ($F(3,8) = 8.13$, $MSE = 1.25_E15$, $p < 0.01$; Figure 3.8). Post hoc analyses were conducted using Tukey's HSD test. There were statistically significant differences between ΔSphZ and: WT ($p < 0.05$) and ΔSphX ($p < 0.01$).

There were statistically significant differences between group means for strains grown in BG-11 – $K_2HPO_4$ after 168 h, determined by one-way ANOVA ($F(3,8) = 37.73$, $MSE = 2.53_E12$, $p < 0.001$; Figure 3.8). Post hoc analyses were conducted using Tukey's HSD test. There were statistically significant differences between WT and: ΔSphZ ($p < 0.001$) and ΔSphX:SphZ ($p < 0.05$). There were also statistically significant differences between ΔSphX and: ΔSphZ ($p < 0.001$) and ΔSphX:SphZ ($p < 0.001$).

These differences could directly be seen from turbidity and coloration (Figure 3.9). The acclimation to phosphate limitation can be seen as a difference in the intensity of color for all strains (Figure 3.9.A) and specifically for WT (Figure 3.9.B) and ΔSphZ (Figure 3.9.C), where the coloration is more pronounced for ΔSphZ in comparison to WT.

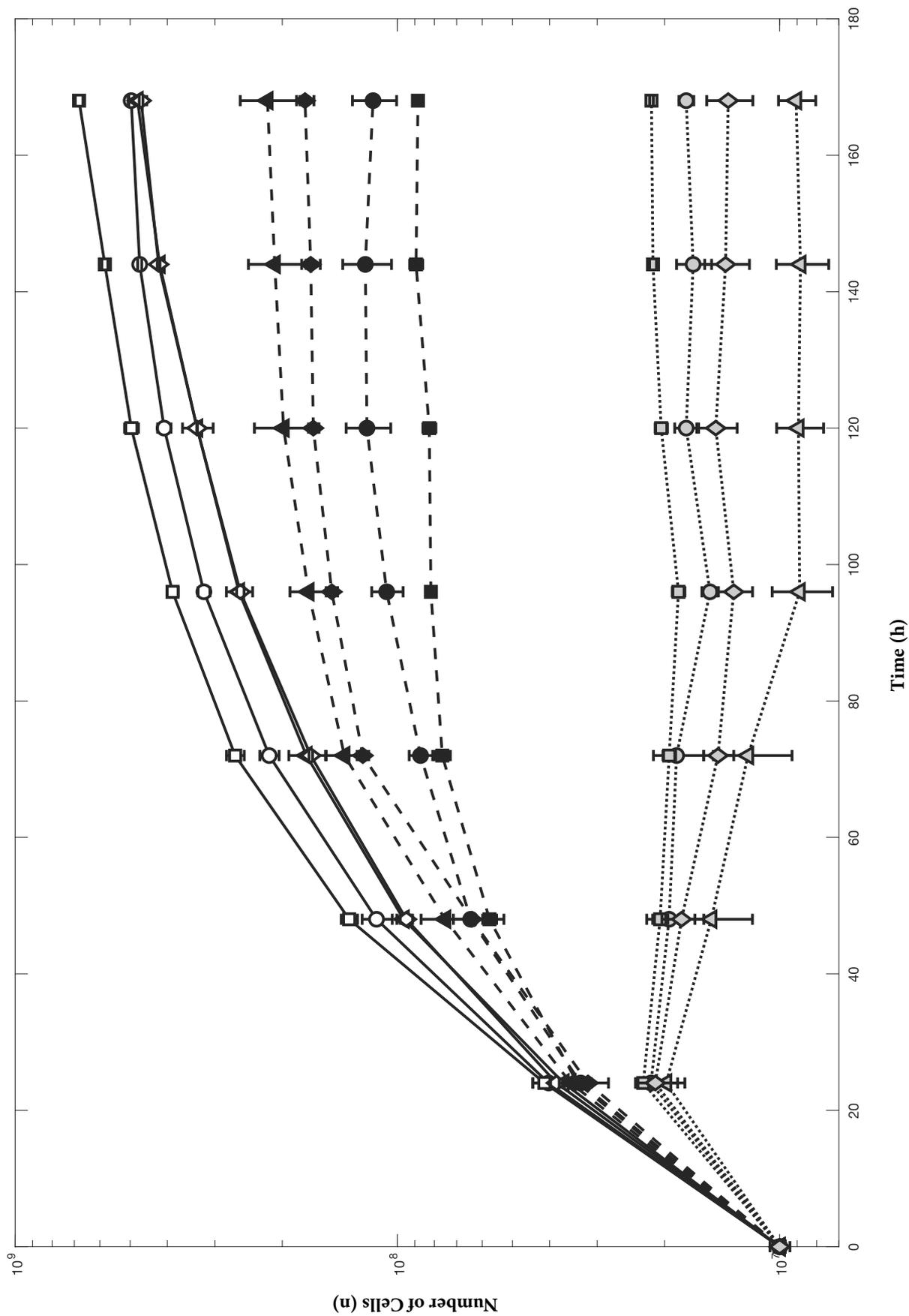

**Figure 3.8 Growth Curves.** Growth over the course of 1 week of WT (circles), ΔSphX (squares), ΔSphZ (triangles), ΔSphX:SphZ (diamonds) grown in BG-11 (white), BG-11 − $P_i$ (black), and BG-11 − $K_2HPO_4$ (grey). Values are means ± S.E. (n = 3).

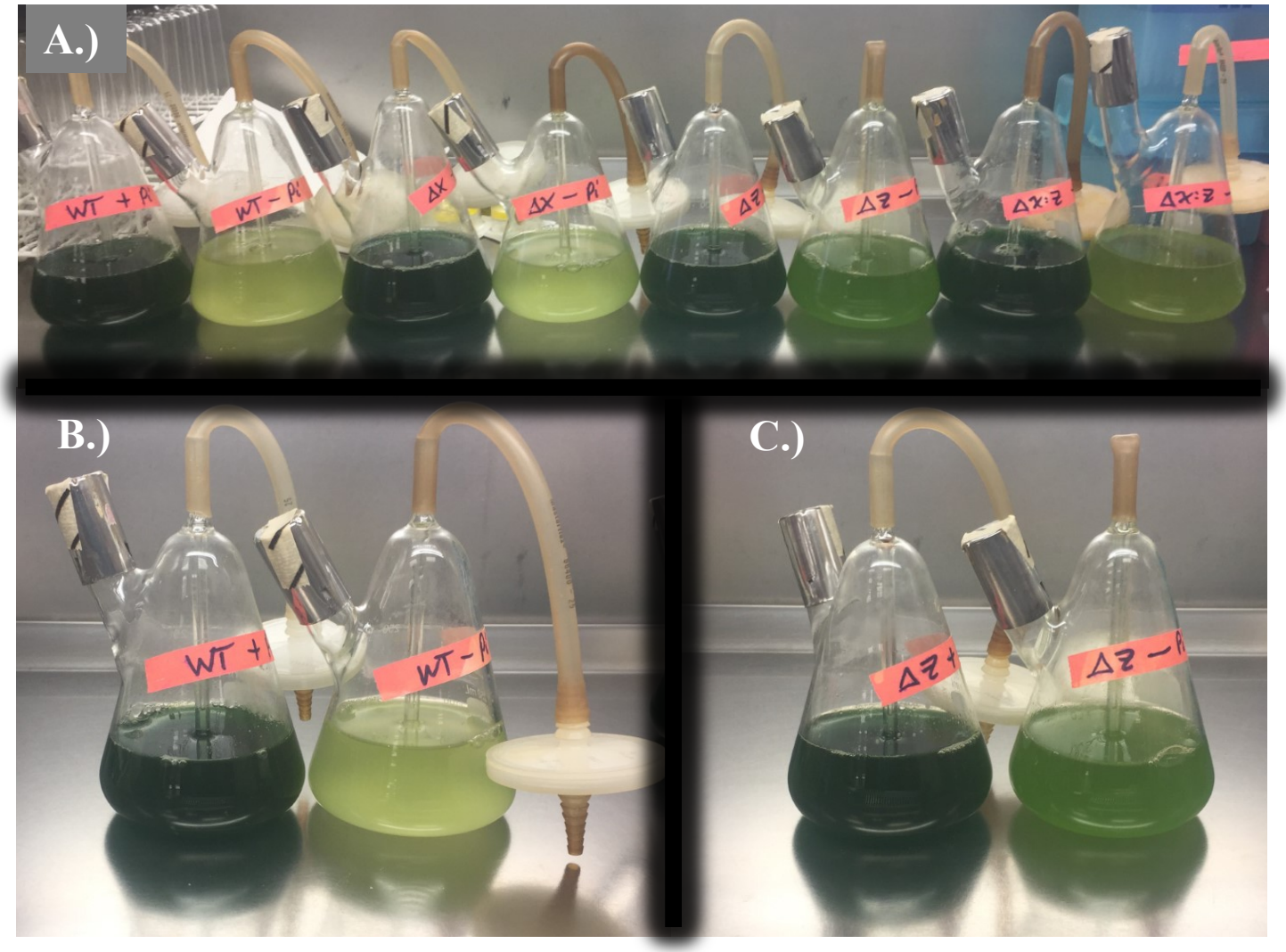

**Figure 3.9 Visual Comparison of Strains Grown with BG-11 and BG-11 − P$_i$.** From left to right in (**A**): WT grown in BG-11; WT grown in BG-11 − P$_i$; ΔSphX grown in BG-11; ΔSphX grown in BG-11 − P$_i$; ΔSphZ grown in BG-11; ΔSphZ grown in BG-11 − P$_i$; ΔSphX:SphZ grown in BG-11; ΔSphX:SphZ grown in BG-11 − P$_i$. (**B**) A closer examination of WT grown in BG-11 (left) and BG-11 − P$_i$ (right). (**C**) A closer examination of ΔSphZ grown in BG-11 (left) and BG-11 − P$_i$ (right). After one week, strains grown with BG-11 − P$_i$ have a lesser intensity of coloration in comparison to each respective strain grown with BG-11 (**A**). However, ΔSphZ and ΔSphX:SphZ have darker coloration in comparison to WT when grown with BG-11 − P$_i$ (**B; C**).

### 3.2.2.1 Modelling the Growth Curves

Information is stored within these growth curves, so mathematical models were devised to best fit each of them (Figure 3.10). Strains grown in BG-11 were modelled through nonlinear least-squares regression in R with the logistic function: $\frac{(a \cdot b)}{(a+(b-a) \cdot e^{(-rt)})}$.

Parameters for WT were estimated with $a = 2.09_E07$ ($p < 0.01$), $b = 5.12_E08$ ($p < 0.001$), and $r = 3.85_E\text{-}02$ ($p < 0.001$). The first and second derivatives indicated that the maximum growth rate and inflection point occurred at ~82 h (Figure 3.10.A1-3). Parameters for ΔSphX were estimated with $a = 3.10_E07$ ($p < 0.05$), $b = 7.08_E08$ ($p < 0.001$), and $r = 3.37_E\text{-}02$ ($p < 0.001$). The first and second derivatives indicated that the maximum growth rate and inflection point occurred at ~92 h (Figure 3.10.A1-3). Parameters for ΔSphZ were estimated with $a = 2.46_E07$ ($p < 0.01$), $b = 5.28_E08$ ($p < 0.001$), and $r = 3.06_E\text{-}02$ ($p < 0.001$). The first and second derivatives indicated that the maximum growth rate and inflection point occurred at ~99 h (Figure 3.10.A1-3). Parameters for ΔSphX:SphZ were estimated with $a = 2.32_E07$ ($p < 0.01$), $b = 5.15_E08$ ($p < 0.001$), and $r = 3.15_E\text{-}02$ ($p < 0.001$). The first and second derivatives indicated that the maximum growth rate and inflection point occurred at ~97 h (Figure 3.10.A1-3).

Strains grown in BG-11 – $P_i$ were modelled through nonlinear least-squares regression in R with a Fourier polynomial: $a_0 + a_k \cos(kx) + b_k \sin(kx) + c_k \cos(2kx) + d_k \sin(2kx)$. Parameters for WT were estimated with $a_0 = 4.15_E07$ ($p < 0.05$), $a_k = 2.05_E06$, $b_k = 2.25_E06$, $c_k = \text{-}3.35_E07$ ($p < 0.05$), $d_k = \text{-}7.33_E07$ ($p < 0.05$), and $k = 1.17$ ($p < 0.001$). The first and second derivatives indicated that the maximum growth rate and inflection point occurred at ~35 h (Figure 3.10.B1-3). Parameters for ΔSphX were estimated with $a_0 = 5.70_E07$ ($p < 0.01$), $a_k = \text{-}3.87_E07$ ($p < 0.01$), $b_k = 9.50_E06$, $c_k = \text{-}8.03_E06$, $d_k = 8.96_E06$, and $k = 1.33$ ($p < 0.001$). The first derivative indicated a global maximum growth rate occurred at ~28 h and a local maximum at ~134 h (Figure 3.10.A1-3). The second derivative indicated inflection points occurred at ~28 h, 103 h, and 134 h (Figure 3.10.B1-3). Parameters for ΔSphZ were estimated with $a_0 = $

1.42$_E$08, $a_k$ = -1.05$_E$08, $b_k$ = -6.42$_E$06, $c_k$ = -2.59$_E$07, $d_k$ = 7.82$_E$06, and $k$ = 1.33 ($p$ < 0.001).

The first and second derivatives indicated that the maximum growth rate and inflection point occurred at ~53 h with a secondary inflection point at ~153 h (Figure 3.10.A1-3). Parameters for ΔSphX:SphZ were estimated with $a_0$ = 1.24$_E$08, $a_k$ = -8.26$_E$07, $b_k$ = -1.38$_E$07, $c_k$ = -3.01$_E$07,

$d_k$ = 5.16$_E$06, and $k$ = 1.33 ($p$ < 0.001). The first and second derivatives indicated that the

maximum growth rate and inflection point occurred at ~53 h with a secondary inflection point

at ~139 h (Figure 3.10.A1-3).

Strains grown in BG-11 – K$_2$HPO$_4$ were modelled through nonlinear least-squares regression in R with a rational polynomial: $\frac{(a_k x^4 + b_k x^3 + c_k x^2 + d_k x + e_k)}{(x + f_k)}$. Parameters for WT were estimated with $a_k$ = -11.25, $b_k$ = 4.65$_E$03, $c_k$ = -6.20$_E$05, $d_k$ = 4.46$_E$07, $e_k$ = 2.19$_E$08, $f_k$ = 21.92.

The model has local maxima at ~25 and 166 h (Figure 3.10.C1). The first derivative had roots

at ~25 h, 106 h, and 166 h (Figure 3.10.C2). The second derivative had roots at ~48 h and 136

h (Figure 3.10.C3). Parameters for ΔSphX were estimated with $a_k$ = -18.51, $b_k$ = 6.86$_E$03, $c_k$ = -

7.90$_E$05, $d_k$ = 5.00$_E$07, $e_k$ = 2.28$_E$08, $f_k$ = 22.75. The model has local maxima at ~24 and 159 h

(Figure 3.10.C1). The first derivative had roots at ~24 h, 84 h, and 158 h (Figure 3.10.C2). The second derivative had roots at ~43 h and 123 h (Figure 3.10.C3). Parameters for ΔSphZ were estimated with $a_k$ = -8.62, $b_k$ = 3.90$_E$03, $c_k$ = -5.73$_E$05, $d_k$ = 3.61$_E$07, $e_k$ = 1.15$_E$08, $f_k$ = 11.45.

The model has local maxima at ~16 and 175 h (Figure 3.10.C1). The first derivative had roots

at ~16 h, 127 h, and 175 h (Figure 3.10.C2). The second derivative had roots at ~39 h and 151 h

(Figure 3.10.C3). Parameters for ΔSphX:SphZ were estimated with $a_k$ = -23.70, $b_k$ = 8.74$_E$03,

$c_k$ = -1.05$_E$06, $d_k$ = 5.54$_E$07, $e_k$ = 2.95$_E$08, $f_k$ = 29.53. The model has local maxima at ~22 h and

143 h (Figure 3.10.C1). The first derivative had roots at ~22 h, 101 h, and 143 h (Figure

3.10.C2). The second derivative had roots at ~44 h and 123 h (Figure 3.10.C3).

The first derivative of the modelled curves indicates the growth rate; whereas, the second derivative indicates intrapopulation response time with the curbing of growth rate to a carrying capacity. When grown in BG-11, ΔSphX had an increased maximum growth rate (Figure 3.10.A2) with a slower response (Figure 3.10.A3) compared to WT. ΔSphZ and ΔSphX:SphZ had a decreased maximum growth rate (Figure 3.10.A2) and the slowest response (Figure 3.10.A3). When grown in BG-11 – $P_i$, ΔSphX accumulated lower biomass (Figure 3.10.B1) with a similar maximum growth rate (Figure 3.10.B2) and faster response time (Figure 3.10.B3) as compared to WT. On the other hand, ΔSphZ and SphX:SphZ accumulated more biomass (Figure 3.10.B1) with an increased maximum growth rate (Figure 3.10.B2) and once again the slowest response time for curbing growth (Figure 3.10.B3). When grown in BG-11 – $K_2HPO_4$, all of the strains were dying, and their responses were ubiquitous (Figure 3.10.C1-3).

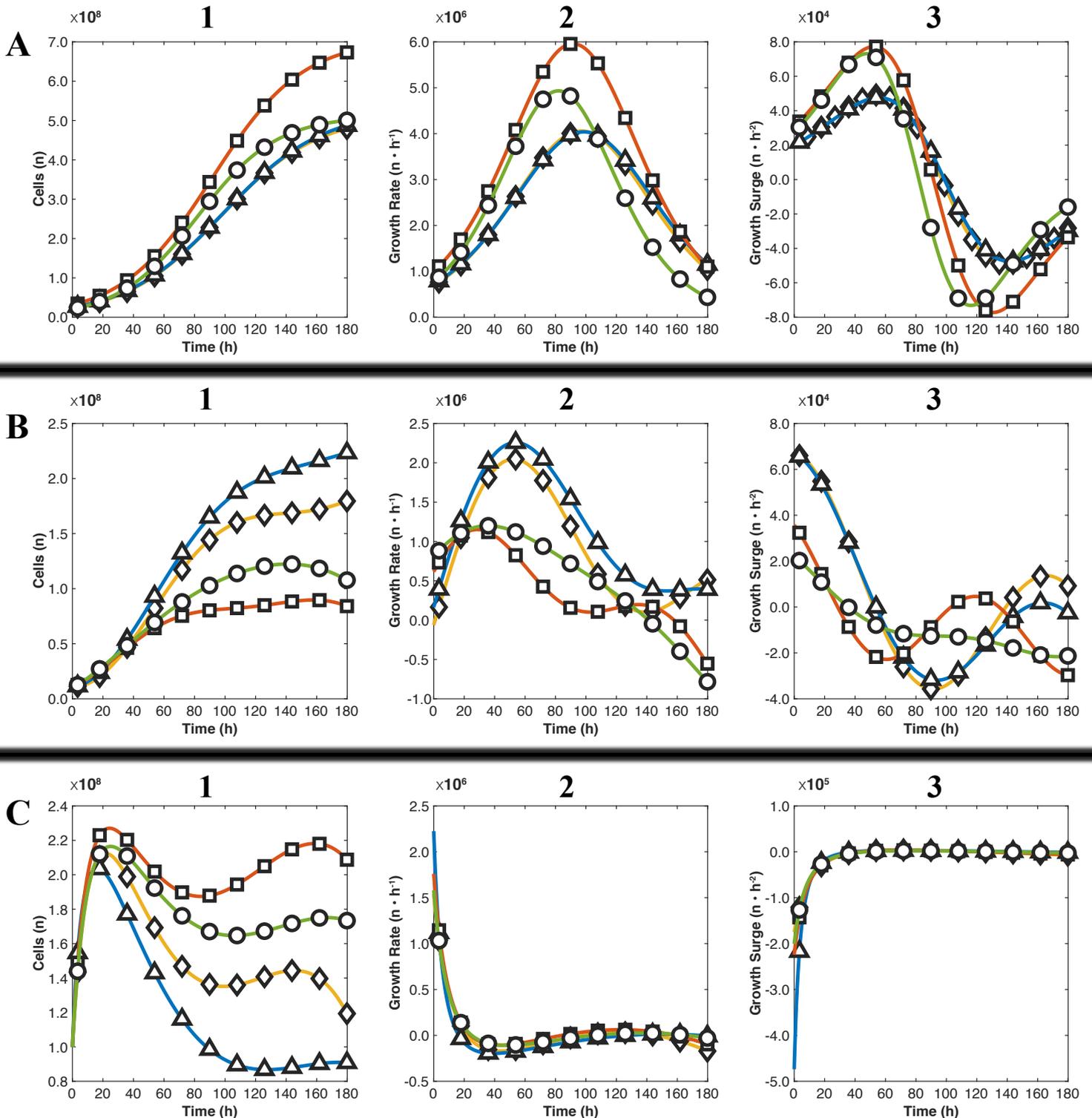

**Figure 3.10 Growth Curve Models.** Mathematical models of growth curves: (**A1**) logistic functions for those grown in BG-11; (**B1**) Fourier series for those grown in BG-11 − $P_i$; (**C1**) rational polynomials for those grown in BG-11 − $K_2HPO_4$. First derivative of the growth models are plotted in the second column (**A2**,**B2**,**C2**). Second derivative of the growth models are plotted in the third column (**A3**,**B3**,**C3**). WT (circles) models are green, ΔSphX (squares) models are orange, ΔSphZ (triangles) models are blue, and ΔSphX:SphZ (diamonds) models are yellow.

### 3.2.4 Measuring Total Phosphorus Over the Course of Growth

Total phosphorus content was measured over the course of a week as strains were grown in various media (Figure 3.11). The BG-11 – $P_i$ and BG-11 – $K_2HPO_4$ media ensured total phosphorus concentration was maintained under 10 µg · $L^{-1}$ after 24 h (Figure 3.11). Phosphorus was depleted throughout BG-11 media over the course of a week (Figure 3.11). Measurements from WT cultures indicated complete depletion of total phosphorus by ~96 h and the mutant strains appeared to be delayed (Figure 3.11).

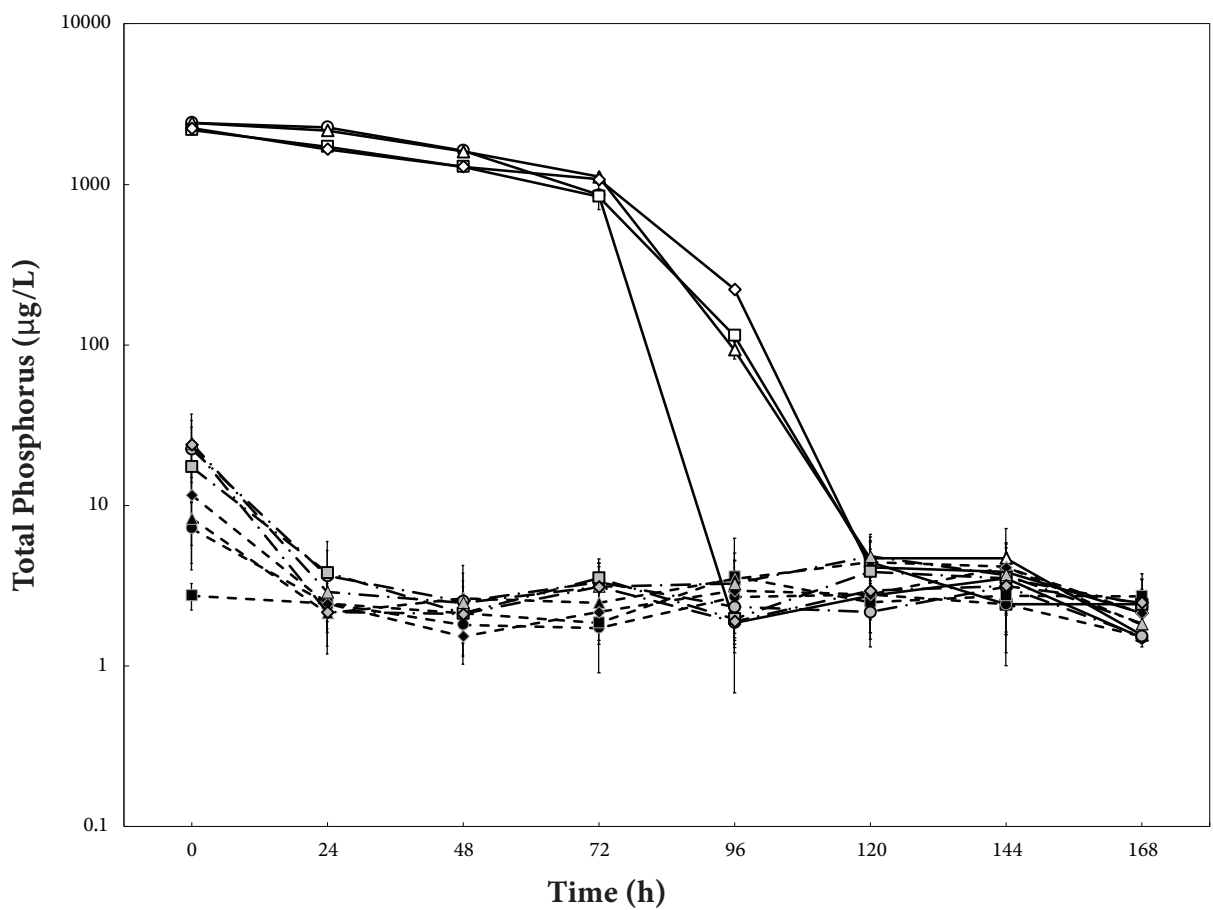

**Figure 3.11 Time Course of Total Phosphorus.** Extracellular total phosphorus was measured every 24 h for strains grown in BG-11 (white), BG-11 – $P_i$ (black), and BG-11 – $K_2HPO_4$ (grey). Values are means for WT (circles), ΔSphX (squares), ΔSphZ (triangles), and ΔSphX:SphZ (diamonds) ± S.E. plotted on a logarithmic y-axis (n=3).

Total phosphorus depletion from BG-11 media is directly attributed to the uptake of phosphate, particularly since the only form of phosphorus supplied in the media is $K_2HPO_4$ as previously stated, so this was further analyzed utilizing the same data (Figure 3.12). The initial concentration of total phosphorus was normalized to 2500 µg · L⁻¹ (Figure 3.12). Over the course of 48 h, the intake of phosphate was faster for ΔSphX and ΔSphX:SphZ; whereas, the ΔSphZ was comparable to WT (Figure 3.12). At 72 h, there appeared to be clustering of ΔSphZ with ΔSphX:SphZ and WT with ΔSphX (Figure 3.12). As previously stated, phosphate was depleted by ~96 h for WT and the mutant strains appeared to lag behind (Figure 3.12).

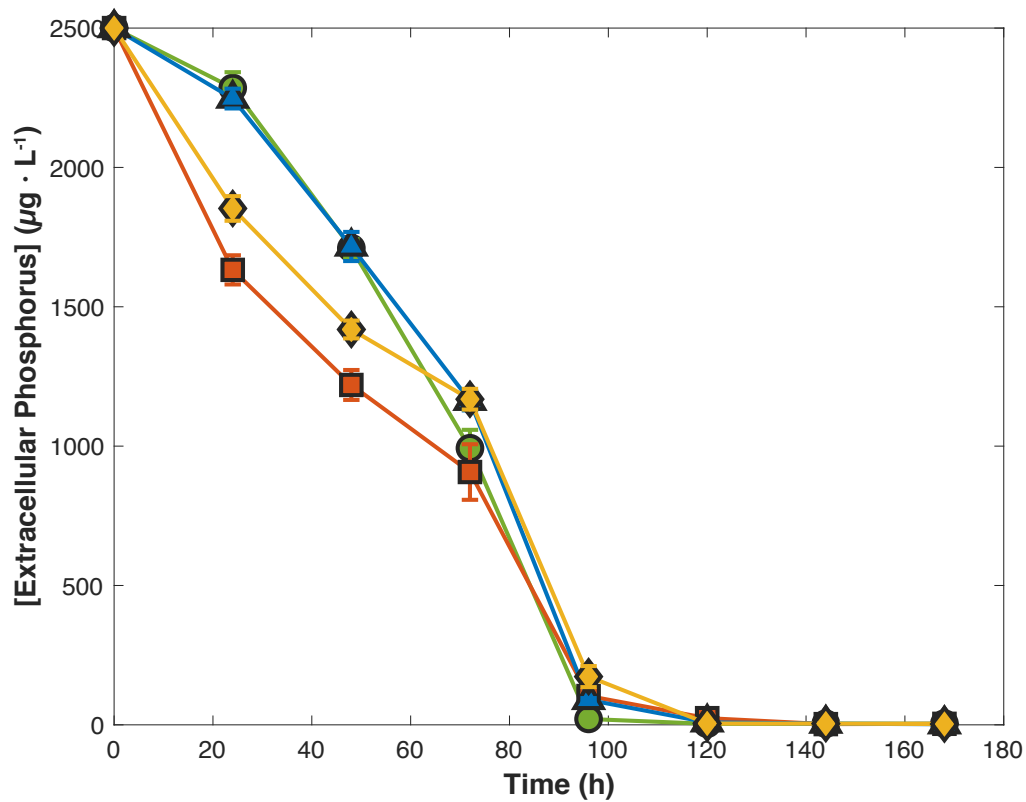

**Figure 3.12 Phosphorus Depletion from BG-11.** Extracellular phosphorus normalized with an initial concentration of 2500 µg · L⁻¹. WT (circles) measurements are indicated in green; ΔSphX (squares) measurements are indicated in orange; ΔSphZ (triangles) measurements are indicated in blue; ΔSphX:SphZ (diamonds) measurements are indicated in yellow. Values are means ± S.E. (n=3).

Since these measurements were taken over the course of the strain's growth, it follows that there may be a connection of phosphate uptake based upon the number of cells present in the media. A parametric curve was then generated to assess this potential with the following vector form:

$$< time, number\ of\ cells, extracellular\ phosphorus >.$$

A biharmonic spline interpolant surface could then be generated for each strain (Figure 3.13.A-D). The WT and ΔSphZ surfaces are similar in form (Figure 3.13.A; Figure 3.13.C). Moreover, the ΔSphX and ΔSphX:SphZ surfaces were also similar (Figure 3.13.B; Figure 3.13.D). It follows to proceed in analyzing the resultant contour plots from the generated surfaces as the height function in this case is attributed to decreasing concentrations of extracellular phosphorus (Figure 3.13.E-H). It can be clearly seen that WT and ΔSphZ have similar contour lines with regard to the interval between them (Figure 3.13.E; Figure 3.13.G). Moreover, ΔSphX and ΔSphX:SphZ appear to have warped contour lines corresponding to fewer cells and within the 72-hour period as compared to WT (Figure 3.13.F; Figure 3.13.H).

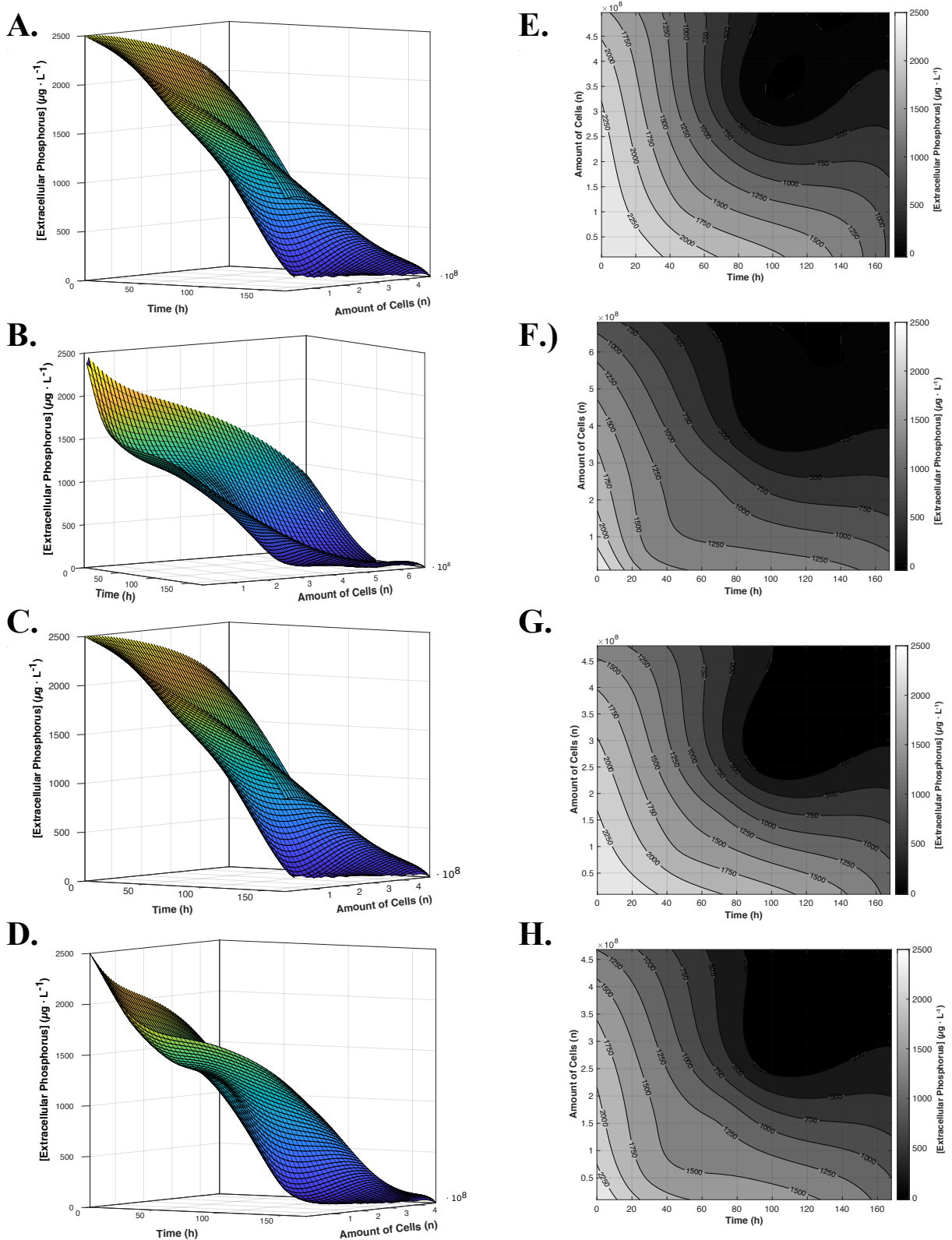

**Figure 3.13 3D Parametric Surfaces.** Biharmonic spline interpolant surfaces generated growth curves and extracellular phosphorus measurements: WT (**A**), ΔSphX (**B**), ΔSphZ (**C**), ΔSphX:SphZ (**D**). Contour plots generated from surfaces: WT (**E**), ΔSphX (**F**), ΔSphZ (**G**), ΔSphX:SphZ (**H**).

Analogous to the projection onto the yz-plane, the number of cells were directly plotted against extracellular phosphorus (Figure 3.14). Mutant strains highlight differences in the depletion of phosphorus from the media in comparison to the linearity showcased by WT (Figure 3.14). $\Delta$SphX and $\Delta$SphX:SphZ, as previously noted, uptake phosphorus at a faster rate as compared to that of WT with fewer cells when extracellular phosphorus is abundant (Figure 3.14). In addition, a nexus was found through the implementation of this parametric mapping, specifically when extracellular phosphorus is at 1160 $\mu$g $\cdot$ L$^{-1}$ which is ~12 $\mu$M of phosphorus, as seen with the intersection of $\Delta$SphX, $\Delta$SphZ, and $\Delta$SphX:SphZ (Figure 3.14). At this point, ΔSphX:SphZ begins to exhibit a similar response as ΔSphZ which is depleting phosphorus with less cells than required by WT (Figure 3.14). On the other hand, ΔSphX exhibits a stagnation – ultimately requiring more cells than that of WT in order to continually deplete phosphorus from the media (Figure 3.14).

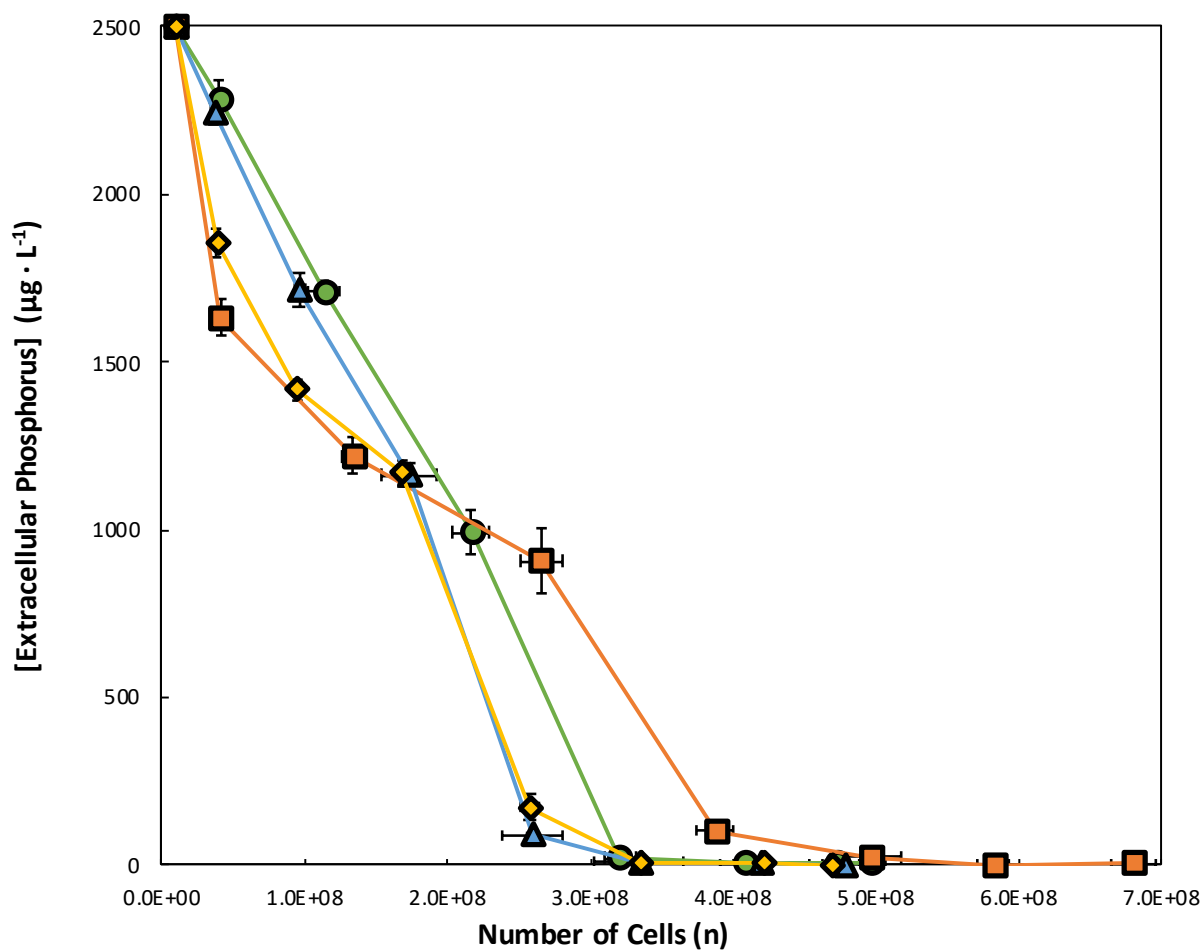

**Figure 3.14 Parametric Analysis of Extracellular Phosphorus against Growth Curves.** Extracellular phosphorus directly plotted against growth curves. WT (circles) are indicated in green, $\Delta$SphX (squares) are indicated in orange, $\Delta$SphZ (triangles) are indicated in blue, and $\Delta$SphX:SphZ (diamonds) are indicated in yellow. Values are means ± S.E. (n=3).

### 3.2.5 Alkaline Phosphatase Activity

The activity of alkaline phosphatase and, by extension, the upregulation of the pho regulon was measured after 36 h (Figure 3.15). The measurements were normalized with the mean value of WT grown in BG-11 – $P_i$ equivalent to 100 units $\cdot$ Min$^{-1}$ $\cdot$ Cell$^{-1}$ (Figure 3.15).

Group means of strains grown in BG-11 after 36 h failed to reject the null hypothesis as determined by one-way ANOVA ($F(3,8) = 2.943$, $MSE = 8.08_E\text{-}16$, $p > 0.05$). However, there were statistically significant differences between group means for strains grown in BG-11 – $P_i$ after 36 h as determined by one-way ANOVA ($F(3,8) = 5.32$, $MSE = 1.71_E\text{-}11$, $p < 0.05$). Post hoc analyses were conducted using Tukey's HSD test. There were statistically significant differences between ΔSphX and: ΔSphZ ($p < 0.05$) and ΔSphX:SphZ ($p < 0.05$). In addition, there were statistically significant differences between group means for strains grown in BG-11 – $K_2HPO_4$ after 36 h as determined by one-way ANOVA ($F(3,8) = 11.70$, $MSE =$ 6.51$_E$-13, $p < 0.005$). Post hoc analyses were also conducted using Tukey's HSD test. There were statistically significant differences between ΔSphX and: WT ($p < 0.05$), ΔSphZ ($p < 0.005$), and ΔSphX:SphZ ($p < 0.005$).

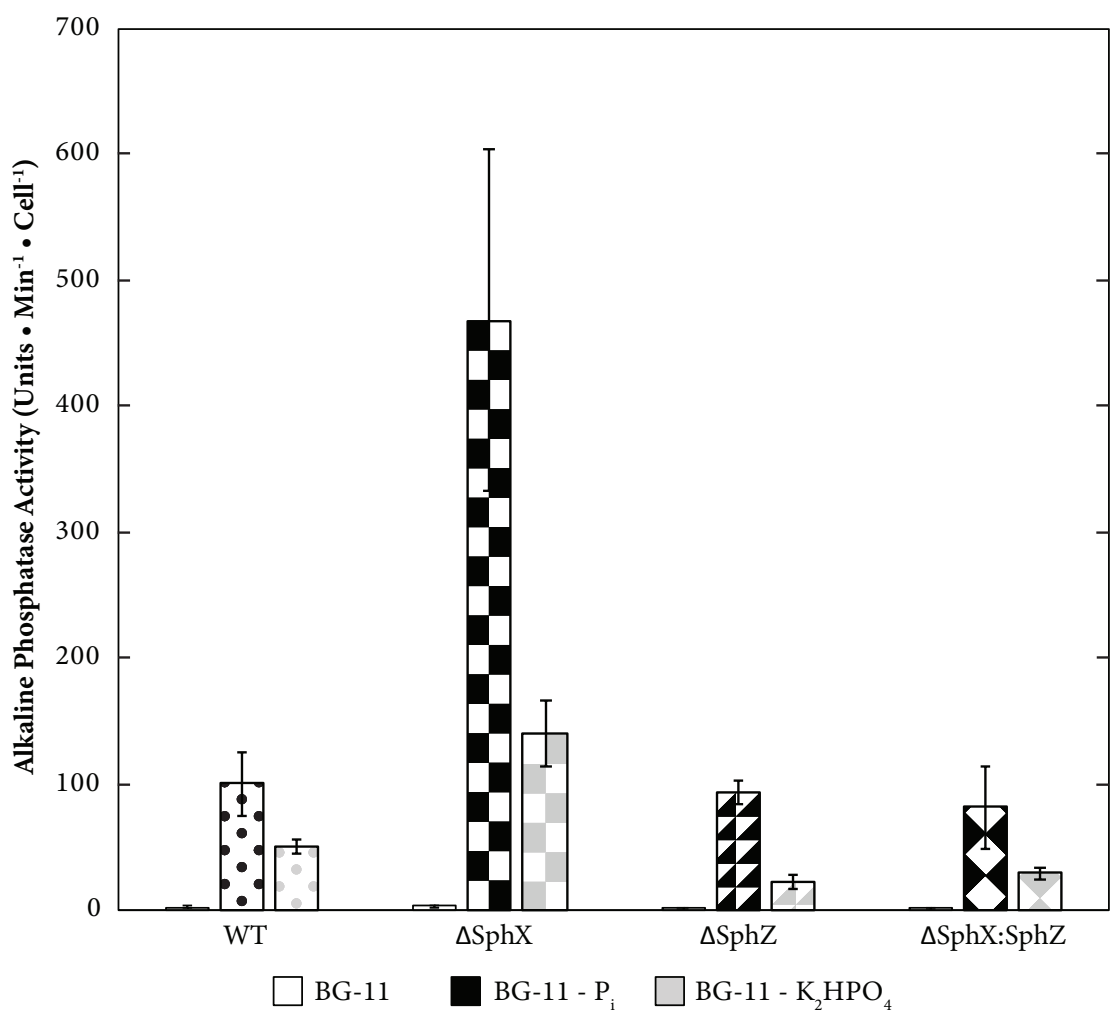

**Figure 3.15 Alkaline Phosphatase Activity.** Alkaline phosphatase activity measured after being grown in BG-11 (white), BG-11 – $P_i$ (black), BG-11 – $K_2HPO_4$ (grey) for 36 h. Values are means ± S.E. (n=3).

### 3.2.3 Whole-Cell Spectra Observations

Whole-cell absorption spectra were measured in order to determine the ratio of chromophores involved in photosynthesis from mutant strains to compare with WT for each environmental condition (Figure 3.16). Absorption maxima of phycobilins and chlorophyll *a* were plotted in MATLAB® for each environmental condition in order to compare the mutant strains: the change of photosynthetic pigments was expressed as a percentage (Table 3.1-3). The average area under the spectra for strains grown in BG-11 was: 239.40 for WT, 329.00 for ΔSphX, 221.00 for ΔSphZ, and 262.50 for ΔSphX:SphZ (Figure 3.16.A). The average area under the spectra for strain grown in BG-11 – P$_i$ was: 131.10 for WT, 203.70 for ΔSphX, 125.60 for ΔSphZ, and 140.50 for ΔSphX:SphZ (Figure 3.16.B). The average area under the spectra was: 184.20 for WT, 269.40 for ΔSphX, 172.40 for ΔSphZ, and 186.00 for ΔSphX:SphZ (Figure 3.16.C).

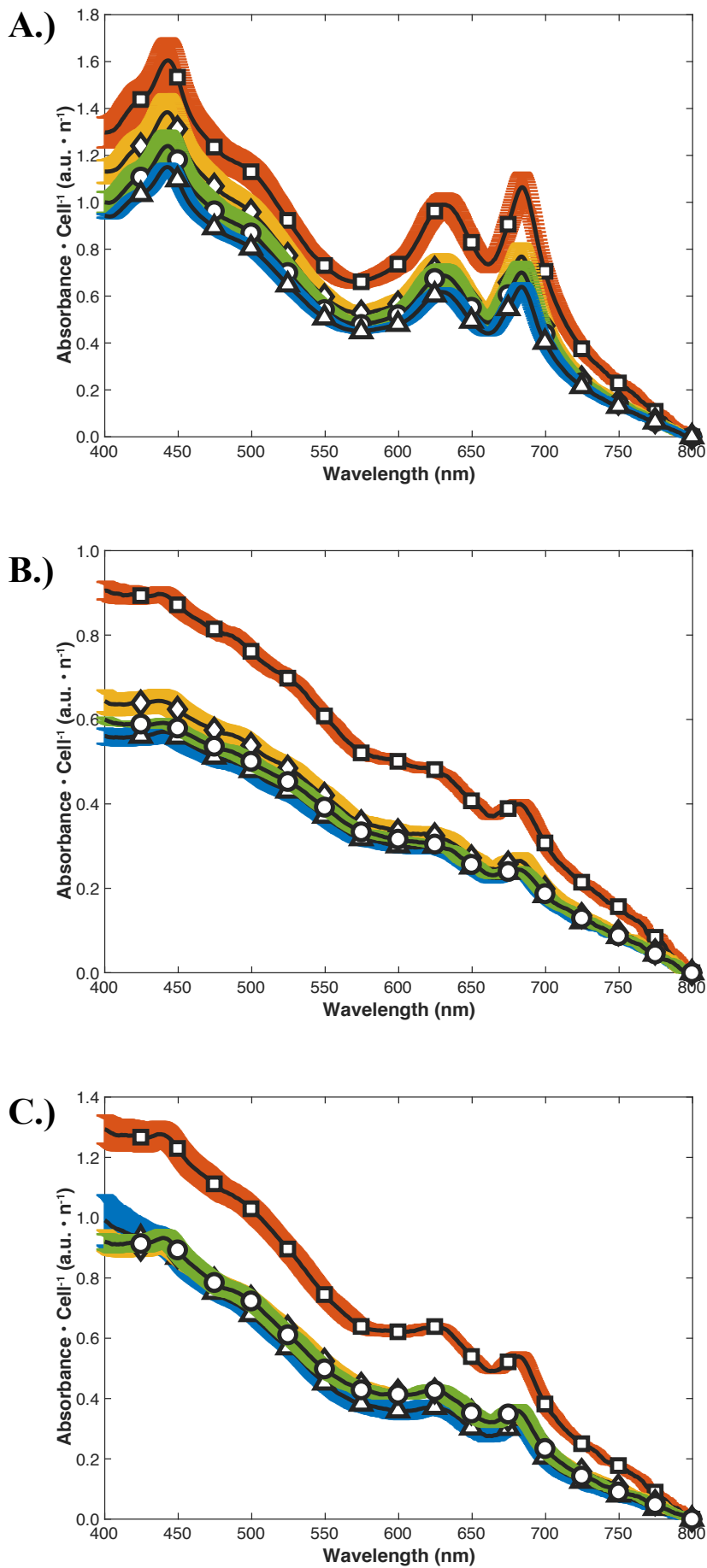

**Figure 3.16 Whole-Cell Absorption Spectral Analysis.** Whole-cell spectra for strains grown in BG-11 (**A**), BG-11 − $P_i$ (**B**), and BG-11 − $K_2HPO_4$ (**C**). Spectral averages are indicated in black. WT (circles) spectral S.E. are indicated in green; ΔSphX (squares) spectral S.E. are indicated in orange; ΔSphZ (triangles) spectral S.E. are indicated in blue; ΔSphX:SphZ (diamonds) spectral S.E. are indicated in yellow (n=3).

*Table 3.1 Photosynthetic Pigments of Mutant Strains Compared to WT Grown in BG-11*

| Peak (nm): | Difference between WT and ΔSphX: | Difference between WT and ΔSphZ: | Difference between WT and ΔSphX:SphZ: |
|---|---|---|---|
| 443 | +29% | -7% | +12% |
| 630 | +44% | -10% | +6% |
| 683.5 | +52% | -9% | +10% |

*Table 3.2 Photosynthetic Pigments of Mutant Strains Compared to WT Grown in BG-11 - $P_i$*

| Peak (nm): | Difference between WT and ΔSphX: | Difference between WT and ΔSphZ: | Difference between WT and ΔSphX:SphZ: |
|---|---|---|---|
| 439.5 | +51% | -4% | +9% |
| 624 | +58% | -2% | +6% |
| 681 | +63% | - | +8% |

*Table 3.3 Photosynthetic Pigments of Mutant Strains Compared to WT Grown in BG-11 − $K_2HPO_4$*

| Peak (nm): | Difference between WT and ΔSphX: | Difference between WT and ΔSphZ: | Difference between WT and ΔSphX:SphZ: |
|---|---|---|---|
| 439.5 | +37% | -1% | -3% |
| 623.5 | +49% | -13% | - |
| 681.5 | +50% | -12% | +6% |

### 3.2.6 Low-Temperature (77 K) Fluorescence Emission Spectroscopy

After one week, fluorescence emission spectroscopy with two excitation wavelengths was used to analyze proteins required in the photosynthetic electron transport chain (Figure 3.17). An excitation wavelength of 440 nm was specifically used to analyze chlorophyll *a*-binding proteins (Figure 3.17.A-C1). Mutant Strains are comparable to WT when grown in BG-11 (Figure 3.17.A1). Strains grown with BG-11 – $P_i$ exhibit a decreased fluorescence emission at 695 nm in comparison to the strains grown with BG-11 (Figure 3.17.B1). However, variability was noted for strains grown in BG-11 – $K_2HPO_4$ (Figure 3.17.C1) The PS II: PS I ratio appears to have decreased for ΔSphX when compared to WT (Figure 3.17.C1). On the other hand, ΔSphZ and ΔSphX:SphZ appear to have increased the PS II: PS I ratio and an increased 685 nm peak relative to the 695 nm peak (Figure 3.17.C1).

Low-temperature fluorescence emission spectroscopy was also used with an excitation wavelength of 580 nm in order to analyze the phycobilisome and its association with the photosystems. Once again, mutant strains grown in BG-11 were comparable to WT (Figure 3.17.A2). When grown in BG-11 – $P_i$, all strains had a sharpened peak at ~685 nm and seemed to lack a 695 nm peak (Figure 3.17.B2). Variability for strains grown in BG-11 – $K_2HPO_4$ was also noted at the 580 nm excitation wavelength (Figure 3.17.C2). $\Delta$SphZ had higher fluorescence at the ~667 nm peak and a collapsed ~650 nm peak as compared to WT (Figure 3.17.C2). $\Delta$SphX and $\Delta$SphX:SphZ also exhibited higher fluorescence at ~667 nm, yet still maintained an ~650 nm peak (Figure 3.17.C2). On the other hand, $\Delta$SphZ and $\Delta$SphX:SphZ had markedly decreased fluorescence at ~725 nm (Figure 3.17.C2).

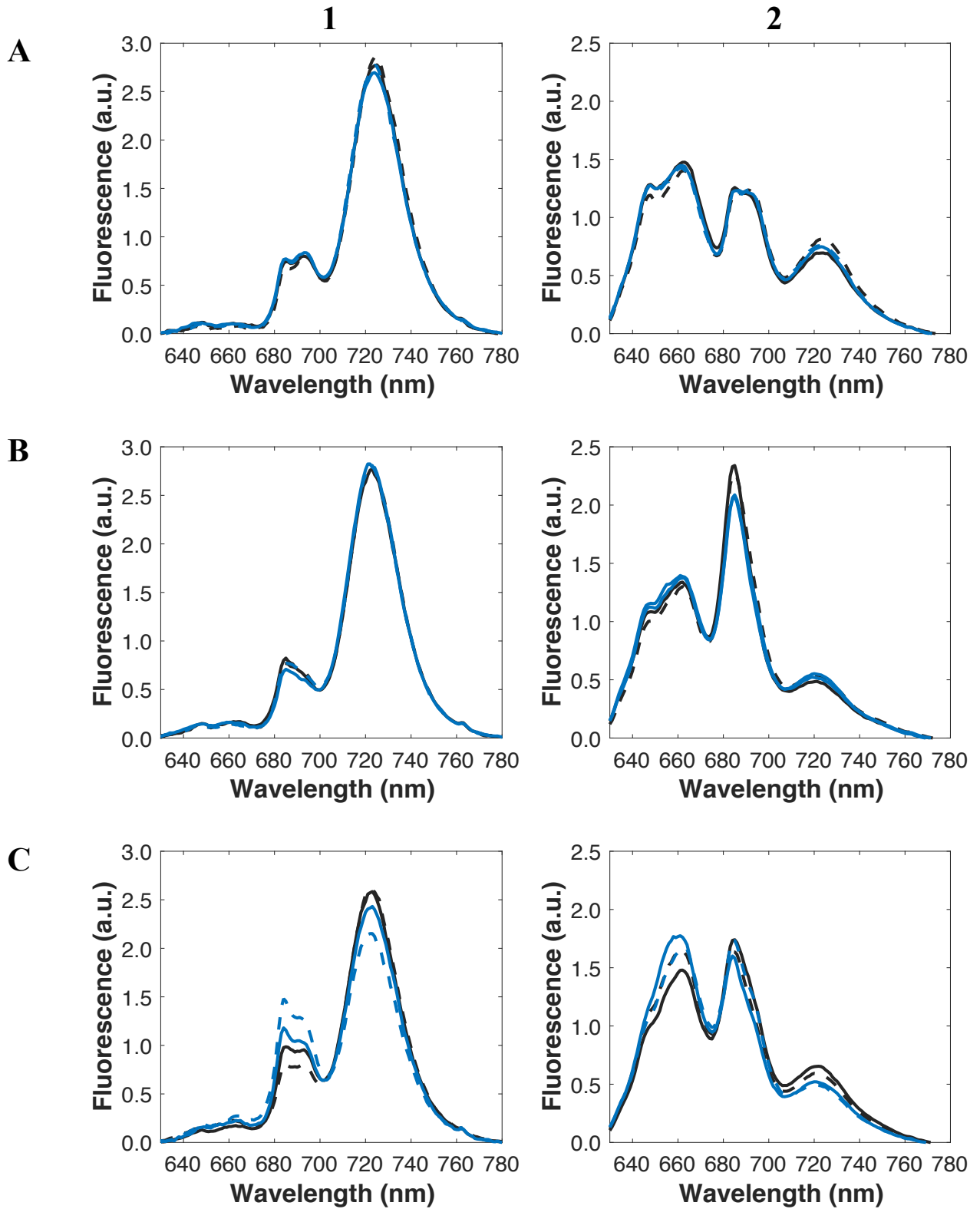

**Figure 3.17 Low-Temperature (77 K) Fluorescence Emission Spectroscopy.** 77 K of strains grown in BG-11 (**A**), BG-11 − P$_i$ (**B**), BG-11 − K$_2$HPO$_4$ (**C**) with an excitation wavelength of 440 nm (**1**) and 580 nm (**2**). WT indicated with a solid, black line; ΔSphX indicated with a dashed, black line; ΔSphZ indicated with a solid, blue line; and ΔSphX:SphZ indicated with a dashed, blue line. Spectral values were an average (n=3).

# Chapter Four: Discussion of SphX and SphZ Characterizations

Acclimation to phosphate limitation is an intricate process that has evolved for millennia to aid in the fitness of a species within a particular environment. In *Synechocystis* 6803, there are two inorganic phosphate transporter systems: the Pst1 system is the main transporter for inorganic phosphate, so why is the Pst2 system needed? It has been shown to be preferentially upregulated under phosphate limitation with higher transcript levels than that of the Pst1 system, so the ratio of Pst1:Pst2 system decreases (Pitt *et al.*, 2010). It was deduced that the Pst2 system along with the extracellular nuclease and alkaline phosphatase are vital in function to aid the cells in acquiring inorganic phosphate during limitation (Pitt *et al.*, 2010). Yet, how is the response regulated? There are similarities in the amino acid sequences of PstS1 and PstS2 PBP homologues as well as SphX and SphZ PBP homologues from the separation into distinct clades. Following the rationale of how PstS1 and PstS2 are similar in structure as well as functional role specific for their respective Pst system, then could there be some similarity for the structure and possible functional role of SphX and SphZ? Moreover, there is a connection between SphX and PstS1 PBP homologues as well as SphZ and PstS2 PBP homologues based on the number of homologues present throughout the cyanobacteria analyzed in this study. The fact that SphX and PstS1 PBP homologues probably have some sort of connection is not surprising since, in at least *Synechocystis* 6803, they are encoded within the same operon. The interesting point is that there may be a connection between PstS2 and SphZ PBP homologues. So, does SphZ play a role in the acquisition of inorganic phosphate? The roles of SphX and SphZ PBPs have remained elusive for quite some time. Differences between these two proteins are presented to contribute to an all-encompassing model for the regulation of the pho regulon within *Synechocystis* 6803, which has possible implications for phosphate-deficiency acclimation within other cyanobacteria as well as other bacterial species.

## 4.1 Differences between Phosphate-Binding Sites

First, there are likely differences in the binding affinity for phosphate due to the volume of the phosphate-binding site, illustrated in the predictive models. The varying position of their aspartic acid residues would contribute to this as well as other amino acid residue differences within the phosphate-binding site of the two proteins. It has been previously indicated that these proteins are more closely related to low-affinity transporter systems as well as differences in the ability to discern binding against arsenate, an analog of phosphate, from phosphate (Yan *et al*., 2017). Lower binding affinity would allow for these proteins to bind and subsequently release anions at a faster rate compared to the high binding affinity protein counterparts, namely PstS1 and PstS2. Kinetics for these proteins should be further analyzed through phosphate uptake measurements with these mutant strains over varying molarity of phosphate or directly with isolated proteins after having been overexpressed within *E*. coli as an example.

## 4.2 SphX Competes for Inorganic Phosphate Against PstS PBPs

SphX hinders luxury uptake via the Pst1 system when extracellular phosphorus is greater than 12 µM and the deletion of *sphX* affects the cell's size probably due to the accumulation of polyPs. The depletion of phosphorus throughout the water column over the course of a week illustrates that the deletion of *sphX* allowed cells to uptake phosphorus faster with less cells required than WT – this illustrates that SphX plays a competitive role against PstS1 for the binding of phosphate so that the nutrient is not depleted from the environment at a fast rate. Moreover, the cell counts revealed how a lesser number of cells from the ΔSphX strain were required for an equivalent optical density with that of WT – this suggests that the cell sizes of the ΔSphX strain were larger in comparison to WT (Stevenson *et al.*, 2016). Measurements of cells should indicate differences in the volume of the ΔSphX cells. In addition, the whole-cell spectra revealed increased absorbance for ΔSphX in comparison to WT under all three environmental conditions. This does showcase an increase in the absorbance from pigments, yet are the pigments changing with that of the volume? Most likely the spectra indicate once again a difference between the sizes of the ΔSphX cells from that of WT and can be confirmed through analyzing the pigments against the volume of cells. Since luxury uptake was ultimately the response under high concentrations of phosphorus in the media when *sphX* was deleted, this indicates that a potential increase in volume is likely due to the accumulation of polyPs – this can be tested through the visualization of polyphosphates with a DAPI-based protocol (Morohoshi *et al.*, 2002; Burut-Archanai *et al.*, 2013; Voronkov and Sinetova, 2019). With this logic though, the growth curves illustrate the concept: the ΔSphX strain was able to increase the carrying capacity due to the ability to uptake and retain phosphate when grown in BG-11. Moreover, during the limitation of $K_2HPO_4$, all cells were dying yet the rate of death is likely contingent upon how much polyPs has been stored prior to subjection of stress. Therefore, increased polyphosphate storage would allow the ΔSphX cells to survive for a longer period of time. This phenomenon has wider implications for optimizing bacteria, beyond the terms of growth rates, with an ability to store reserves of excess energy by which the population could utilize the resource as environmental nutrients are depleted and, therefore, offer a competitive edge against other species. The SphX protein likely fulfills this niche role so that inorganic phosphate is not depleted from the environment at a rate greater than required for cellular metabolism.

## 4.3 SphX and SphZ Aid in Switching Between the Pst Systems

As previously mentioned, there is likely roles shared between SphX and SphZ. Based upon the parametric analysis of phosphorus depletion in BG-11 media against the number of cells, a nexus formed between the mutant strains at 12 µM of phosphorus – this likely indicates the switching between the Pst1 and Pst2 systems. The competition of binding to inorganic phosphate between the PBPs plays an integral role in the scenario: deletion of *sphX* hindered luxury uptake at high phosphorus concentrations; whereas, the interruption of *sphZ* caused a faster uptake rate for lower concentrations of phosphorus as compared to WT. Moreover, the interruption of *sphZ* delayed intrapopulation dynamics for curbing growth rates. When the $\Delta$SphZ strain was subjected to BG-11 – $P_i$, the rate of cell growth was faster than that of WT yet the cells exhibited the longest delay in curbing the growth rate and, by extension, exhibited less stress in terms of the rate of growth. This likely illustrates how SphZ is required in order for cells to oscillate between stressed and non-stressed conditions: as seen with the limitation of $K_2HPO_4$, $\Delta$SphZ cells were the most ill-prepared for survival yet would continually grow when supplied with equimolar of potassium. Under phosphate-deficiency, polyPs are mobilized; however, these reserves within the $\Delta$SphZ strain were inadequate and hence the inability to sustain a steady-state above the initial biomass after one week.

## 4.4 Cross-Regulation with the Photosynthetic Apparatus

The low-temperature (77 K) fluorescence emission spectra are similar, as to be expected since the photosynthetic apparatus was not targeted directly, yet the response is ubiquitous for phosphate limitation. However, there are differences under $K_2HPO_4$ limitation. The $\Delta$SphZ strain exhibited increased monomers of PS II and less PS I, as this is seen in both spectra, in comparison to WT. There could be cross regulation of the pho regulon and photosynthetic apparatus – this could be further tested through real-time reverse transcription PCR (RT-qPCR) targeting *psbA1, psbA2, psbB, psbC, psaA, and psaC* as examples. Decreased fluorescence emission from phycocyanin is also of interest as there may be an association between the pho regulon and auxiliary proteins of the phycobilisome: once again this can be tested via RT-qPCR targeting *cpcC1, cpcC2, cpcA, cpcB, apcA, apcB, and apcE* as examples.

## 4.5 The Big Picture

To ascertain a role with the SphS-SphR system is to phosphate-stress the cells without the cells responding as if being stressed. So, the system is either off or on, whereby various events need to occur for the system to turn on. Phosphate uptake and regulation of the pho regulon rely innately on each other in a negative feedback loop. Bacterial response to phosphate-deficiency is typically analyzed through alkaline phosphatase activity since the protein is encoded by a gene within the pho regulon and would, therefore, be upregulated due to the SphS-SphR system properly functioning.

When SphS or SphR are removed, then alkaline phosphatase activity is not measured and, thus, indicates that the system remains off even when cells are subjected to phosphate-stress (Hirani *et al.*, 2001; Suzuki *et al.*, 2004; Juntarajumnong *et al.*, 2007; Burut-Archanai *et al.*, 2009; Kimura *et al.*, 2009). SphU is a "negative regulator" for the SphS-SphR system because the protein inhibits the system (Juntarajumnong *et al.*, 2007). When the protein is removed, the cells exhibit constitutive expression of the pho regulon regardless of the environment and, therefore, the cells respond as if constantly phosphate deficient (Juntarajumnong *et al.*, 2007). So, the system is always on. SphU interacts with ATPase subunits and, therefore, likely hinder the transport of inorganic phosphate into the cytosol – this is would be an evolutionary trait needed to ensure that the cells are continually supplied with inorganic phosphate without expending the reserve in order to transport inorganic phosphate into the cell. Moreover, phosphorus was rapidly depleted from BG-11 when *sphU* was deleted – this illustrates that SphU oscillates between both Pst systems. Marine species, particularly *Prochlorococcus* species, usually lack SphU homologues and rely solely on Pst1-like systems for transporting inorganic phosphate (Juntarajumnong *et al.*, 2007). There is therefore an evolutionary connection between SphU and Pst2-like systems.

The three aforementioned proteins, namely SphS, SphR, and SphU, have been localized to the cytosol. However, information regarding phosphate depletion must be conveyed across the inner membrane in order for the SphS-SphR system to function. This is seen in the fact that the N-terminus of SphS is required as a functional sensory component within *Synechocystis* 6803; however, an association with an auxiliary protein is also needed for function (Burut-Archanai *et al.*, 2009). Therefore, there is a fundamental difference in the removal of PBPs as these proteins reside in the periplasmic space. It was originally seen that the removal of the Pst1 system, later discovered due to the removal of specifically PstS1, also contributed to constitutive alkaline phosphatase activity. So, the system once again was always on. To reiterate if the SphS-SphR system is disrupted, then alkaline phosphatase activity should not be measured. If there is constitutive upregulation by the removal of any of the constituents within the periplasmic space, then this actually suggests that the SphS-SphR system is not reliant on the protein(s) being knocked out. Instead, the system is reliant on the properly functioning proteins present within that deletion strain. In order for these observations to coincide, the regulation of these genes must utilize the system that is present; therefore, the Pst2 system becomes a key target for regulation of these genes.

## 4.6 SphZ is the Sensor Protein for the SphS-SphR System

A model is hence presented to incorporate SphX and SphZ into the phosphate acquisition system (Figure 4.1; Figure 4.2). SphS and SphZ both have transmembrane domains with few residues residing in the periplasmic space or cytosol, respectively. Interactions via the transmembrane domains play an integral role in the individual and collective function. This model proposes that phosphate is bound to SphZ in the native state under phosphate replete conditions (Figure 4.1). Phosphate when discharged from the protein causes a conformational change and, thus, alters the transmembrane-transmembrane interaction to relay information across the inner membrane to SphS, initiating the SphS-SphR system (Figure 4.2). SphZ is once again not acting alone. It is proposed that SphS is also working in tandem with the Pst2 system via SphU (Figure 4.1). This creates a safe-guard, so that if one isn't properly functioning, then the other would ensure that the SphS-SphR system would still function to acclimate to the environment. Hence, when the Pst2 system was solely removed, SphZ would still allow SphS to properly function and, hence, why there was alkaline phosphatase activity under limiting conditions. The current ΔSphZ strain interrupted the phosphate-binding region of the protein and not the transmembrane domain, so the functioning of SphS-SphR system was probably due to alleviation of the interaction between SphU and SphS. As for SphX, the competition for inorganic phosphate with PstS1 was incorporated into the phosphate replete conditions; however, SphX would also compete with PstS2 for inorganic phosphate under phosphate-stress conditions. This is seen by how the deletion of *sphX* caused higher alkaline phosphatase activity during phosphate limitation since the probability of inorganic phosphate binding to PstS2 would increase and, therefore, would allow a faster rate of inorganic phosphate being transported into the cell via the Pst2 system. In general, the competition of SphX for $P_i$ likely is an evolutionary trait retained to ensure that phosphate is not depleted at a faster rate than required for metabolic efficiency. Under phosphate-stress conditions, SphU is proposed to have an association to disrupt the transport of phosphate across the Pst1 system due to its removal having increased the removal of phosphate. Does this mean that the schematic is complete? PstB1 and PstB1' are indicated with the Pst1 system; however, it has not been verified if the subunits are natively in a heterodimeric form or if the genes are differentially regulated over the course of phosphate depletion and actually form homodimeric associations with the Pst1 system – this should be further analyzed via RT-qPCR. The entire model itself should be met with scrutiny and must withstand the rigor of science. A time course for alkaline phosphatase activity should be measured over the course of phosphorus depletion from BG-11 for strains moving forward in order to test this model, for if the removal of a protein were to inhibit the response of the SphS-SphR system, then it would suggest that the protein is essential.

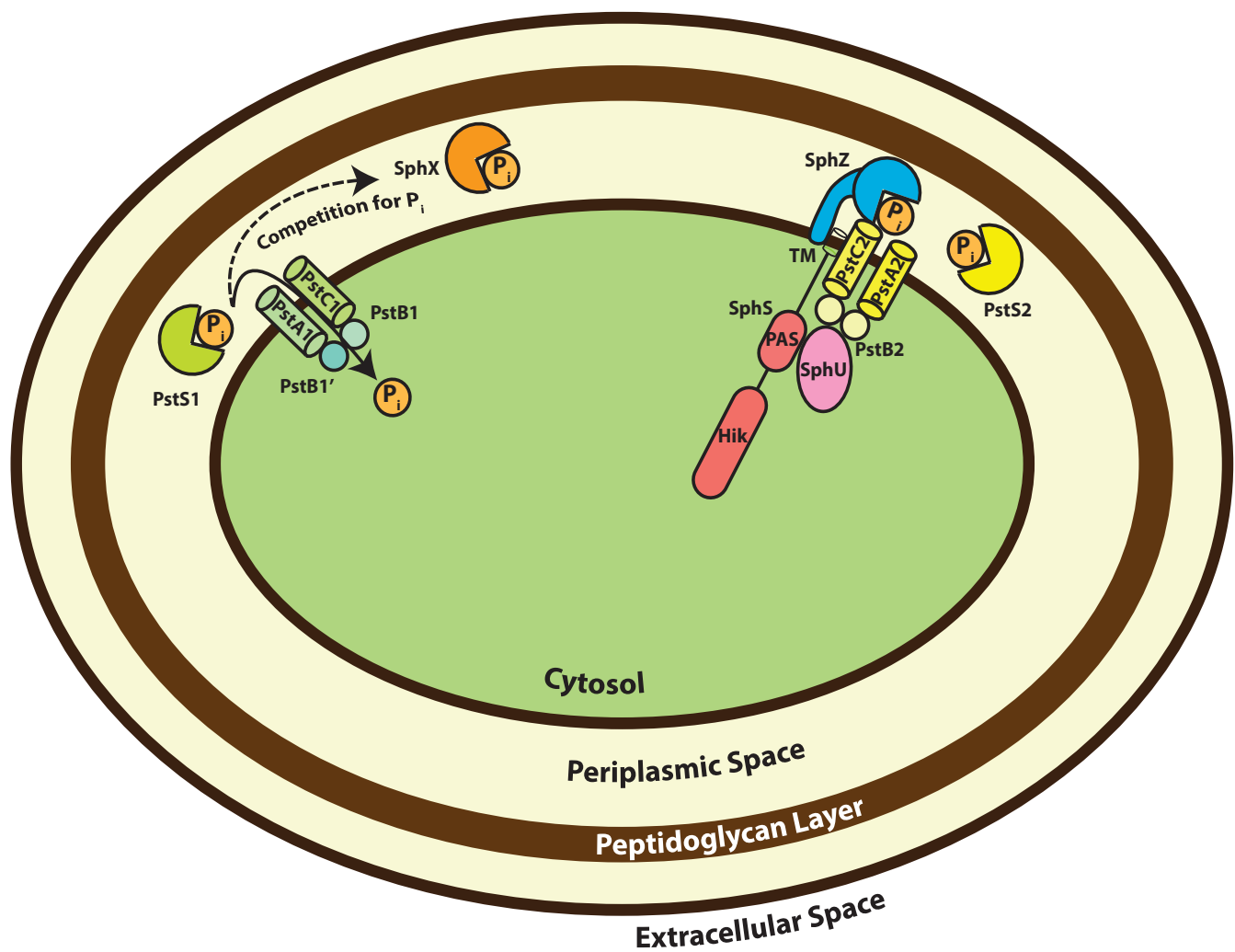

**Figure 4.1 Model of *Synechocystis* 6803 under Phosphate-Replete Conditions.** Inorganic phosphate (P$_i$) is translocated from the periplasmic space via the Pst1 system (green). Competition between SphX (orange) and the periplasmic PstS1 subunit of the Pst1 system for the binding of P$_i$ hinders luxury uptake. The SphS histidine kinase (red) interacts with the SphZ sensor (blue) and the Pst2 system (yellow) via its hydrophobic membrane-associated domain (TM) as well as an interaction between the Per-Arnt-Sim (PAS) domain and PstB2 ATPase subunits of the Pst2 system via the SphU negative regulator (pink).

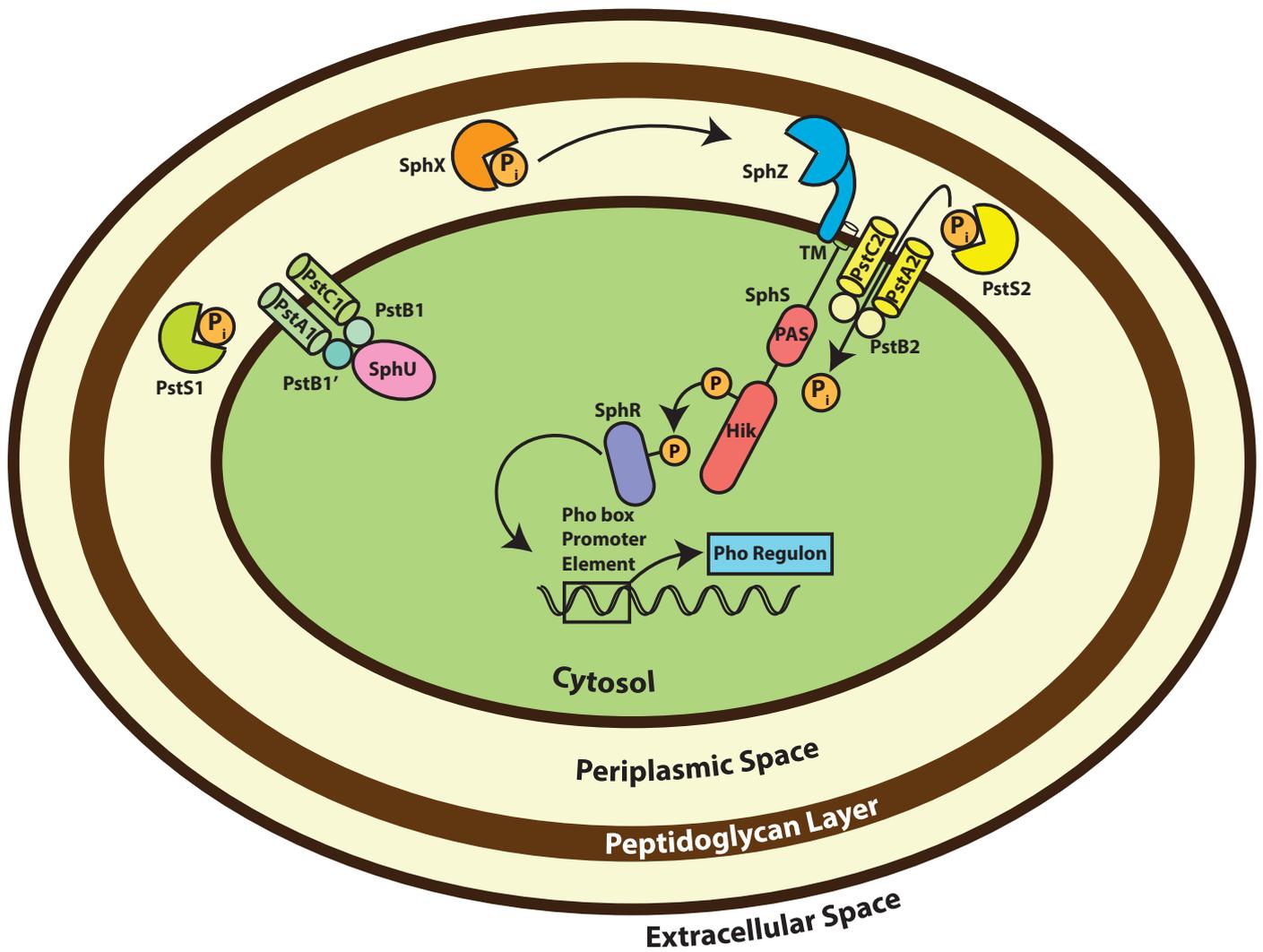

**Figure 4.2 Model of *Synechocystis* 6803 under Phosphate-Deficient Conditions.** Inorganic phosphate ($P_i$) is translocated from the periplasmic space via the Pst2 system (yellow). A phosphoryl group from SphS histidine kinase (red) is phosphorelayed to the SphR response regulator (purple), which then upregulates specific genes – the pho regulon (light blue). The SphU negative regulator (pink) hinders the Pst1 system. The SphZ sensor (blue) and hydrophobic membrane-associated domain (TM) interaction remains to ensure adequate regulation of the negative feedback loop. It is proposed that SphX may be necessary for the sensory switch.

# Chapter Five: Acclimation of WT to Phosphate and Iron Limitation

## 5.1 WT Response Over the Course of a Week to Phosphate Limitation

### 5.1.1 Chl *a* Biosynthesis

Since the photosynthetic apparatus was being altered, the concentration of chlorophyll *a* was measured over the course of WT growth in BG-11 – $P_i$ (Figure 5.1). There was a decrease throughout the week with periodicity of 48 h after the initial 24 h growth period (Figure 5.1). There was high variability in the measurements taken after 96 h of phosphate limitation (Figure 5.1).

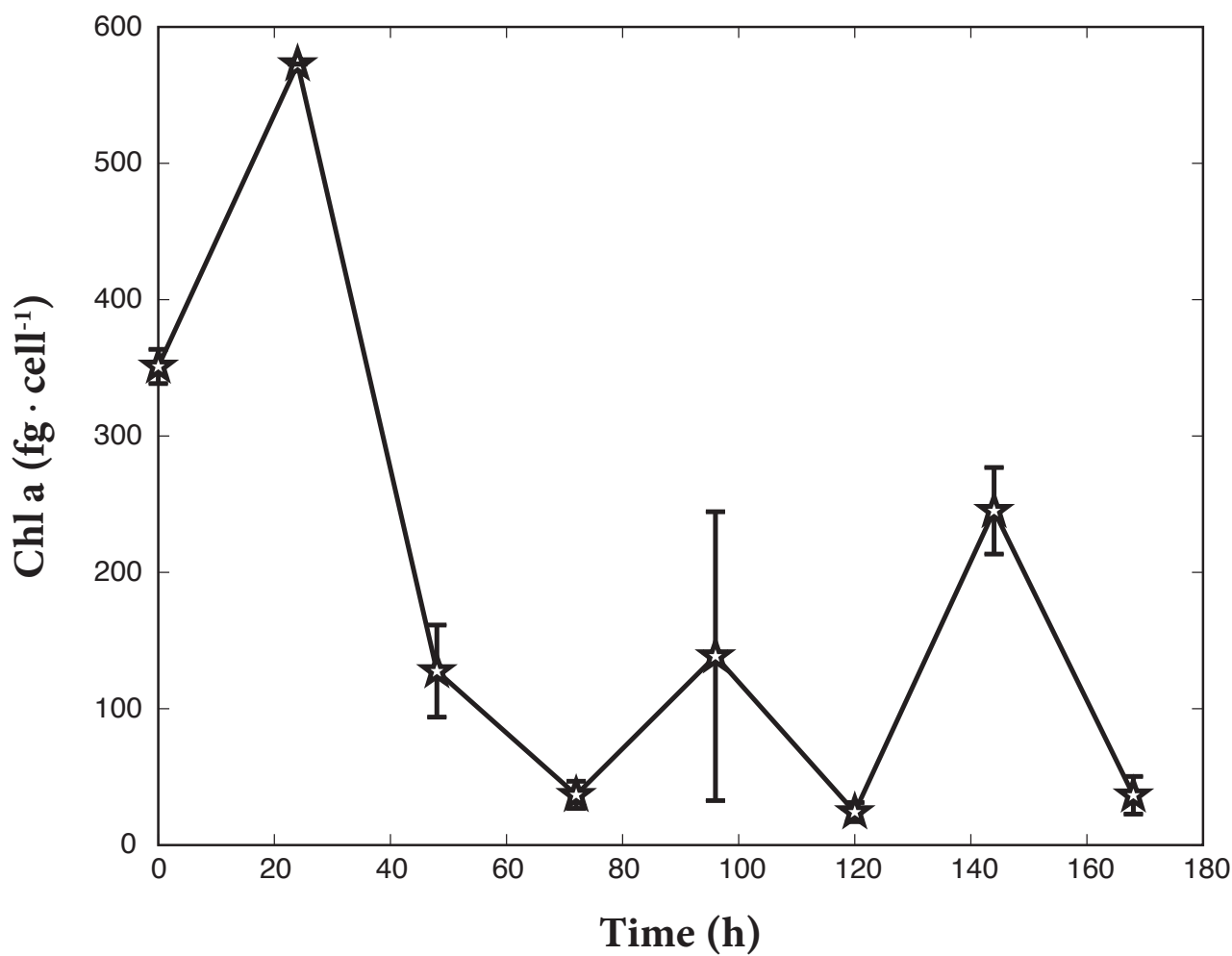

**Figure 5.1 Time Course of Chlorophyll *a* Biosynthesis.** Chl *a* measured every 24 h as WT was grown in BG-11 – $P_i$ for one week.

**5.1.2 Reconstruction of the Photosynthetic Apparatus**

In order to further analyze the time course of the WT response to phosphate limitation, low-temperature fluorescence emission spectroscopy was used to measure the fluorescence of proteins comprising the photosynthetic electron transport chain every 24 h as the cultures were grown (Figure 5.2). For the spectra obtained with an excitation wavelength of 440 nm, two paraboloids fused by a hyperbolic paraboloid formed by which the saddles were located between ~685/695 nm peaks and ~725 nm peak (Figure 5.2.A). An initial linear increase of the ~685/695 nm peaks with the ~725 nm peak was observed for the first 48 h of increasing biomass (Figure 5.2.A-B). Then, a collapse of the spectral peaks for ~48 h after 72 h of phosphate limitation (Figure 5.2.A-B). The second hyperbolic paraboloid was formed over the course of ~48 h after 96 h of phosphate limitation, yet there was a decrease of the ~685/695 nm peaks (Figure 5.2.A-B). A second collapse of the photosynthetic apparatus occurred at the measurements taken after 144 h of phosphate limitation (Figure 5.2.A-B). After 168 h of phosphate limitation, the ~685/695 nm and ~725 nm peaks reemerged (Figure 5.2.A-B).

For the spectra obtained with an excitation wavelength of 580 nm, the spectral data were found to reinforce the scenario observed from the measurements taken with an excitation wavelength of 440 nm (Figure 5.2.C-D). The two main paraboloids occurred at ~667 and ~685 nm and were also fused by a hyperbolic paraboloid (Figure 5.2.C-D). The ~650, ~667, and ~685 nm peaks appeared to increase in a linear manner for the first 48 h (Figure 5.2.C-D). Then, there is a collapse of the spectral peaks for ~48 h after the 72 h measurements (Figure 5.2.C-D). When the paraboloids began increasing for ~48 h after the 96-h measurement, the ~650 nm peak was diminished (Figure 5.2.C-D). Moreover, the ratio between the ~667:685 nm peaks had decreased (Figure 5.2.C-D). After the second collapse of the spectral peaks at the 144 h measurements, the 168 h measurements indicated an even greater decrease of the ~667:685 nm peaks (Figure 5.2.C-D).

**Figure 5.2 Time Course of Low-Temperature (77 K) Fluorescence Emission Spectroscopy.** 77 K measurements every 24 h of WT being grown in BG-11 – P, over one week with an excitation wavelength of 440 nm (**A**) and 580 nm (**C**). Corresponding contour plots are found in the same row (**B**,**D**).

## 5.2 Analyzing WT response to Iron, Phosphate, and Potassium-Phosphate Limitation

In order to further analyze the physiological responses *Synechocystis* 6803 exhibits under phosphate limitation, the limitation was directly analyzed against iron-deficiency.

### 5.2.1 Whole-cell Spectra

Whole-cell absorption spectra were measured in order to determine the ratio of chromophores involved in photosynthesis for WT subjected to each environmental condition (Figure 5.3). Absorption maxima of phycobilins and chlorophyll *a* were plotted in MATLAB® for each environmental condition, noted for differences at the nm position for which the maxima occurred, and the change of photosynthetic pigments was expressed as a percentage (Table 5.1). The average area under the curve for: WT grown in BG-11 was 64.46; WT grown in BG-11 $+ \frac{1}{10^{th}}$ Fe$^{3+}$ was 58.94; WT grown in BG-11 $-$ P$_i$ was 35.29; and WT grown in BG-11 $-$ K$_2$HPO$_4$ was 44.43 (Figure 5.3).

*Table 5.1 Overview of Changes Observed from Whole-Cell Spectra in Limiting Media*

| Environmental Condition: | Peak (nm): | Blue-shifted by ___ nm: | Difference from WT: |
|---|---|---|---|
| BG-11 | 443 | - | - |
| BG-11 $+ \frac{1}{10^{th}}$ Fe$^{3+}$ | 440 | 3 | -7% |
| BG-11 $-$ P$_i$ | 439.5 | 3.5 | -52% |
| BG-11 $-$ K$_2$HPO$_4$ | 439.5 | 3.5 | -31% |
| BG-11 | 630 | - | - |
| BG-11 $+ \frac{1}{10^{th}}$ Fe$^{3+}$ | 628.5 | 1.5 | -16% |
| BG-11 $-$ P$_i$ | 624 | 6 | -57% |
| BG-11 $-$ K$_2$HPO$_4$ | 623.5 | 6.5 | -47% |
| BG-11 | 683.5 | - | - |
| BG-11 $+ \frac{1}{10^{th}}$ Fe$^{3+}$ | 678 | 5.5 | -11% |
| BG-11 $-$ P$_i$ | 681 | 2.5 | -66% |
| BG-11 $-$ K$_2$HPO$_4$ | 681.5 | 2 | -55% |

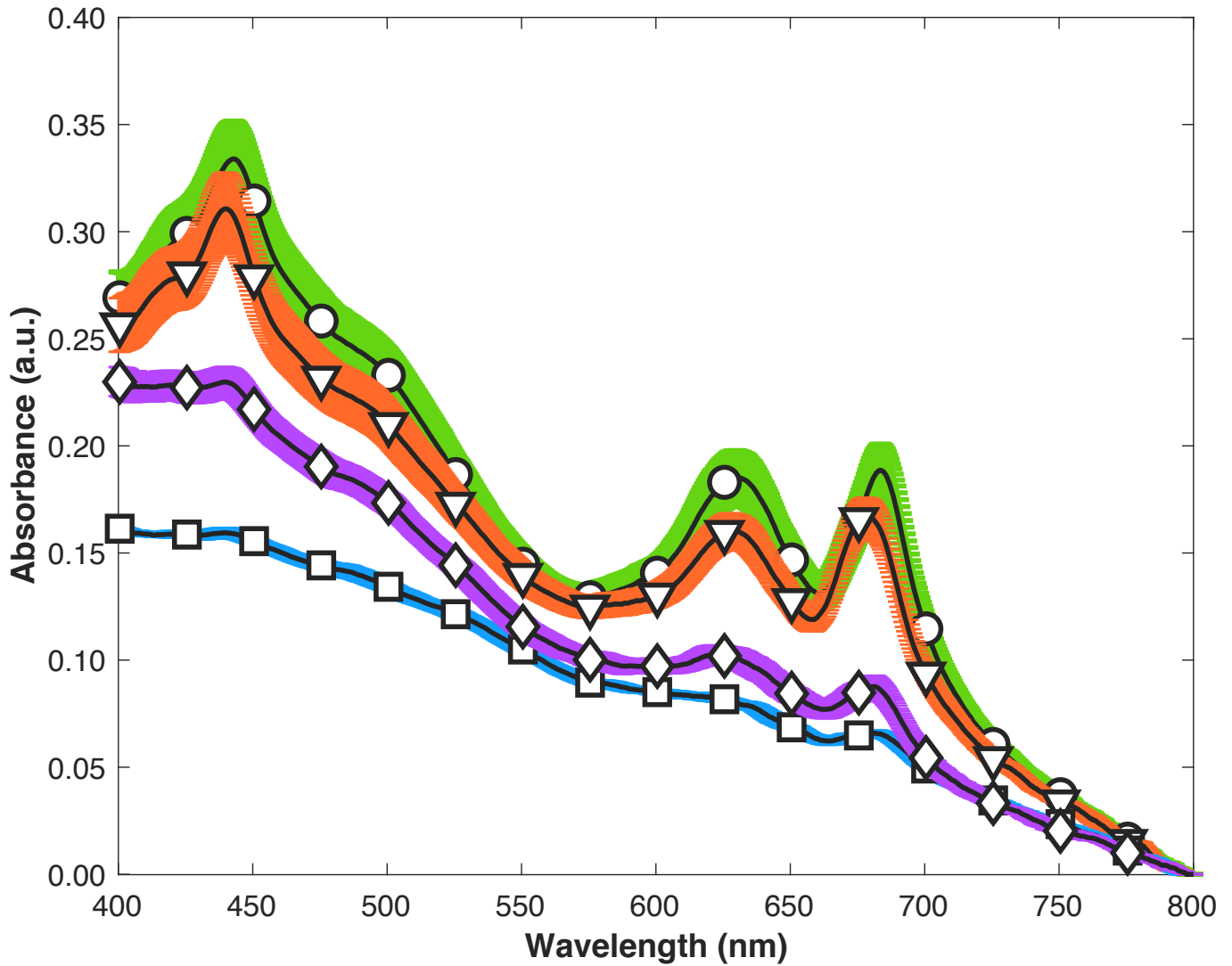

**Figure 5.3 Whole-Cell Spectral Analysis.** Whole-cell spectra for WT grown in BG-11 (circles), BG-11 supplemented with $\frac{1}{10th}$ of iron (inverted triangles), BG-11 − $P_i$ (squares), and BG-11 − $K_2HPO_4$ (diamonds) for one week. Spectral averages are indicated in black. BG-11 spectral S.E. are indicated in green; BG-11 supplemented with $\frac{1}{10th}$ of iron spectral S.E. are indicated in orange; BG-11 − $P_i$ spectral S.E. are indicated in blue; BG-11 − $K_2HPO_4$ spectral S.E. are indicated in purple.

## 5.2.2 77 K Measurements with an Excitation Wavelength of 440 of WT under Varying Environmental Conditions

The traces produced by WT grown in BG-11 are considered the typical ratios for the photosynthetic apparatus within *Synechocystis* 6803 (Figure 5.4). The traces exhibit increased fluorescence at ~685 nm, as previously stated, when grown in BG-11 + $\frac{1}{10^{th}}$ Fe$^{3+}$ (Figure 5.4). There was also a decrease of fluorescence at ~725 nm and a 3-nm shift when compared to WT grown in BG-11 (Figure 5.4). WT grown in BG-11 – P$_i$ exhibited a decrease of fluorescence at ~ 685 nm with comparable emission from ~725 nm whereby the emission was blue-shifted by 2 nm (Figure 5.4). WT grown in BG-11 – K$_2$HPO$_4$ exhibited higher fluorescence from ~685 and 695 nm peaks and lower fluorescence from ~725 nm and blue-shifted by 2 nm when compared to WT grown in BG-11 (Figure 5.4).

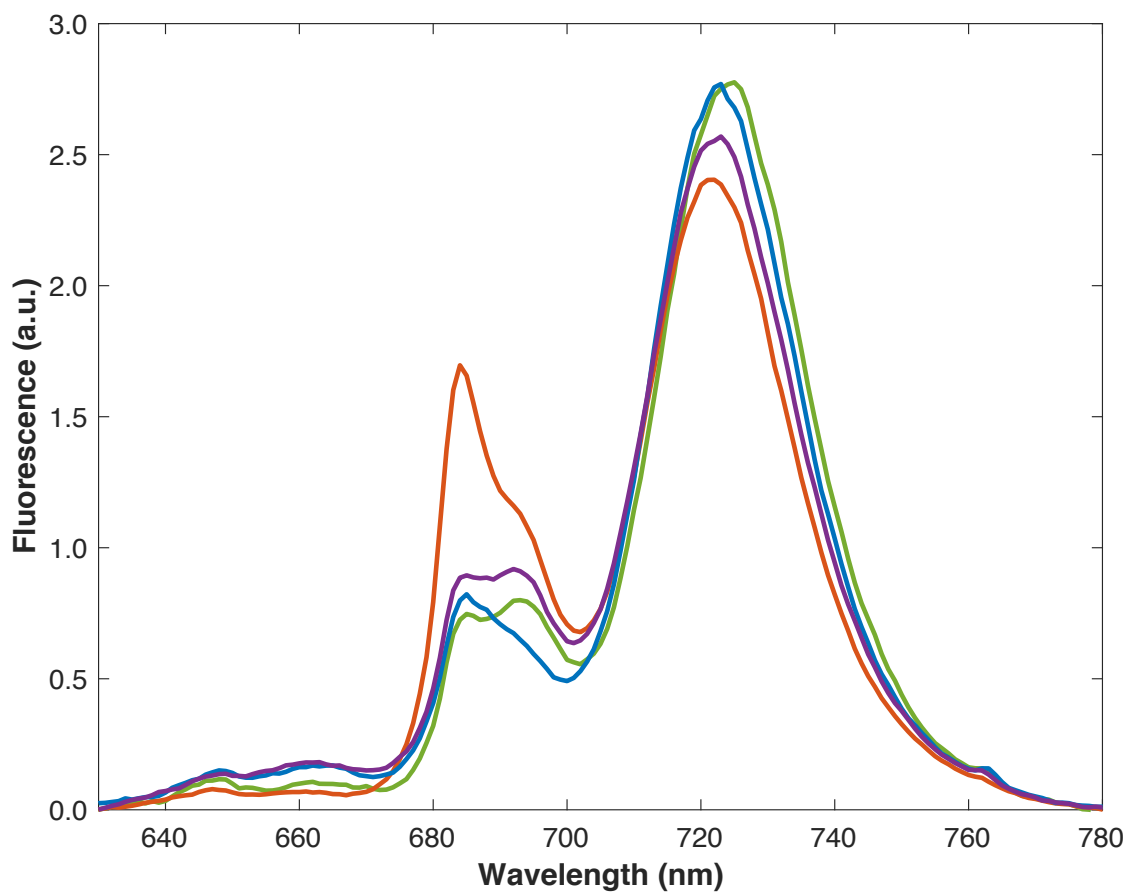

**Figure 5.4 Low-Temperature (77 K) Fluorescence Emission Spectroscopy with 440 nm Excitation.** 77 K measurements with an excitation wavelength of 440 nm for WT grown in BG-11 (green), BG-11 + $\frac{1}{10^{th}}$ Fe$^{3+}$ (orange), BG-11 – P$_i$ (blue), and BG-11 – K$_2$HPO$_4$ (purple) for one week.

### 5.2.3 77 K Measurements with an Excitation Wavelength of 580 of WT under Varying Environmental Conditions

Once again, the traces produced by WT grown in BG-11 are considered the typical ratios for the photosynthetic apparatus within *Synechocystis* 6803 (Figure 5.5). The fluorescence emission at ~645 nm and ~660 nm decreased for BG-11 + $\frac{1}{10th}$ $Fe^{3+}$ and BG-11 – $P_i$ as compared to WT grown in BG-11 (Figure 5.5). On the other hand, traces from WT grown in BG-11 – $K_2HPO_4$ produced comparable fluorescence at the ~600 nm peak; however, the ~645 nm peak had markedly decreased (Figure 5.5). Fluorescence at ~685 nm increased for WT grown in all limiting media when compared to that of BG-11 (Figure 5.5). The ~695 nm peak collapsed for WT grown in BG-11 − $P_i$ and BG-11 − $K_2HPO_4$; however, there was a shoulder present in the traces for WT grown in BG-11 + $\frac{1}{10^{th}}$ $Fe^{3+}$ (Figure 5.5). The traces also indicate decreased fluorescence at ~725 nm for WT grown in BG-11 − $P_i$ and BG-11 − $K_2HPO_4$ compared to WT grown in BG-11; whereas, the ~725 nm fluorescence for WT grown in BG-11 + $\frac{1}{10^{th}}$ $Fe^{3+}$ was comparable to that of WT grown in BG-11 (Figure 5.5). Moreover, the ~725 nm fluorescence peaks for WT grown in all limiting media resulted in a blue-shift by 3 nm for WT grown in both phosphate-limiting media and 4 nm for WT grown in iron-limiting media (Figure 5.5).

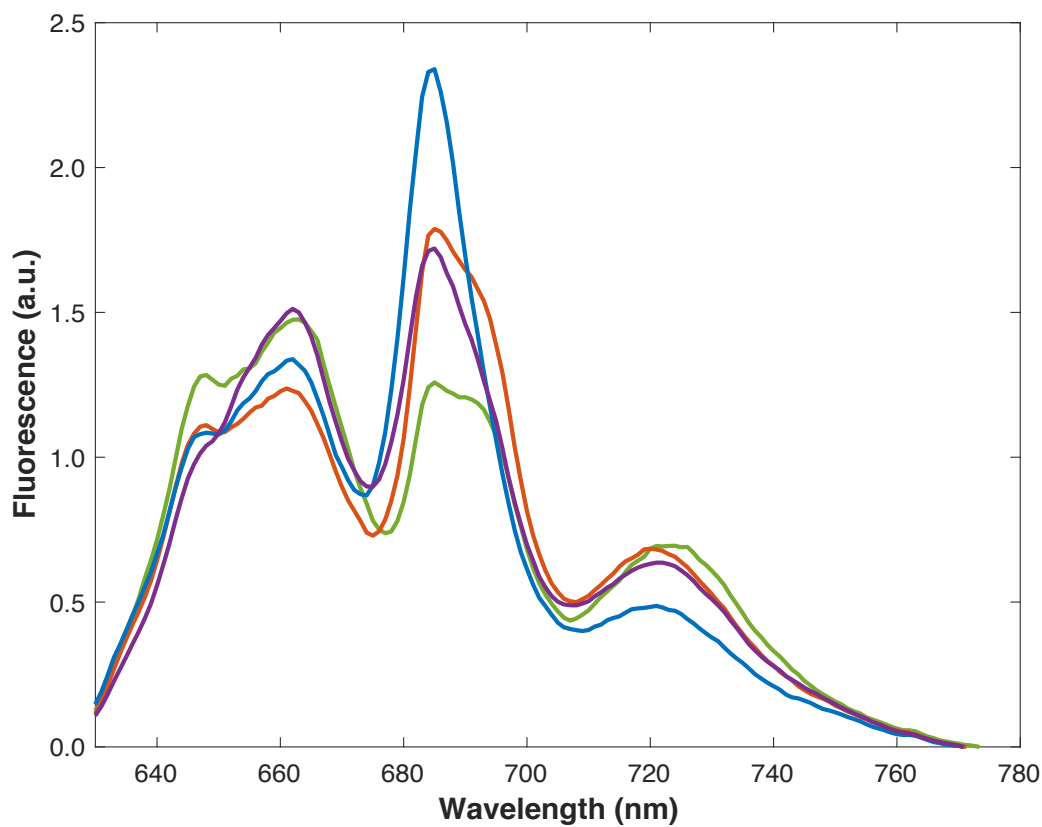

**Figure 5.5 Low-Temperature (77 K) Fluorescence Emission Spectroscopy with 580 nm Excitation.** 77 K measurements with an excitation wavelength of 580 nm for WT grown in BG-11 (green), BG-11 + $\frac{1}{10th}$ $Fe^{3+}$ (orange), BG-11 – $P_i$ (blue), and BG-11 – $K_2HPO_4$ (purple) for one week.

### 5.2.4 440 vs. 580 77 K Spectral Measurements

Traces were then parameterized within $\mathbb{R}^3$ in order to elucidate the importance between Chl *a*- and phycobilisome-associated complexes (Figure 5.6). Differences were readily highlighted, and projection onto the yz-plane resulted in the direct comparison of the spectra (Figure 5.6). However, is this parameterized method reproducible?

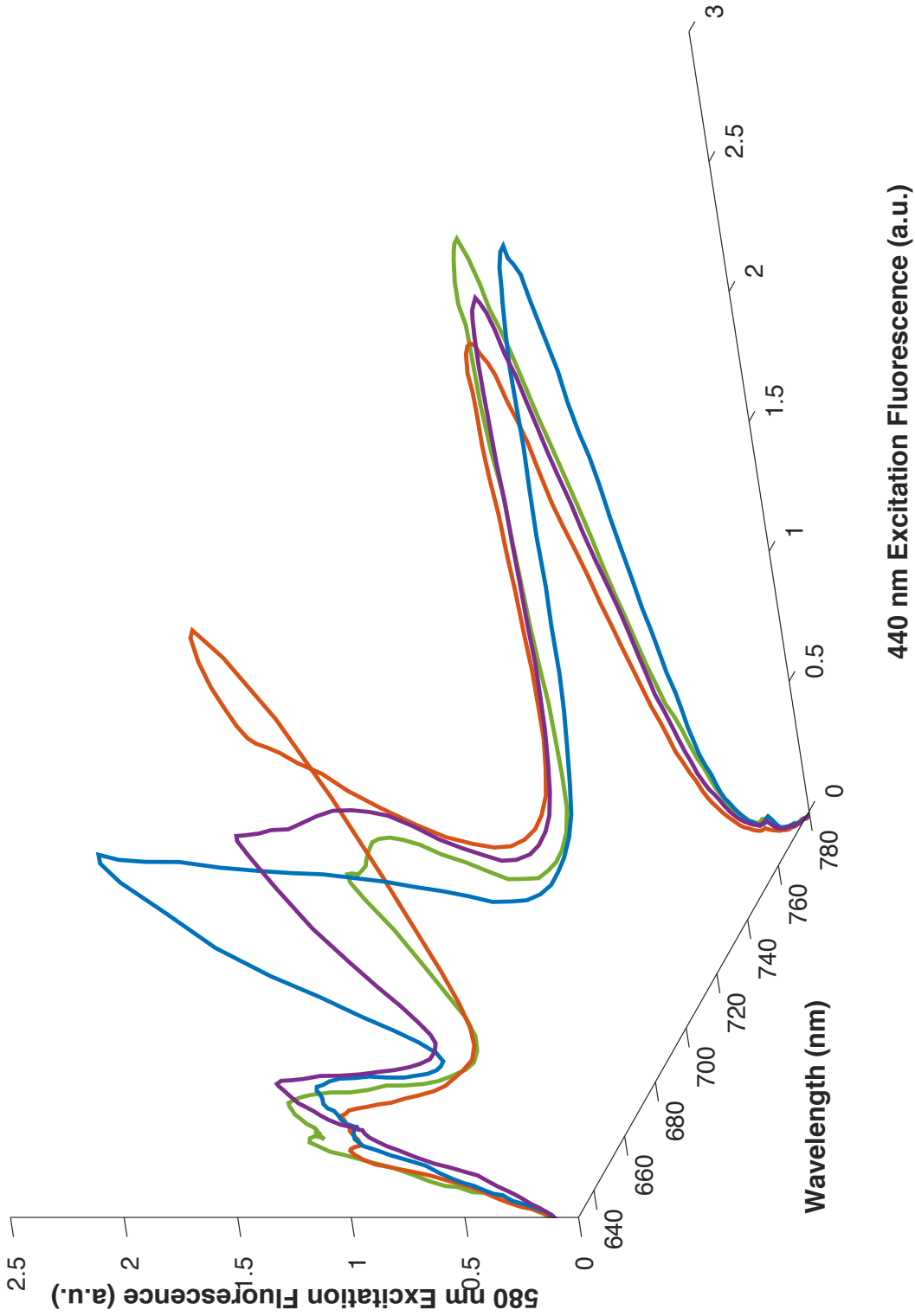

**Figure 5.6 3D Parametric Coordinate System Utilizing 77 K Spectra.** 77 K measurements with an excitation wavelength of 580 nm for WT grown in BG-11 (green), BG-11 + $\frac{1}{10th}$ Fe$^{3+}$ (orange), BG-11 – P$_i$ (blue), and BG-11 – K$_2$HPO$_4$ (purple) for one week.

Therefore, as stated, the total possible combination of 440 and 580 nm spectra was exhausted in order to showcase the reproducibility of this parametric analysis (Figure 5.7). The vector traces of the spectra indicate differences of the angles produced between the parameterized axes (Figure 5.7). The three major complexes involved in photosynthesis, namely the phycobilisome, PS II, and PS I, are clearly illustrated in the parametric curves from left to right, respectively (Figure 4.2.4.1). Positional differences were noticed for the region corresponding to the fluorescence at 685 nm wavelength, so k-means clustering analysis was

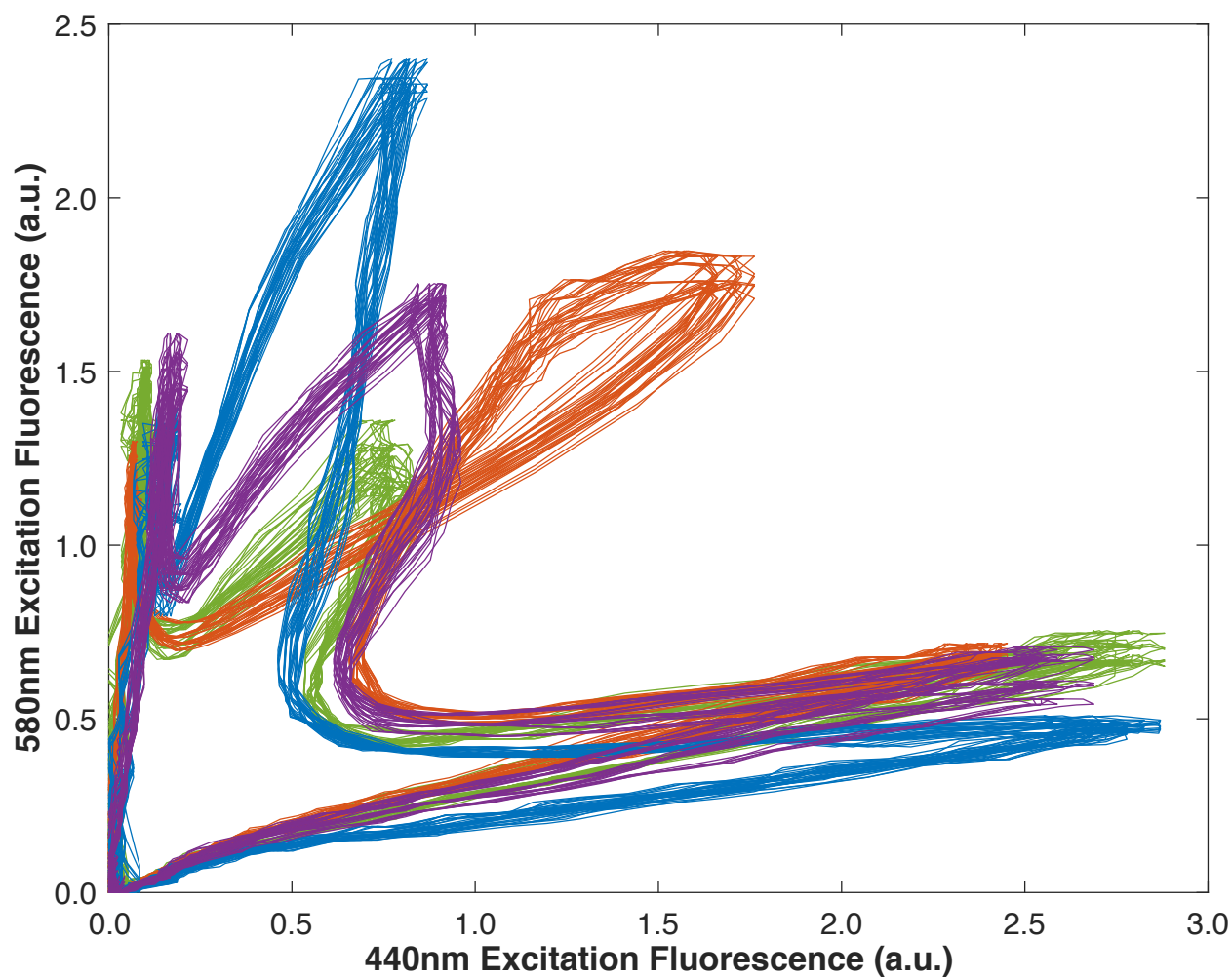

**Figure 5.7 Parametric Curves Generated from 77 K Spectral Data.** 77 K measurements with an excitation wavelength of 580 nm directly plotted against 77 K measurements with an excitation wavelength of 440 nm for WT grown in BG-11 (green), BG-11 + $\frac{1}{10th}$ $Fe^{3+}$ (orange), BG-11 − $P_i$ (blue), and BG-11 − $K_2HPO_4$ (purple) for one week.

carried out in the R environment across the span of $683 - 687$ nm (Figure 5.8). This component analysis showcased that the centroids for the four environmental conditions were in fact differentiable in the parametric space (Figure 5.8). The ($F_{440}$, $F_{580}$) coordinates for the centroids were as follows: (0.72, 1.23) for WT grown in BG-11; (1.59, 1.74) for WT grown in BG-11 +

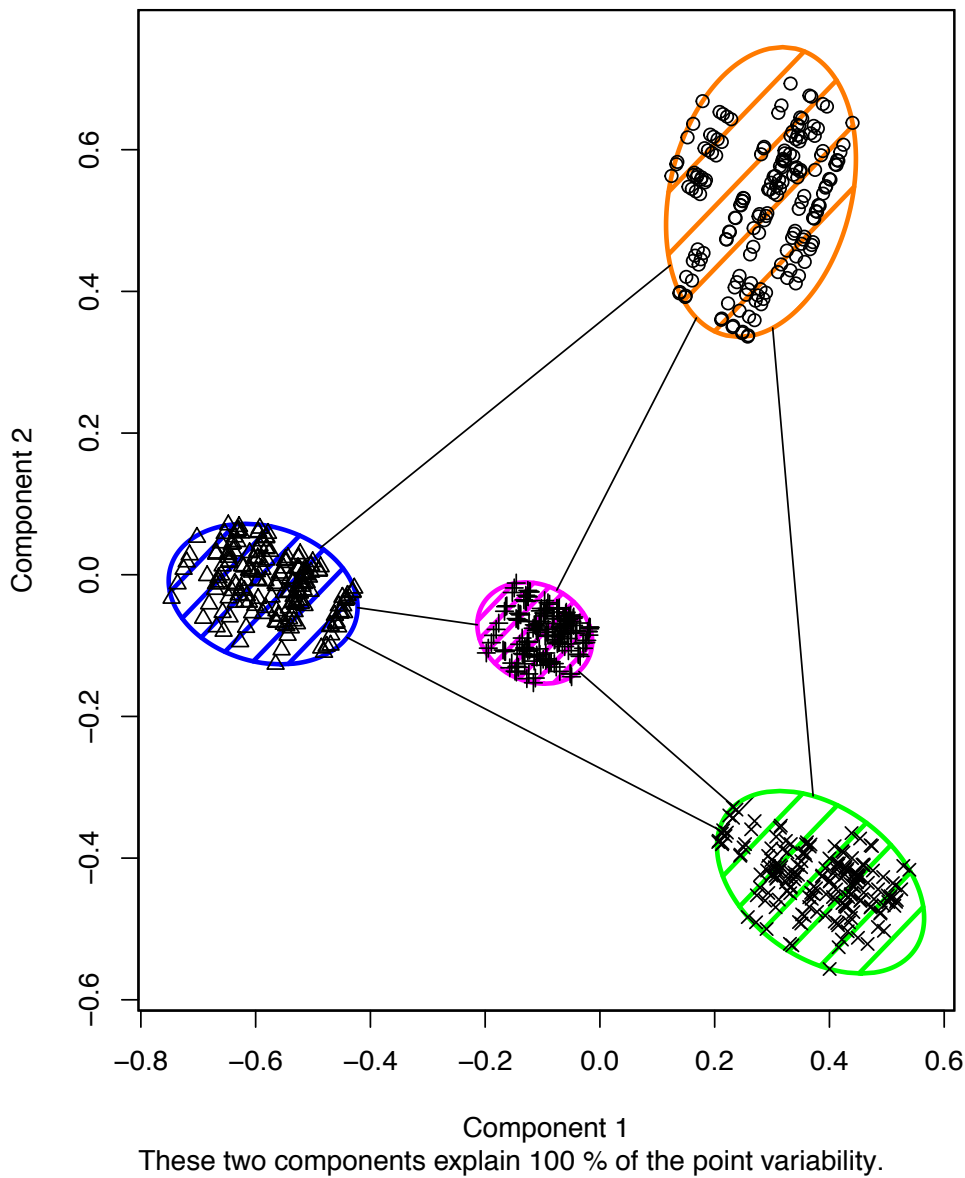

These two components explain 100 % of the point variability.

**Figure 5.8 K Means Clustering Analysis of Parametric Curves.** K means analysis of 683-687 nm fluorescence from 77 K measurements with an excitation wavelength of 580 nm plotted directly against an excitation wavelength of 440 nm of WT grown in BG-11 (green), BG-11 + $\frac{1}{10th}$ $Fe^{3+}$ (orange), BG-11 − $P_i$ (blue), and BG-11 − $K_2HPO_4$ (purple) for one week.

$\frac{1}{10th}$ $Fe^{3+}$; (0.78, 2.27) for WT grown in BG-11 – $P_i$; and (0.88, 1.68) for WT grown in BG-11 – $K_2HPO_4$ (Figure 5.8).

## 5.3 Protein Analysis under Phosphate and Iron Limitation

Fluorescence at 685 nm from the 77 K measurements of WT grown in BG-11 – $P_i$ along with blue-shifts in the spectra warrant the investigation of the possibility that CP43' may play a role in the response of *Synechocystis* 6803.

### 5.3.1 SDS-PAGE of *Synechocystis* 6803 Replete with Phosphate, Replete with Phosphate and Iron-Limited, or Phosphate-Limited

Immunoblotting was performed in order to differentiate the presence of the photosystems and CP43'. The immunoblots indicate a decrease of PsaA for WT grown in BG-11 + $\frac{1}{100th}$ $Fe^{3+}$ and WT grown in BG-11 – $P_i$ seemed comparable to WT grown in BG-11 (Figure 5.9). Moreover, the CP43 subunit has decreased for WT grown in BG-11 + $\frac{1}{100th}$ $Fe^{3+}$ and BG-11 – $P_i$ (Figure 5.9). CP43' was present for WT grown in BG-11 + $\frac{1}{100th}$ $Fe^{3+}$ as well as cells replete with phosphate in BG-11.

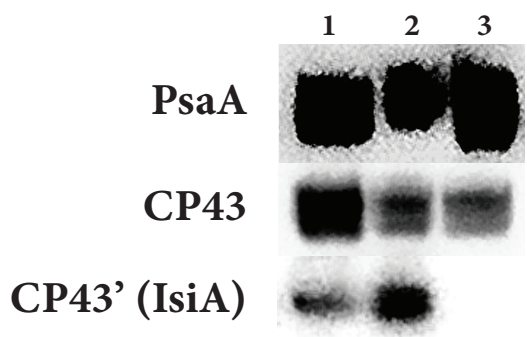

**Figure 5.9 Immunoblotting from SDS-PAGE.** Immunoblotting analyses of SDS-PAGE of WT grown in BG-11 (**1**), BG-11 + $\frac{1}{100th}$ $Fe^{3+}$ (**2**), and BG-11 – $P_i$ (**3**).

### 5.3.2 Blue NATIVE-PAGE

In order to confirm SDS-PAGE results, BN-PAGE was also performed (Figure 5.10). The immunoblots indicate decreased trimeric, dimeric and monomeric PS I from $\alpha$-PsaA for WT grown in BG-11 + $\frac{1}{100th}$ Fe$^{3+}$ (Figure 5.10.A-B). However, $\alpha$-PsaA was found to bind to a region comprised of light-harvesting complexes for WT grown in BG-11 + $\frac{1}{100th}$ Fe$^{3+}$ (Figure 5.10.A-B). Moreover, $\alpha$-PsaA indicated increased monomeric PS I for WT grown in BG-11 − P$_i$ (Figure 5.10.A-B). The binding of $\alpha$-CP43 indicated that PS II dimers and monomers had decreased for WT grown in BG-11 − P$_i$ (Figure 5.10.A; Figure 5.10.C). It was also noted that $\alpha$-CP43 was bound to the PS I region for WT replete with phosphate and iron-limited (Figure 5.10.A; Figure 5.10.C). Finally, $\alpha$-CP43' was only present in BG-11 $+ \frac{1}{100th}$ Fe$^{3+}$ and cells replete with phosphate in BG-11 (Figure 5.10.A; Figure 5.10.D).

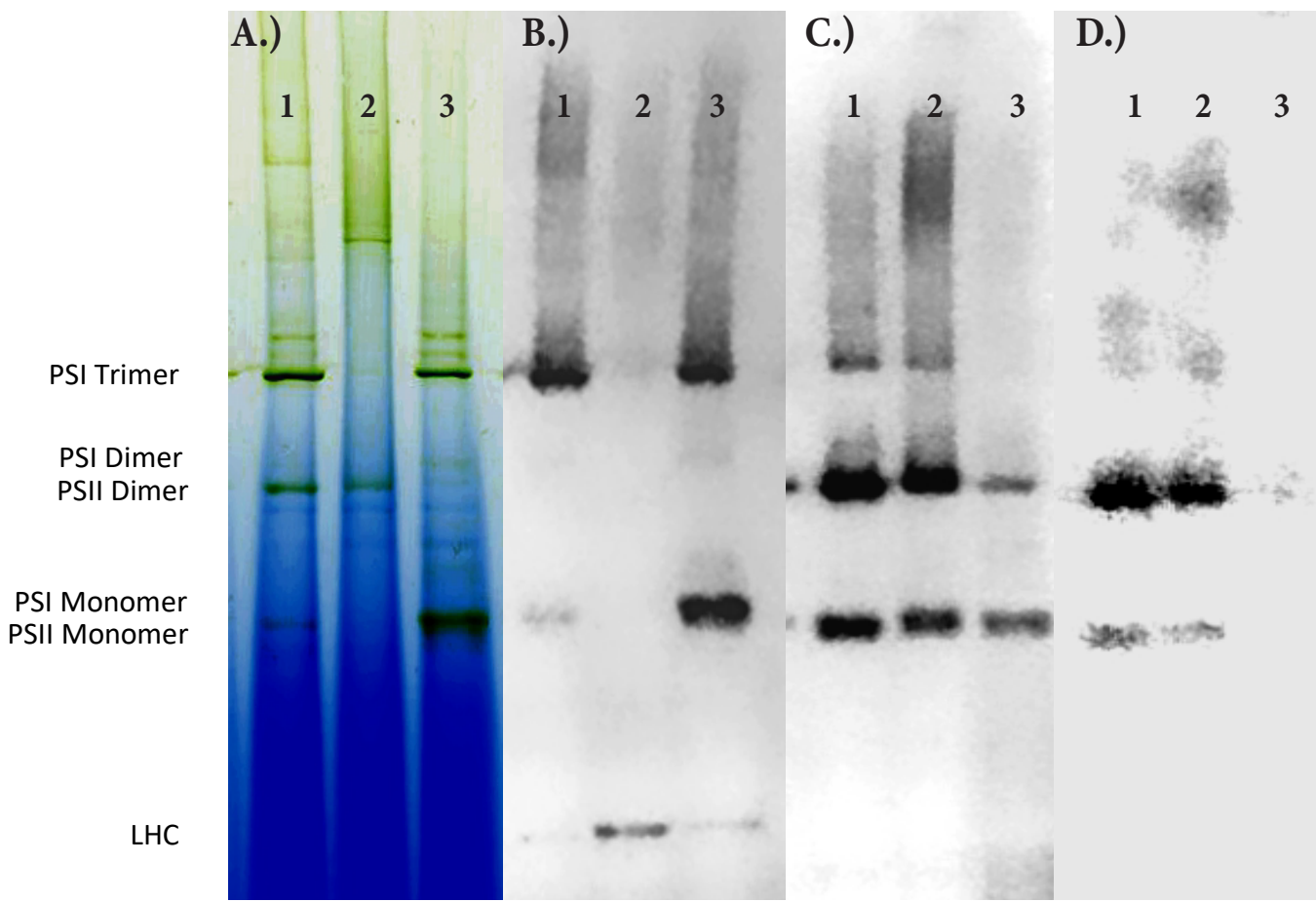

**Figure 5.10 BN-PAGE and Immunoblotting.** Blue native-PAGE (**A**) of WT grown in BG-11 (**1**), BG-11 + $\frac{1}{100th}$ Fe$^{3+}$ (**2**), and BG-11 − P$_i$ (**3**). Immunoblots of the blue native-PAGE gels incubated with $\alpha$-PsaA (**B**), $\alpha$-CP43 (**C**), and $\alpha$-CP43' (**D**).

### 5.3.3 77 K Measurements of Isolated Thylakoid Membranes or Solubilized Native Proteins

In order to further analyze the SDS- and BN-PAGE gels along with their subsequent immunoblotting, the fluorescence was measured from intact isolated thylakoid membranes as well as the sole native proteins utilized in the BN-PAGE itself (Figure 5.11). The 440 nm excitation wavelength measurements indicated differences in the ratio of ~685:725 nm peaks (Figure 5.11.A-B). Isolated thylakoid membranes from WT grown in BG-11 indicated that that the ratio of ~685:725 nm vastly increase when proteins are solubilized for BN-PAGE (Figure 5.11.A-B). Moreover, WT grown with limiting conditions follow the trend; however, the traces from WT grown in BG-11 $+ \frac{1}{100th}$ Fe$^{3+}$ indicated that the initial ratio of ~685:725 nm peaks was lower compared to that of WT grown in BG-11 with intact thylakoid membranes and, once removed, the ratio relied more heavily on the 685 nm peak (Figure 5.11.A-B). The traces produced by WT grown in BG-11 – P$_i$ were quite messy with intact thylakoid membranes but became clearer after their removal with a comparable ~725 nm peak with that of WT grown in BG-11 (Figure 5.11.A-B). Yet, the 685 nm peak was widened and blue-shifted by ~6 nm for WT grown in BG-11 – P$_i$ as compared to WT grown in BG-11 (Figure 5.11.B).

As for the fluorescence measurements obtained with an excitation wavelength of 580 nm, most of the emission was measured from 630-660 nm (Figure 5.11.C-D). The traces from WT grown in BG-11 indicate that the ~645:660 nm ratio decreased after the removal of the thylakoid membranes (Figure 5.11.C-D). Traces for WT grown in BG-11 $+ \frac{1}{100th}$ Fe$^{3+}$ indicated a similar ratio for the ~645:660 nm peaks; moreover, fluorescence was emitted at ~685 nm for intact thylakoids and subsequently decreased after their removal (Figure 5.11.C-D). Traces of WT grown in BG-11 $-$ $P_i$ indicated an initial decrease of the ~645 and ~660 nm peaks from intact thylakoids as compared to WT grown in BG-11 (Figure 5.11.C-D). The ratio of ~645:660 increased for WT grown in BG-11 $-$ $P_i$ after the removal of thylakoid membranes, which was in contrast to WT grown in BG-11 (Figure 5.11.D). In addition, traces obtained from WT grown in BG-11 $-$ $P_i$ emitted fluorescence at ~725 nm for both intact thylakoids and native proteins as well as the emergence of a peak at ~671 nm (Figure 5.11.C-D).

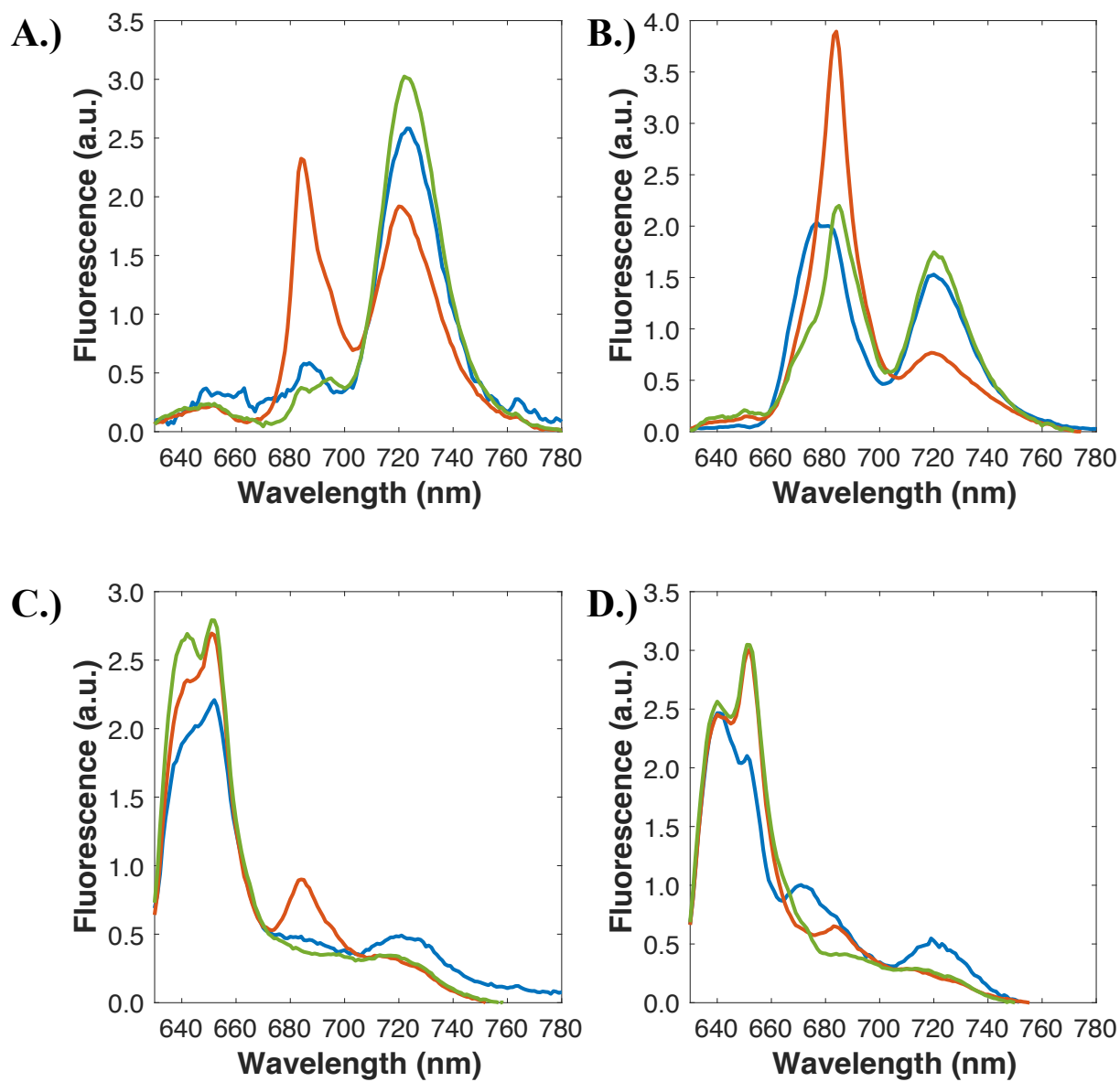

**Figure 5.11 Low-Temperature (77 K) Fluorescence Emission Spectroscopy of Isolated Thylakoids and Solubilized Proteins.** 77 K measurements of WT grown in BG-11 (green), BG-11 + $\frac{1}{100th}$ Fe$^{3+}$ (orange), and BG-11 − P$_i$ (blue) with an excitation wavelength of 440 nm (**A-B**) of isolated thylakoid membranes (**A**) and solubilized native proteins (**B**). 77 K measurements with an excitation wavelength of 580 nm (**C-D**) of isolated thylakoid membranes (**C**) and solubilized native proteins (**D**).

### 5.3.4 Deconvolution of Spectral Measurements with an Excitation Wavelength of 440 nm from Isolated Thylakoids and Native Proteins

To further analyze the 77 K measurements from intact thylakoid membranes and native proteins, the spectra were deconvoluted to indicate the emergent property of the measurements from peaks assuming that they follow a single gaussian (Figure 5.12). The traces were fitted with 4 gaussians and produced fitted models with an $r^2 \geq 0.98$ for each environmental condition. Individual peaks were summarized for each environmental condition (Table 5.2).

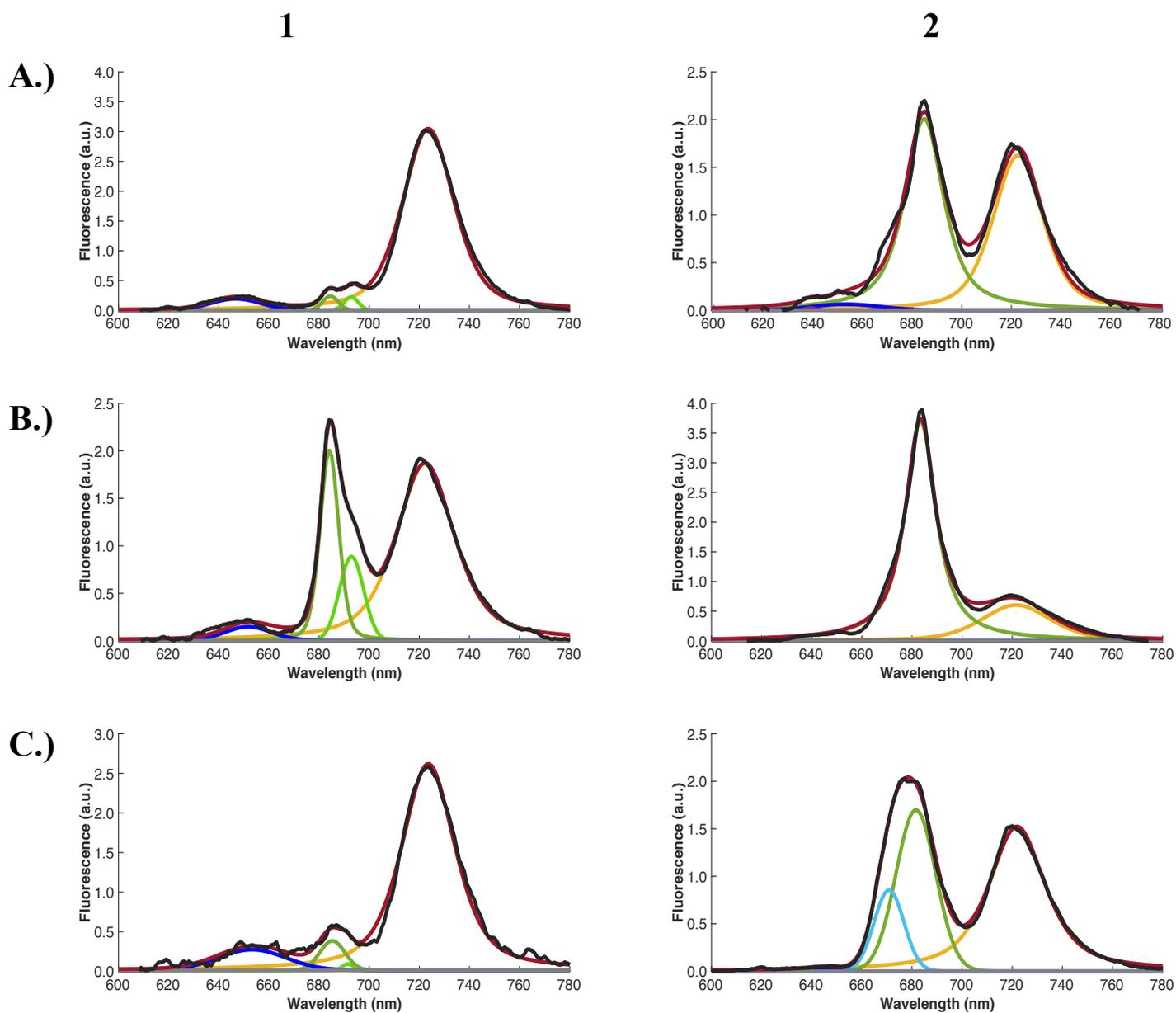

**Figure 5.12 Deconvolution of Low-Temperature (77 K) Fluorescence Emission Spectra with 440 nm Excitation.** Deconvolution of 77 K measurements with an excitation wavelength of 440 nm taken from WT grown in BG-11 (**A**), BG-11 + $\frac{1}{100th}$ $Fe^{3+}$ (**B**), and BG-11 − P$_i$ (**C**) of isolated thylakoids (**1**) and native proteins (**2**).

*Table 5.2 Overview of the Peaks Obtained from Spectral Deconvolution of Traces Measured with an Excitation Wavelength of 440 nm*

| Environmental Condition from which Thylakoids were Isolated: | Peak Number: | Center of Gaussian (nm): | Width of Gaussian (nm): | Area of Gaussian: | Environmental Condition from which Native Proteins were Isolated: | Peak Number: | Center of Gaussian (nm): | Width of Gaussian (nm): | Area of Gaussian: |
|---|---|---|---|---|---|---|---|---|---|
| BG-11 | 1 | 646.74 ± 2.36 | 23.83 ± 6.65 | 4.93 | BG-11 | 1 | 638.73 ± 8.93 | 4.98 ± 25.59 | 0.14 |
| BG-11 + $Fe^{3}$ | 1 | 651.73 ± 1.93 | 21.98 ± 5.52 | 3.77 | BG-11 + $Fe^{3}$ | 1 | 638.67 | 0.44 | 0.0005 |
| BG-11 − $P_i$ | 1 | 653.25 ± 3.90 | 31.97 ± 9.84 | 8.92 | BG-11 − $P_i$ | 1 | 643.18 ± 9.39 | 12.86 ± 25.00 | 0.27 |
| BG-11 | 2 | 684.54 ± 2.14 | 7.42 ± 4.76 | 1.9 | BG-11 | 2 | 653.25 ± 35.49 | 33.28 ± 54.37 | 2.28 |
| BG-11 + $Fe^{3}$ | 2 | 684.15 ± 0.40 | 8.81 ± 0.51 | 21.34 | BG-11 + $Fe^{3}$ | 2 | 646.75 | 9.74E-05 | 1.02E-11 |
| BG-11 − $P_i$ | 2 | 685.30 ± 7.11 | 11.81 ± 11.21 | 4.7 | BG-11 − $P_i$ | 2 | 670.85 ± 4.02 | 13.98 ± 3.27 | 12.74 |
| BG-11 | 3 | 692.82 ± 2.16 | 6.65 ± 4.50 | 1.59 | BG-11 | 3 | 684.79 ± 0.39 | 18.78 ± 1.20 | 54.45 |
| BG-11 + $Fe^{3}$ | 3 | 692.97 ± 1.22 | 11.41 ± 1.96 | 10.79 | BG-11 + $Fe^{3}$ | 3 | 683.42 ± 0.12 | 14.73 ± 0.44 | 82.81 |
| BG-11 − $P_i$ | 3 | 691.53 ± 4.82 | 5.08 ± 20.92 | 0.44 | BG-11 − $P_i$ | 3 | 681.58 ± 5.20 | 18.69 ± 5.70 | 33.82 |
| BG-11 | 4 | 723.60 ± 0.18 | 24.45 ± 0.57 | 97.75 | BG-11 | 4 | 722.68 ± 0.48 | 23.90 ± 1.44 | 50.76 |
| BG-11 + $Fe^{3}$ | 4 | 722.34 ± 0.28 | 27.60 ± 0.83 | 70.37 | BG-11 + $Fe^{3}$ | 4 | 721.59 ± 1.15 | 32.59 ± 3.24 | 25.12 |
| BG-11 − $P_i$ | 4 | 723.60 ± 0.27 | 25.44 ± 0.87 | 89.89 | BG-11 − $P_i$ | 4 | 721.90 ± 0.33 | 26.97 ± 0.96 | 58.29 |

## 5.3.5 Deconvolution of Spectral Measurements with an Excitation Wavelength of 580 nm from Isolated Thylakoids and Native Proteins

Deconvolution was next performed on the spectra obtained with an excitation wavelength of 580 nm (Figure 5.13). The traces were also fitted utilizing 4 gaussians and produced fitted models with an $r^2 \geq 0.98$ for each environmental condition. Individual peaks deconvoluted from spectral measurements were summarized (Table 5.3).

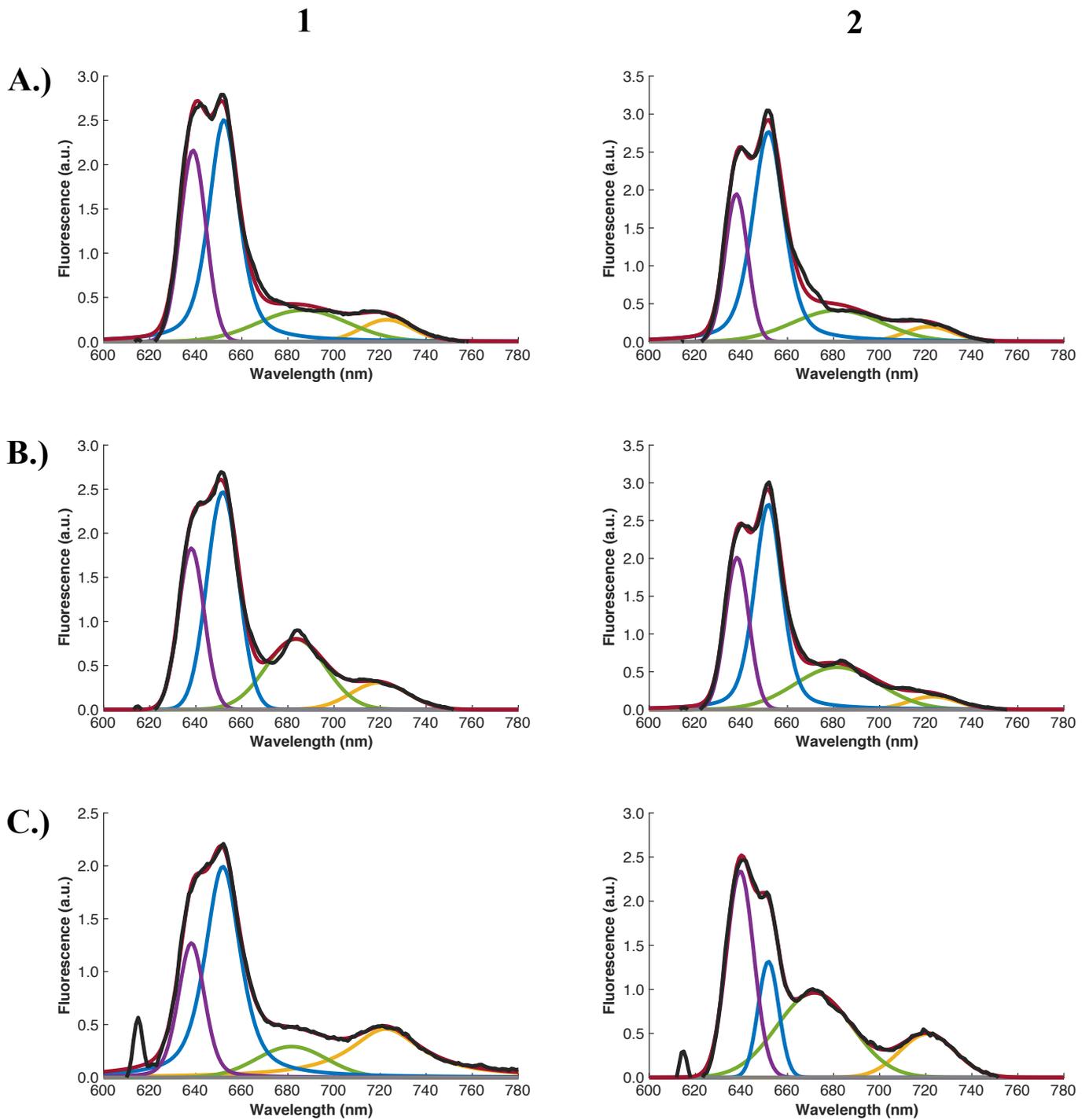

**Figure 5.13 Deconvolution of Low-Temperature (77 K) Fluorescence Emission Spectra with 580 nm Excitation.** Deconvolution of 77 K measurements with an excitation wavelength of 580 nm taken from WT grown in BG-11 (**A**), BG-11 + $\frac{1}{100th}$ Fe$^{3+}$ (**B**), and BG-11 − P$_i$ (**C**) of isolated thylakoids (**1**) and native proteins (**2**).

Table 5.3 Overview of the Peaks Obtained from Spectral Deconvolution of Traces Measured with an Excitation Wavelength of 580 nm

| Environmental Condition from which Thylakoids were Isolated: | Peak Number: | Center of Gaussian (nm): | Width of Gaussian (nm): | Area of Gaussian: | Environmental Condition from which Native Proteins were Isolated: | Peak Number: | Center of Gaussian (nm): | Width of Gaussian (nm): | Area of Gaussian: |
|---|---|---|---|---|---|---|---|---|---|
| BG-11 | 1 | $638.79 \pm 0.62$ | $13.24 \pm 1.05$ | 30.47 | BG-11 | 1 | $637.82 \pm 0.49$ | $11.56 \pm 1.11$ | 23.96 |
| BG-11 + Fe$^3$ | 1 | $638.05 \pm 0.85$ | $13.07 \pm 1.04$ | 25.47 | BG-11 + Fe$^3$ | 1 | $638.14 \pm 0.37$ | $12.03 \pm 0.78$ | 25.76 |
| BG-11 − P$_i$ | 1 | $637.82 \pm 1.198$ | $13.35 \pm 2.85$ | 21.92 | BG-11 − P$_i$ | 1 | $639.46 \pm 0.63$ | $13.91 \pm 1.17$ | 34.67 |
| BG-11 | 2 | $652.25 \pm 0.68$ | $15.53 \pm 3.77$ | 51.39 | BG-11 | 2 | $651.73 \pm 0.52$ | $16.27 \pm 2.82$ | 57.76 |
| BG-11 + Fe$^3$ | 2 | $651.73 \pm 0.75$ | $16.12 \pm 2.12$ | 42.34 | BG-11 + Fe$^3$ | 2 | $651.68 \pm 0.34$ | $14.28 \pm 1.31$ | 48.58 |
| BG-11 − P$_i$ | 2 | $651.50 \pm 1.158$ | $18.16 \pm 4.66$ | 45.06 | BG-11 − P$_i$ | 2 | $651.73 \pm 0.82$ | $10.54 \pm 2.05$ | 14.76 |
| BG-11 | 3 | $686.55 \pm 8.73$ | $45.50 \pm 129.56$ | 17.19 | BG-11 | 3 | $681.58 \pm 11.97$ | $45.50 \pm 86.48$ | 19.99 |
| BG-11 + Fe$^3$ | 3 | $683.40 \pm 1.74$ | $30.48 \pm 12.95$ | 25.91 | BG-11 + Fe$^3$ | 3 | $681.58 \pm 4.12$ | $42.86 \pm 28.33$ | 25.59 |
| BG-11 − P$_i$ | 3 | $680.08 \pm 15.45$ | $37.71 \pm 45.05$ | 12.8 | BG-11 − P$_i$ | 3 | $671.68 \pm 3.88$ | $38.39 \pm 8.03$ | 39.25 |
| BG-11 | 4 | $722.86 \pm 10.97$ | $26.07 \pm 31.32$ | 6.85 | BG-11 | 4 | $721.35 \pm 11.25$ | $24.64 \pm 29.35$ | 5.18 |
| BG-11 + Fe$^3$ | 4 | $719.57 \pm 9.12$ | $28.87 \pm 16.09$ | 9.36 | BG-11 + Fe$^3$ | 4 | $723.08 \pm 8.43$ | $25.11 \pm 20.02$ | 4.6 |
| BG-11 − P$_i$ | 4 | $722.48 \pm 7.52$ | $37.77 \pm 15.52$ | 27.3 | BG-11 − P$_i$ | 4 | $720.75 \pm 2.23$ | $28.09 \pm 5.23$ | 14.78 |

# Chapter Six: Discussion of Photosynthetic Apparatus Response to Phosphate and Iron Limitation

## Key Targets of Chlorophyll Biogenesis and Photosynthetic Apparatus of *Synechocystis* 6803 under Phosphate-Deficiency

By analyzing the fluctuations of Chlorophyll *a* as well as the photosynthetic apparatus over the course of one week, it was seen that the tetrapyrrole biogenesis pathway was being hindered or altered. Tetrapyrrole biosynthesis in the early stages is identical for both the heme and phycobilin biosynthetic pathway (Beale, 1993). Since there are fluctuations in Chlorophyll *a* and 77 K measurements at both 440 and 580 nm excitation wavelengths, then this likely indicates that these early stages are being targeted – this can be confirmed via a time course of RT-qPCR of *slr1790*, a Protox homologue, as well as other genes encoding proteins essential for the initial stages of protoporphyrin IX biosynthesis (Bollivar and Beale, 1996; Kato *et al.*, 2010). In addition, a time course of the photosynthetic apparatus from *Synechocystis* 6803 could be analyzed via BN-PAGE or at least at the indicated time points to further analyze the "collapse" of the photosynthetic apparatus.

## Potassium Required for Phosphate-Deficiency Responses Previously Established

Cations, such as $K^+$, are vital for cellular integrity as they are cofactors for reactions, ligands utilized in protein stability, and maintain homeostasis (Berry *et al.*, 2003; Matsuda and Uozumi, 2006; Nanatani *et al.*, 2015). When grown in BG-11 – $K_2HPO_4$, *Synechocystis* 6803 responded differently than that of being grown in BG-11 – $P_i$, from whole-cell spectral analysis, low-temperature (77 K) fluorescence emission spectroscopy, growth rates, and even alkaline phosphatase activity. These results have broader implications than will be mentioned, at least philosophically, yet how or why these responses happen is only speculative at most in this thesis. This further perpetuates the need of scientific study within this area.

**Low-Temperature (77 K) Fluorescence Emission Spectroscopy**

Photosynthesis is typically analyzed via two independent pigment associations, namely Chl *a* and phycobilins. In order to differentiate the interactions between Chl *a*-binding proteins and phycobilisome-associations with these complexes, the spectra were parameterized within $\mathbb{R}^3$ as this would allow for the curves produced by three-dimensional vectors to warp within the 3D space so that clear differences could be noted between the innate duality between the two spectra. This parameterization produced notable differences, especially for the key CP43 and CP47 subunits of PS II with emission at 685 and 695 nm, respectively. The 695 nm emission is of particular importance as it corresponds to functional PS II and is sometimes masked when collecting the spectra – the parameterization allows a clear indication of the fluorescence emission through the bisecting angle formed specifically between the 580 nm excitation wavelength and 440 nm excitation wavelength axes. There are other nuance differences, yet the main point for this study was to indicate the differences between the phosphate and iron deficiency.

**Reconstruction of the Photosynthetic Apparatus in *Synechocystis* 6803 under Phosphate-Deficiency**

Though there are nuances, the main responses of *Synechocystis* 6803 are the uncoupling of the phycobilisome and potential reliance on cyclic electron transport or cellular respiration – all of which likely stems from hinderance of phosphatidylglycerol (PG) synthesis. This can be analyzed through RT-qPCR of *pgsA* once again over a phosphate-deficient time course. Though there are blue-shifts in the absorption and fluorescence spectra, these observations were not due to CP43' associated with PS I. Deconvolution revealed decreased phycocyanin and a peak at 670 nm in the 580 nm excitation spectra – this is most likely due to differences arising in the phycobilisome architecture of the core and terminal emitters when proteins were

**114**

solubilized. Moreover, it should be stated that the process of isolating native proteins does alter ratios between PS II: PS I – this should be further analyzed as well since the spectra were obtained from a singular isolate of thylakoid membranes.

## Chapter 7: Concluding Remarks

Metabolic processes are typically analyzed as if mutually exclusive – this is referring to how the transcription of a gene of interest is disrupted in some way to elucidate the protein's function within a particular process. However, processes are interconnected through downstream implications due to the alterations of probabilities associated with protein-protein interactions or with various molecules. When cells are regarded in terms of energetic efficiency in maintaining homeostasis, each of the enzymatic processes contribute to the organism's overall fitness. When transcription is disrupted or the gene is altered, the lack of or mutated protein sometimes affects not only one process but also others, giving rise to an entire cascade of metabolic changes so as to ensure homeostasis in terms of energetic expenditure and, thus, survival.

Phosphate is an essential nutrient, necessary for cellular integrity as bonds from $P_i$ moieties serve as an energetic currency. ATP production relies on photosynthesis within *Synechocystis* 6803, so the rationale follows that acclimatization to phosphate deficiency would affect the photosynthetic apparatus through some form of interconnected regulation.

Photosynthesis is initiated via photonic capture. The network of chlorophyll *a* associations within PS II and PS I can be utilized directly. However, photonic capture is optimized within cyanobacteria and red algae as the phycobilisome, associated with phycobilins, increases the volume, area, and, by extension, the probability of interacting with photons as well as extends the energetic range of utilized photons which are subsequently channeled down the supercomplex architecture and transferred to PS II for the extraction of $e^-$ from water. As $e^-$ flow through the photosynthetic ETC, a proton motive force is utilized by ATP synthase due the generation of a $H^+$ gradient produced across the thylakoid membrane. If $P_i$ is depleted from the environment, then after a period of time, the reserve of $P_i$ shall also be depleted within each of the cells. As $P_i$ is transferred to ADP in the generation of ATP via ATP synthase, utilizing the aforementioned proton motive force, hinderance of photosynthetic efficiency and subsequent oxidative stress has to be alleviated. The acclimatization strategy of the photosynthetic apparatus was determined ubiquitous as in other forms of oxidative stress, by which the phycobilisome was uncoupled from the thylakoid membrane and the dimerization of PS II was found to decrease: cyclic electron transport was, therefore, utilized. Even though transcription levels were not analyzed in this study, there is cross regulation of the pho regulon and the photosynthetic apparatus – this idea stems from how both SDS-PAGE and BN-PAGE have showcased decreased intensity of CP43 and both PS II dimers and monomers, respectively, in tandem with 77 K fluorescence emission spectroscopy obtained from the removal of the auxiliary sensor SphZ under $K_2HPO_4$ limitation. However, the reader must derive her, his, or their own conclusion.

As for the regulation of the pho regulon, the auxiliary sensor SphZ was determined essential for the SphS-SphR two component system response under phosphate limitation. To overcome the literal barrier in conveying information regarding phosphate depletion from the environment, a periplasmic PBP equipped with a transmembrane is required. The phosphate-binding site was mutated; therefore, definitive results were not provided as the transmembrane and cytosolic motif remained intact. SphX was determined to facilitate in the limitation of $P_i$ luxury uptake as an evolutionary safeguard regarding the community rather than each cell, particularly for cyanobacterial species which reside in freshwater systems. Once again, the reader must derive her, his, or their own conclusion.

Finally, all of the sciences are derived to analyze and differentiate between natural phenomena of interest within a particular area: mathematics has laid the foundation for such implementation. Transitivity was provided by the Greek mathematician Εὐκλείδης (c. 325 – c. 265 BC) as the first common notion:

"Things which are equal to the same thing are also equal to one another."

Parameterized datasets are often used within the biological sciences, such as heat maps, principal component analysis, and Ramachandran plots. Data which are analyzed in all of the natural sciences are bounded: conceptualization of higher mathematical constructs, such as infinity, have lost purpose. Collective thinking predominantly has resided on some form of a plane, stemming back to Εὐκλείδης. So, perhaps linearity of these planes should be altered as the projective plane within the natural sciences is also bounded.

*Chapter 7*

## Appendices
## Appendix 1: GT-O1 Background Confirmation.
Differences between the GT-O1 and GT-O2 background.

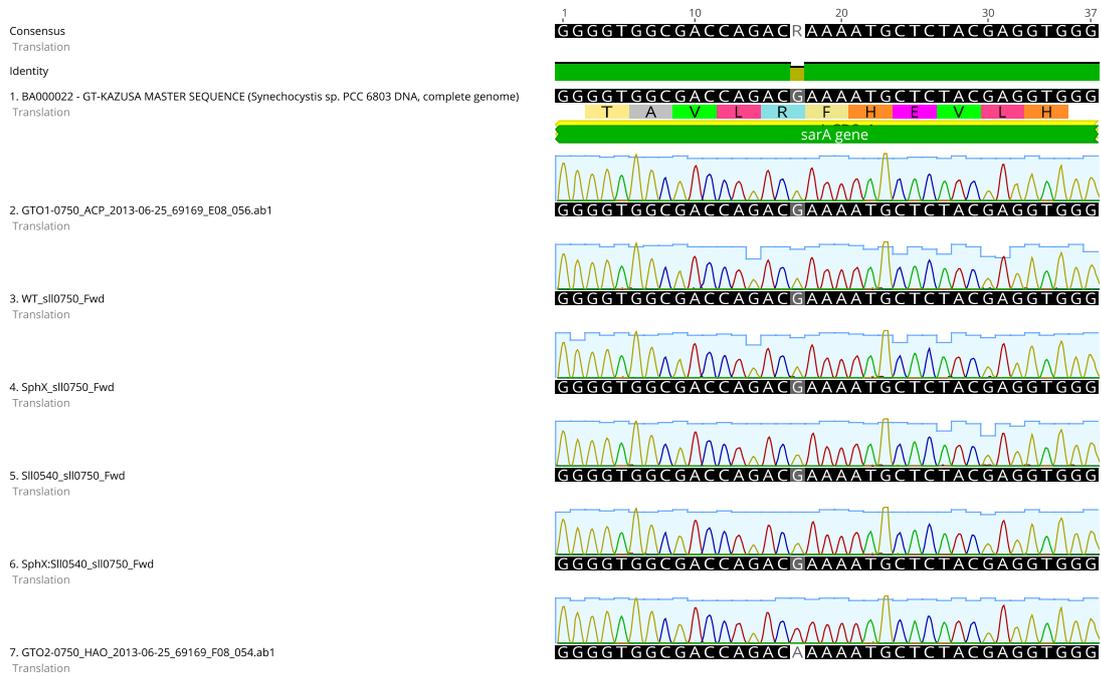

**Figure A.1A.** The *sarA* gene in the GT-O2 background contains a SNP at position 17.

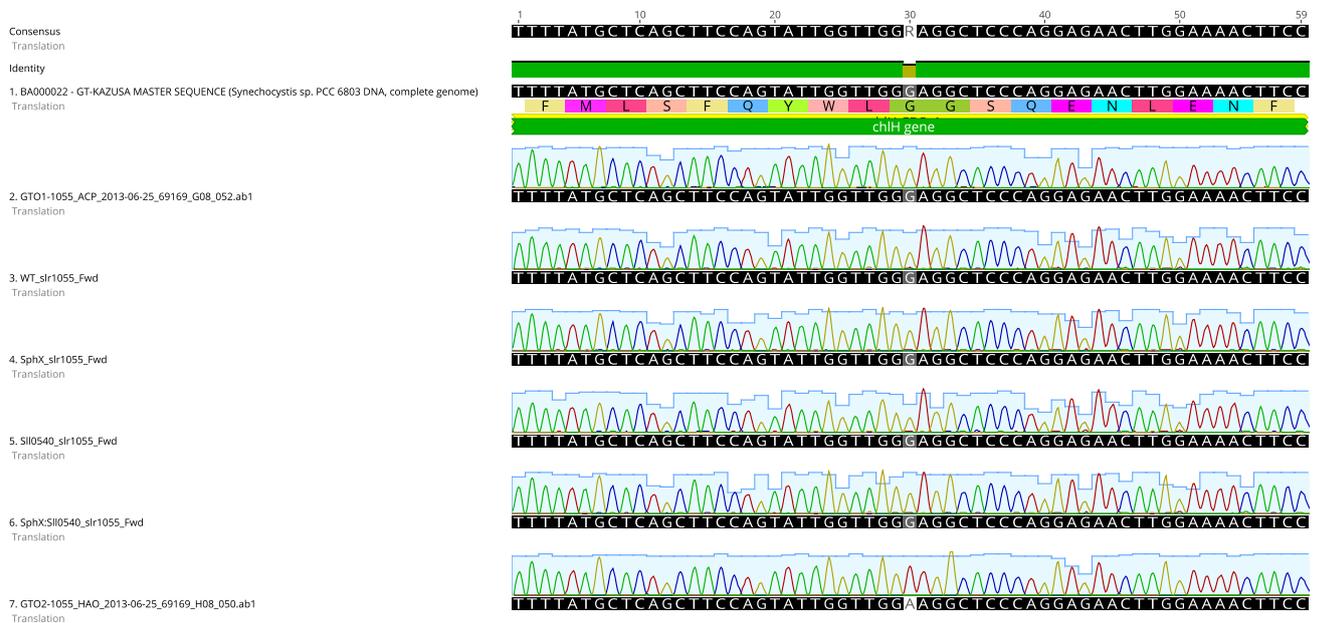

**Figure A.1B.** The *chlH* gene in the GT-O2 background contains a SNP at position 30.

## Appendix 2: Construction of the Δ*sphZ* and Δ*sphX* plasmids

    **Construction of the *sll0540 (sphZ)* interruption strain.** The *sphZ* gene was amplified with the forward primer 5'-GTACCGCAATATGAACTAATTAGACC-3' and the reverse primer 5'-CTAAATAAGCACTATTTGCTCACGGAGCGG-3'. The polymerase chain reaction (PCR) product was ligated into pGEM-T Easy plasmid via TA cloning into a *LacZ* gene for blue-white screening. The *sphZ* gene was then interrupted by the insertion of a chloramphenicol-resistance cassette after a partial restriction digest with SpeI and confirmed via PCR and subsequent sequencing with complementation of the original forward primer with the reverse primer 5'-CGAACCAGGGCAGTTACCTC-3' and the original reverse primer with the forward primer 5'-GGAACTAGCAGATAGATCAACTTTGC -3' (Figure A1). The resulting construct was transformed into *Synechocystis* PCC6803 to produce the Δ*sphZ* deletion strain utilizing established protocols (Williams, 1988; Eaton-Rye and Vermaas, 1991).

    **Construction of the *sll0679* (*sphX*) deletion strain.** A SmaI restriction site was introduced 146 nucleotides upstream of *sphX* start codon utilizing the Quik-Change™ Site-Directed Mutagenesis Kit protocol (Stratagene, La Jolla, CA) with the mutagenic forward primer 5'-CGGCACGATCAA<u>CCCGGG</u>TCTCCAGCCCTTTAACC-3' and the mutagenic reverse primer 5'-GGTTAAAGGGCTGGAGA<u>CCCGGG</u>TTGATCGTGCCG-3'. A kanamycin-resistance cassette was inserted into the introduced SmaI restriction site (Figure A1). Subsequently, the *sphX* start codon was mutated utilizing the Quik-Change™ Site-Directed Mutagenesis Kit protocol (Stratagene, La Jolla, CA) with the mutagenic forward primer 5'-GGAGAAGTAAAA<u>TAG</u>TTTGATTTAAGTCG-3' and the mutagenic reverse primer 5'-CGACTTAAATCAAA<u>CTA</u>TTTTACTTCTCC-3'. This construct was transformed into the ΔPst1 mutant previously described (Burut-Archanai *et al.*, 2009).

*sll0540 (sphZ)*

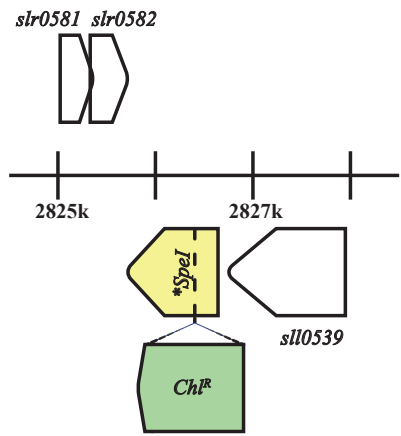

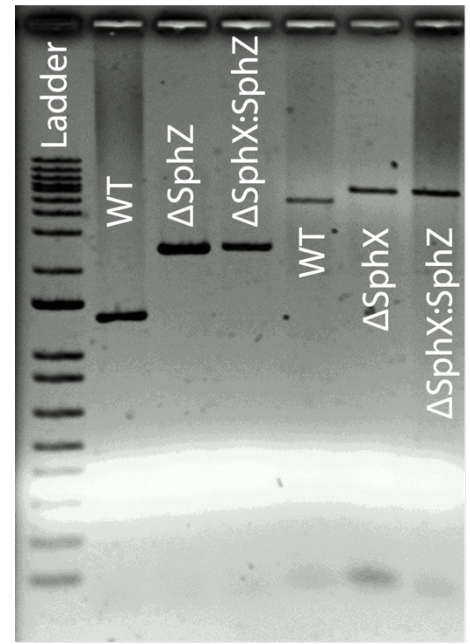

*pst-1*

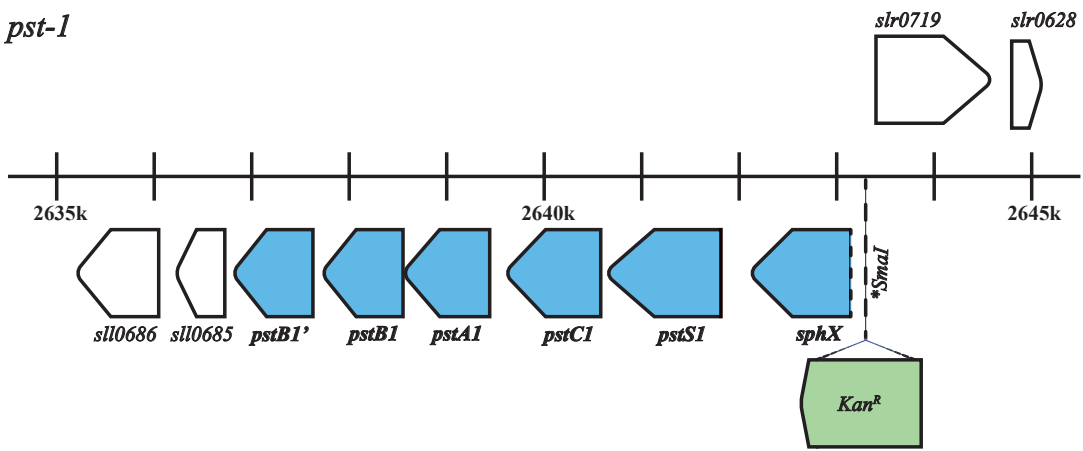

**Figure A.2 Gene Knock-out Constructions.** Gene-knockout construction of *sll0540* (*sphZ*) with an inserted chloramphenicol-resistance cassette via restriction digest of *SpeI* as well as the mutated start codon of *sphX* in the *pst-1* operon with selection pressure asserted by the insertion of a kanamycin-resistance cassette via restriction digest of *SmaI* into a *pst-1* deletion strain. The Pho regulon is represented in blue, antibiotic-resistance cassettes in green, and genes not within the Pho regulon in white.

## Appendix 3: Standard Curve for Total Phosphorus

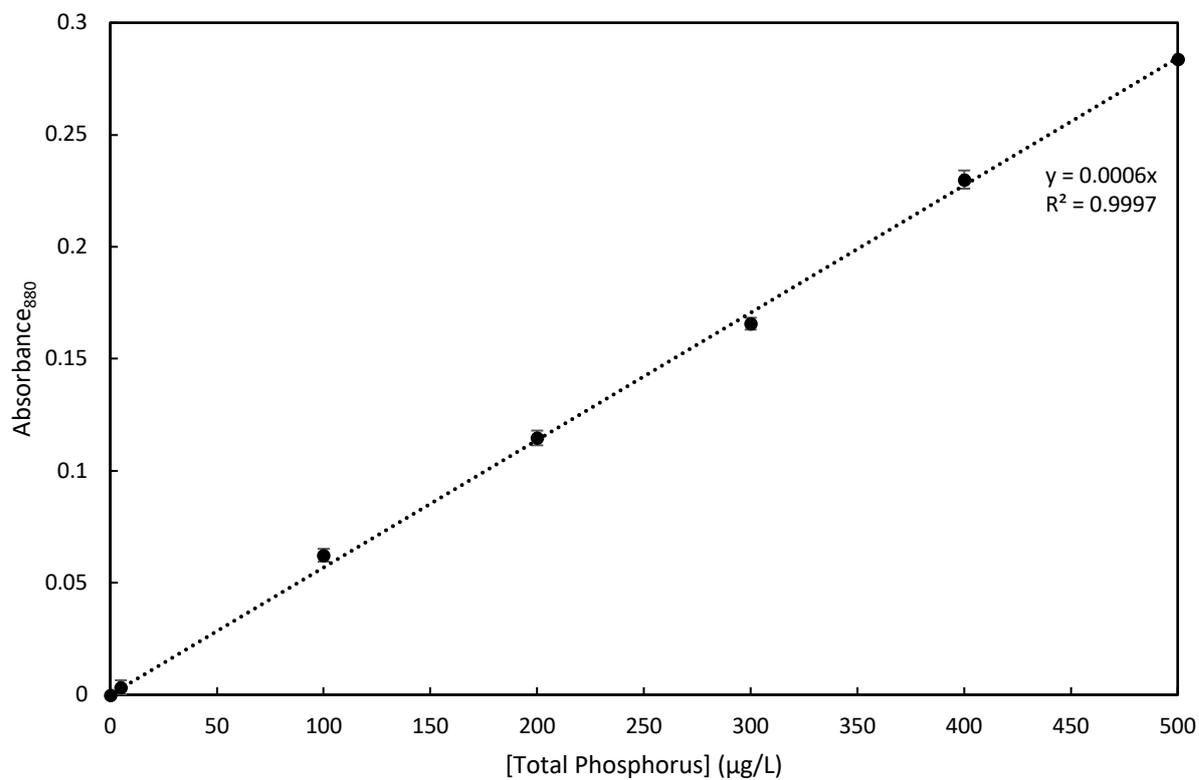

**Figure A.3 Standard Curve for Total Phosphorus.** A standard solution containing 2.5 $\mu g \cdot mL^{-1}$ was diluted to obtain the indicated conentrations. All spectrophotometric measurements at 880 nm were carried out with a Jasco V-550 UV/VIS spectrophotometer (Jasco Inc., USA) in a 1 cm cuvette.

**Appendix 4: Phylogenetic tree of cyanobacterial PBPs, rooted to PstS from *E.coli***

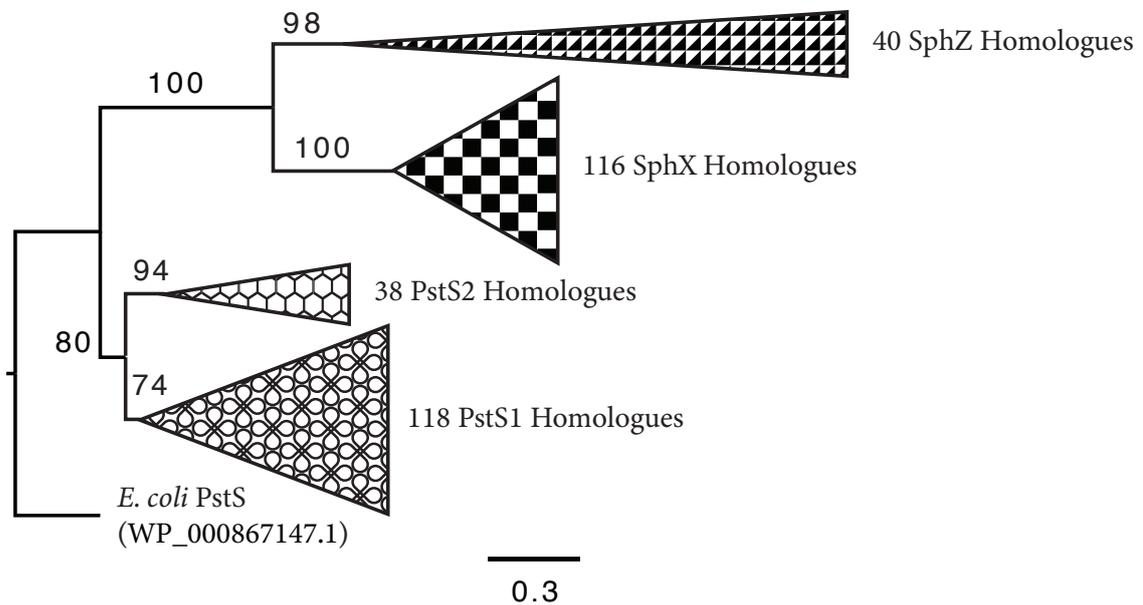

**Figure A.4 Phylogenetic Tree of Cyanobacterial PBP Homologues.** Phylogenetic tree constructed via PhyML software in Geneious on a Clustal Omega alignment of cyanobacterial periplasmic phosphate-binding proteins, rooted with *E. coli* PstS sequence. Indicated values are bootstrap support above 70. Parentheses indicate NCBI accession numbers.

## Appendix 5: Annotated Phylogenetic Tree

**Figure A.5 Phylogenetic Tree of Cyanobacterial PBP Homologues with Extended Branches.** Phylogenetic tree constructed via PhyML software in Geneious on a Clustal Omega alignment of cyanobacterial periplasmic phosphate-binding proteins, rooted with *E. coli* PstS sequence. Indicated values are bootstrap support above 70. Parentheses indicate NCBI accession numbers. Colors indicate morphology and habitat from which the cyanobacterium was isolated.

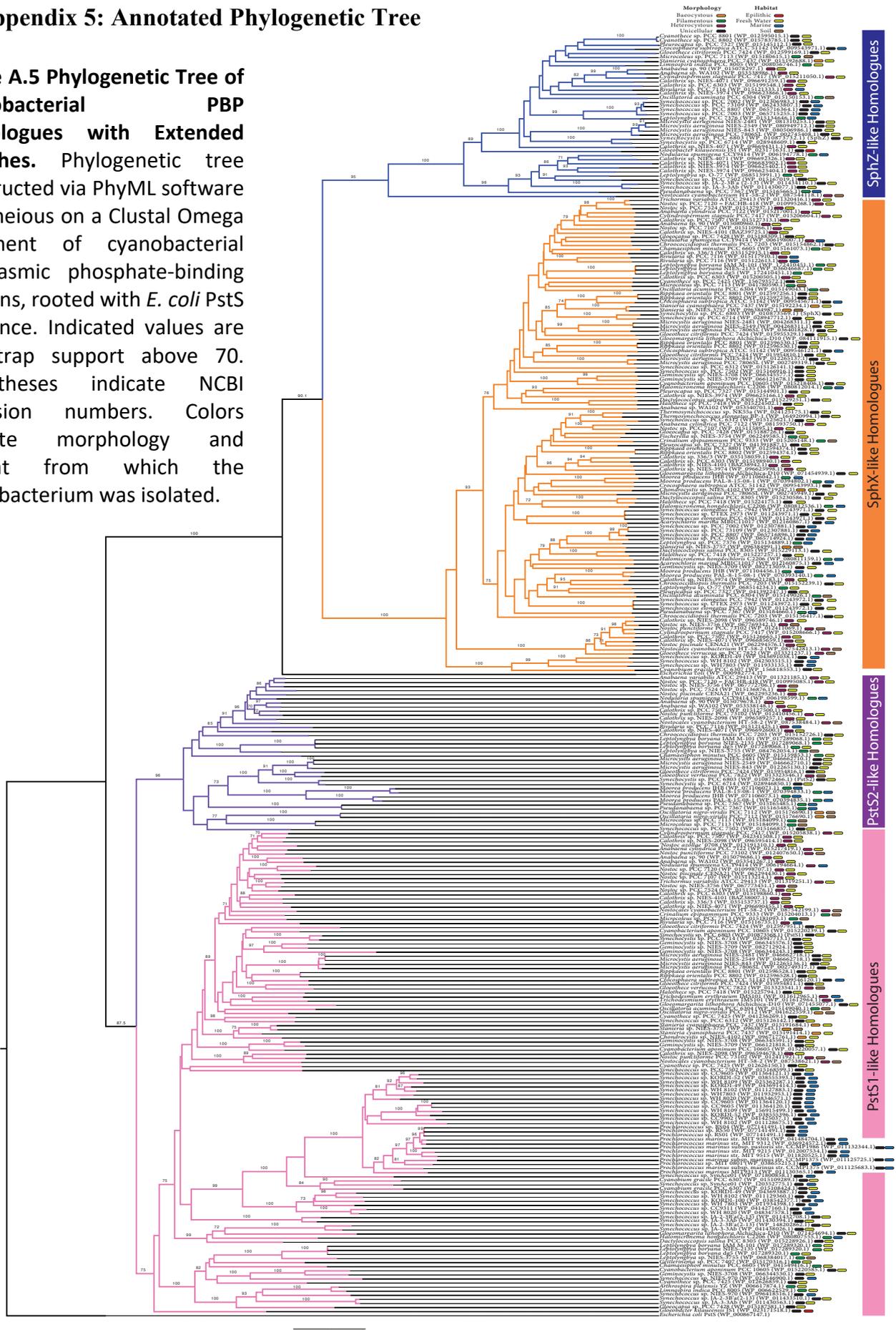

**Appendix 6: Ramachandran Plots**

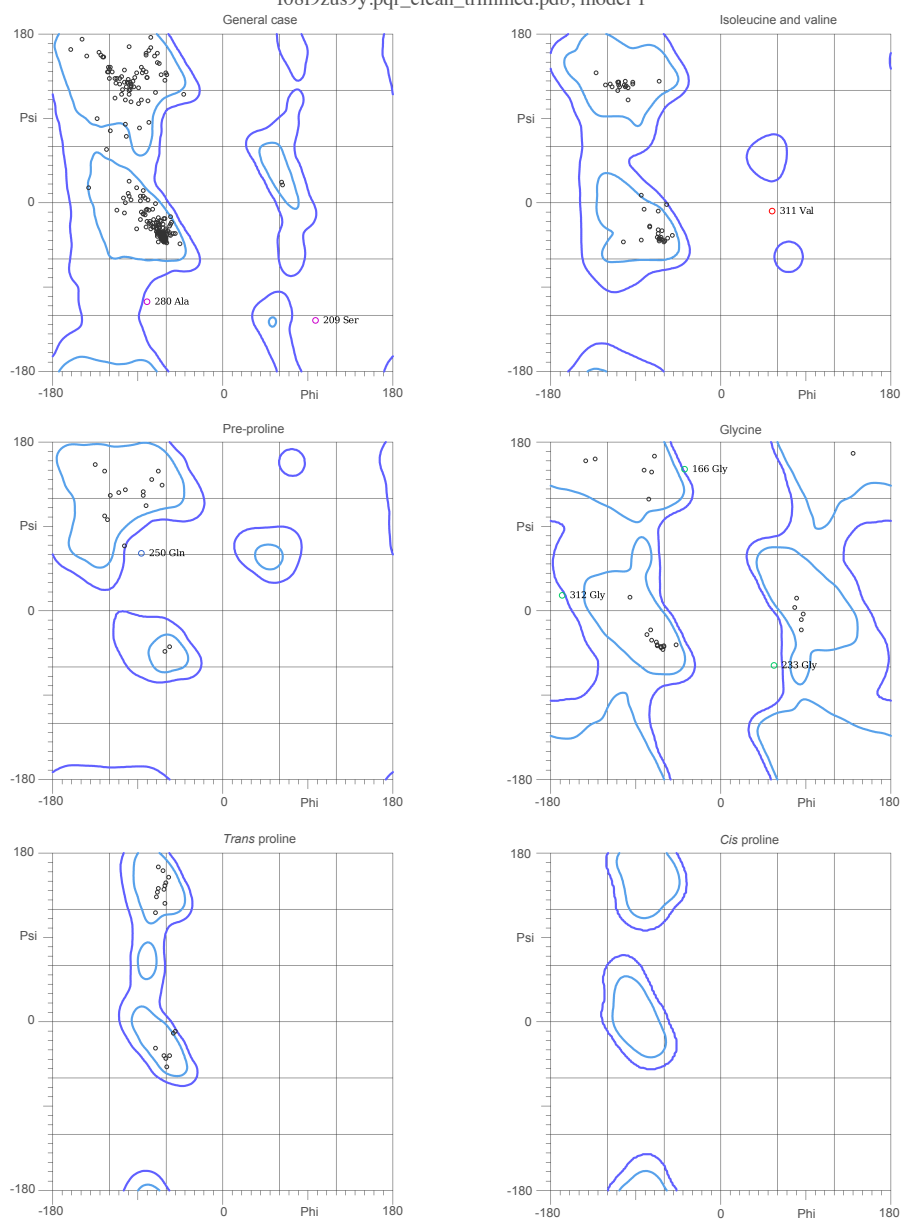

Figure A.6A. Ramachandran plot for the SphX predicted model generated from Raptor X.

## MolProbity Ramachandran analysis

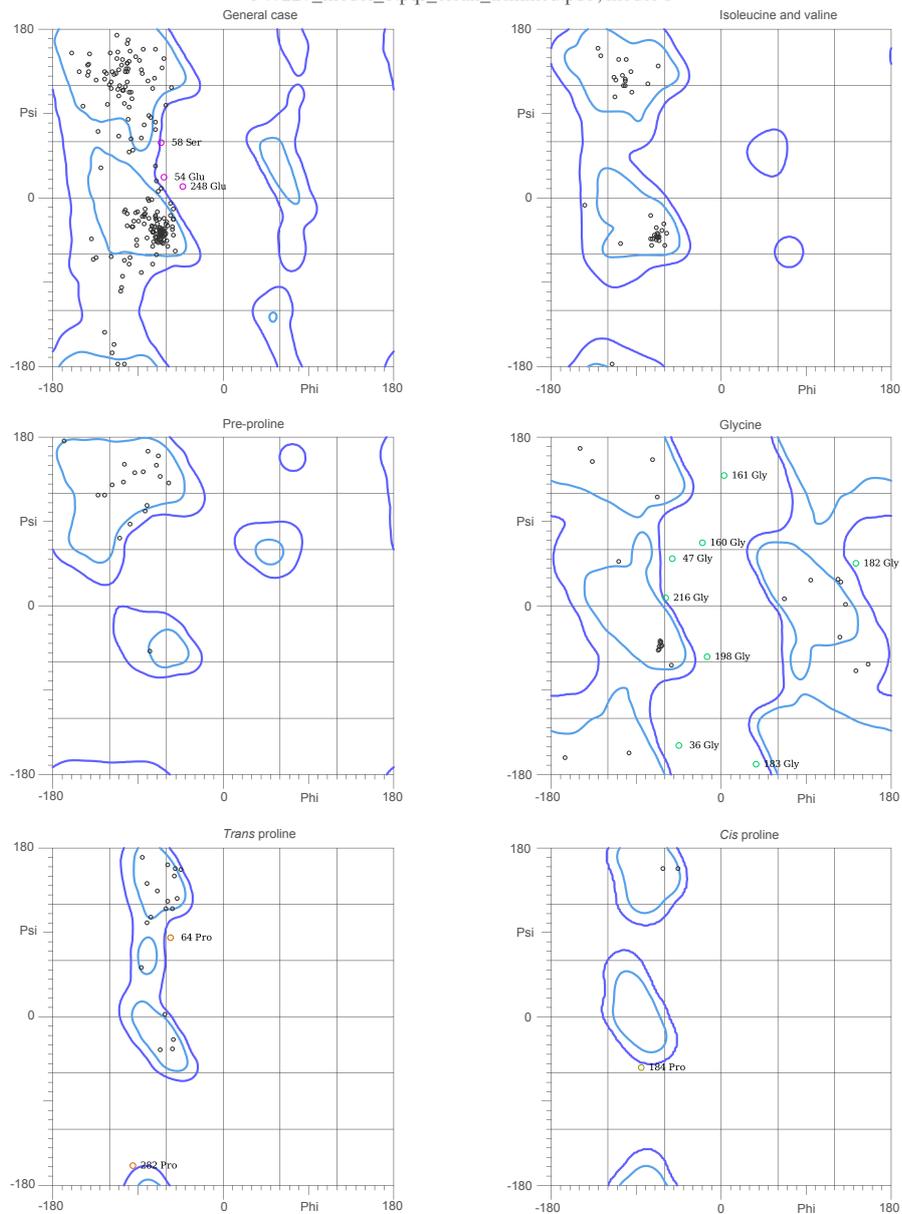

547227_model_1.pqr_clean_trimmed.pdb, model 1

81.0% (247/305) of all residues were in favored (98%) regions.
95.4% (291/305) of all residues were in allowed (>99.8%) regions.

There were 14 outliers (phi, psi):

| | |
|---|---|
| 36 Gly (-45.2, -149.5) | 183 Gly (37.2, -169.8) |
| 47 Gly (-52.5, 51.5) | 184 Pro (-85.4, -55.0) |
| 54 Glu (-63.8, 23.0) | 198 Gly (-15.8, -54.1) |
| 58 Ser (-66.3, 59.7) | 216 Gly (-59.6, 9.1) |
| 64 Pro (-56.8, 85.1) | 248 Glu (-43.6, 12.8) |
| 160 Gly (-20.0, 68.2) | 282 Pro (-96.5, -159.7) |
| 161 Gly (3.9, 140.2) | |
| 182 Gly (143.1, 46.9) | |

**Figure A.6B.** Ramachandran plot for the SphZ predicted model generated from Raptor X.